\pdfoutput=1
\documentclass[12pt,letterpaper]{article}
\usepackage{jheppub}
\usepackage{journals}
\usepackage{graphicx}
\usepackage{nicefrac}
\usepackage{bm}
\usepackage{wasysym}

\def\Nfour	{\mathcal{N}\,{=}\,4}
\def\Nc		{N_{\rm c}}

\def\half	{\tfrac {1}{2}}
\def\eq		{\,{=}\,}
\def\ghat	{\widehat{g}}
\def\Ytilde	{\widetilde{Y}}

\begin		{document}

\title		{Asymmetric shockwave collisions in $\text{AdS}_{\bm 5}$}

\author[a]{Sebastian Waeber,}
\author[a]{Andreas Rabenstein,}
\author[a]{Andreas Sch\"afer,}
\author[b]{Laurence G.~Yaffe}
\affiliation[a]
    {Institute for Theoretical Physics, University of Regensburg,
    D-93040 Regensburg, Germany}
\affiliation[b]
    {Department of Physics, University of Washington, Seattle WA 98195-1560,
    USA}
\emailAdd	{sebastian.waeber@physik.uni-regensburg.de}
\emailAdd	{andreas.rabenstein@physik.uni-regensburg.de}
\emailAdd	{andreas.schaefer@physik.uni-regensburg.de}
\emailAdd	{yaffe@phys.washington.edu}

\keywords	{holography, gravitational shockwaves, quark-gluon plasmas, heavy ion collision, numerical relativity}
\abstract
    {%
    Collisions of asymmetric planar shocks in maximally
    supersymmetric Yang-Mills theory are studied via their
    dual gravitational formulation in asymptotically anti-de Sitter
    spacetime.
    The post-collision hydrodynamic flow is found to be very well
    described by appropriate means of the results of
    symmetric shock collisions.
    This study extends, to asymmetric collisions,
    previous work of Chesler, Kilbertus, and van der Schee
    examining the special case of symmetric collisions
    \cite{Chesler:2015fpa}.
    Given the universal description of hydrodynamic flow produced by
    asymmetric planar collisions one can model,
    quantitatively, non-planar, non-central collisions of
    highly Lorentz contracted projectiles without the need
    for computing, holographically, collisions of finite size
    projectiles with very large aspect ratios.
    This paper also contains a pedagogical description
    of the computational methods and software used to compute
    shockwave collisions using pseudo-spectral methods,
    supplementing the earlier overview of
    Chesler and Yaffe~\cite{Chesler:2013lia}.
    }

\maketitle
\section{Introduction and summary}

Despite the fact that QCD is not conformal,
supersymmetric,
or infinitely strongly coupled,
and has only a small number ($N \eq 3$) of colors,
the comparison of heavy ion phenomenology
with predictions based on
AdS/CFT duality (of ``holography'')
has turned out to be quite fruitful
\cite{Chesler:2010bi,
    Chesler:2013lia,
    Chesler:2015fpa,
    Heller:2012km,
    Chesler:2015wra,
    Heller:2012je,
    Casalderrey-Solana:2013aba,
    Buchel:2015saa,
    1307.2539,
    1507.08195,
    1607.05273,
    1609.03676,
    1506.02209,
    1604.06439,
    1601.01583}.
At temperatures above the QCD phase transition the lack of supersymmetry is
of minor importance and effects caused by the other differences
can be described perturbatively,
either on the QCD or gravity side of the duality.
For example, corrections due to large but finite values of the 't Hooft coupling
$\lambda= g_{\rm YM}^2N$ relevant for QCD can be calculated perturbatively on the
gravity side,
while the effects of non-conformality can be studied within QCD
either perturbatively or using lattice gauge theory.
Hence, it has been possible to identify which results from holographic
modeling of heavy ion collisions should be more, or less,
applicable to real QCD.
Examples of observables with relatively modest corrections
due to finite coupling and non-conformality effects include the
viscosity to entropy density ratio \cite{Buchel:2004di},
$
    4\pi\eta/s=1+15\, \zeta(3)\, \lambda^{-3/2}\approx 1.4
$
for $\lambda \approx 12$,
and the short hydrodynamization time predicted
by AdS/CFT duality based on calculations of the lowest
quasinormal mode (QNM) frequency \cite{Waeber:2018bea}.
For the latter quantity, finite coupling corrections are
larger than for $\eta/s$, but not so much as to
change the picture qualitatively.

In this paper we study the
hydrodynamic flow resulting from asymmetric collisions
of planar shocks in strongly coupled, maximally supersymmetric
Yang-Mills theory.
Our work extends previous work on planar shock collisions
\cite{Chesler:2010bi,
    Heller:2012je,
    Heller:2012km,
    Casalderrey-Solana:2013aba,
    Chesler:2013lia,
    Buchel:2015saa}
and, in particular, the observation by
Chesler, Kilbertus, and van~der~Schee of ``universal'' flow with
simple Gaussian rapidity dependence in the special case
of symmetric collisions of planar shocks \cite{Chesler:2015fpa}.
For such symmetric collisions,
the authors of ref.~\cite{Chesler:2015fpa} found that on a post-collision
surface of constant proper time lying within the hydrodynamic regime,
$\tau = \tau_{\rm init} \gtrsim \tau_{\rm hydro} \approx 2/\mu$,
the fluid 4-velocity is very well described by boost invariant flow,
\begin{equation}
    u^\tau = 1 \,,\quad
    u^\xi = \bm u^\perp = 0 \,,
\label{eq:boostinvflow}
\end{equation}
(with $ds^2 \equiv -d\tau^2  +\tau^2 \, d\xi^2 + d\bm x_\perp^2$),
while the proper energy density is well described by a Gaussian
in spacetime rapidity,
\begin{equation}
    \epsilon(\xi, \tau_{\rm init})
    =
    \mu^4 \, A(\mu w) \, e^{-\frac 12 \, \xi^2/\sigma(\mu w)^2} \,.
\label{eq:Gaussian}
\end{equation}
This proper energy density $\epsilon$ is defined as the
timelike eigenvalue of the rescaled stress-energy tensor,
\begin{equation}
    \widehat T^{\mu\nu}
    \equiv
    \frac {2\pi^2}{\Nc^2} \,
    T^{\mu\nu} \,,
\label{eq:That}
\end{equation}
so $\widehat T^{\mu\nu} \, u_\nu = -\epsilon \, u^\mu$.
The energy scale $\mu$ characterizes the transverse
energy density of each incoming shock and
is defined by the longitudinally integrated (rescaled)
energy density of either incoming shock,
\begin{equation}
    \mu^3 \equiv
    \int dz \> \widehat T^{00}(z\pm t)_{\rm incoming-shock} \,.
\end{equation}
The longitudinal width $w$ of the incoming shocks is
defined as the energy density weighted rms width \cite{Chesler:2015fpa}.
For the specific choice $\tau_{\rm init} = 3.5/\mu$,
Ref.~\cite{Chesler:2015fpa} found
\begin{subequations}\label{eq:Aandw}
\begin{align}
    A(\mu w) &\approx 0.14 + 0.15 \, \mu w - 0.025 \, (\mu w)^2 \,,
\\
    \sigma(\mu w) &\approx
    0.96 - 0.49 \, \mu w + 0.13 \, (\mu w)^2 \,.
\end{align}
\end{subequations}

For studying asymmetric planar shock collisions, we
choose to work in the center-of-momentum (CM) frame in which the
transverse energy densities of the incoming shocks are equal,
\begin{equation}
    \mu \equiv \mu_+ = \mu_- \,.
\end{equation}
In this frame the two incoming shocks will have widths
$w_+$ and $w_-$, and physical results may now depend on two
independent dimensionless combinations which we take to be
$\mu w_+$ and $\mu w_-$.

Over a substantial range of incoming shock widths $\{ w_+, w_- \}$
ranging from $0.35/\mu$ down to $0.075/\mu$,
we find that the spacetime region in which hydrodynamics is
applicable has little or no dependence on the shock widths,
or their asymmetry,
and is sensitive only to the initial energy scale $\mu$.
Using the same definition of a hydrodynamic residual and the
15\% figure of merit chosen in Ref.~\cite{Chesler:2015fpa},
we find that the boundary of the hydrodynamic region of validity
remains at
\begin{equation}
    \mu \, t_{\rm hydro} \approx 2 \,,
\label{eq:thydro}
\end{equation}
even for highly asymmetric collisions.

Similarly, the fluid 4-velocity resulting from asymmetric collisions
remains very close to ideal boost invariant flow (\ref{eq:boostinvflow}),
while the post-collision proper energy density $\epsilon$ remains
well-described by a Gaussian.
However, the amplitude $A$, mean $\bar\xi$, and width $\sigma$
of the Gaussian rapidity dependence are now functions of both
incoming shock widths,
\begin{equation}
    \epsilon(\xi, \tau_{\rm init})
    =
    \mu^4 \, A(\mu w_+, \mu w_-) \,
    e^{-\frac 12 (\xi-\bar\xi(\mu w_+,\mu w_-))^2/\sigma(\mu w_+, \mu w_-)^2}
    \,.
\label{eq:epsilon-asym}
\end{equation}
For asymmetric collisions, the outgoing energy density peaks
at a non-zero mean rapidity $\bar\xi$ which is well-described
 by
\begin{equation}
    \bar\xi(\mu w_+, \mu w_-)
    \approx
   \Xi \> \frac{w_+-w_-}{w_++w_-} \, ,
\label{eq:xibar-asym}
\end{equation}
where the coefficient $\Xi$ is constant for $\tau>2$
(as shown below in Fig.~\ref{Xi_bar})
and has the value $\Xi \approx 7 \times 10^{-2}$.
We find that the amplitude $A$ is well-described by the geometric mean of the symmetric collision results,
\begin{equation}
    A(\mu w_+, \mu w_-) \approx
    \sqrt {A(\mu w_+) \, A(\mu w_-) } \,.
\label{eq:A-asym}
\end{equation}
In fact, after shifting the rapidity by $\bar \xi$,
we find that the geometric mean of the full symmetric collision rapidity
distributions provides a good approximation to the asymmetric collision
results.
For the width of the rapidity distribution, this implies that
\begin{equation}
    \sigma(\mu w_+, \mu w_-) \approx
    \left[ \half \sigma(\mu w_+)^{-2} + \half \sigma(\mu w_-)^{-2}
    \right]^{-1/2}
    \,.
\label{eq:sigma-asym}
\end{equation}

For asymmetric collisions,
the fit to the data provided by
the this Gaussian model is good,
as may be seen below in Fig.~\ref{plots},
but is not quite as perfect as for symmetric collisions.
A more elaborate model, discussed in section \ref{Hydrodynamic flow},
involves a weighted geometric mean of the symmetric collision profiles
and provides an even better description,
valid over a wider range of rapidity.

Given the above extension of the ``universal'' flow resulting from planar
shock collisions to the asymmetric case,
we now have the ingredients needed to predict initial
conditions for the hydrodynamic flow resulting from collisions
of bounded projectiles with finite transverse extent,
provided the transverse size of the incident projectiles is
large compared to their (Lorentz contracted) longitudinal widths,
so that spatial gradients in transverse directions are small
compared to longitudinal gradients.
The following algorithm provides the leading term in an expansion in
transverse gradients:

\begin{itemize}
\item
    Regard the colliding system as composed of independent subregions
    in the transverse plane, or ``pixels'', with each pixel having
    a size $\delta \equiv 1/Q_s$ which is small compared to the
    transverse extent of the projectiles, but large compared to
    their longitudinal widths.
\item
    Let $j$ label independent transverse-plane pixels, with
    $p^\pm_z(j)$ the portion of the longitudinal momentum of each
    incident projectile residing within pixel $j$.
\item
    For each pixel $j$, transform to the CM frame in which the
    total longitudinal momentum within the pixel vanishes, and
    evaluate the resulting energy scale $\mu(j)$ and incident projectile
    widths $w_\pm(j)$ for this pixel.
    Explicitly, $\mu(j)^6 = 4 \, p^+_z(j) \, p^-_z(j) / \delta^4$.
\item
    Use the planar shock results
    (\ref{eq:boostinvflow}),
    (\ref{eq:epsilon-asym})--(\ref{eq:sigma-asym}),
    plus the constitutive relation for a conformal fluid (\ref{hydro_approx}),
    to construct each pixel's stress-energy tensor $T^{\mu\nu}(j)$
    at the initial proper time $\tau_{\rm init}$.
\item
    Transform each pixel's stress-energy tensor $T^{\mu\nu}(j)$
    from its CM frame back to the original (lab) frame.
\end{itemize}
The result is a representation of the full system's
stress-energy tensor on the $\tau_{\rm init}$ initial surface,
with transverse variation on the pixel scale $\delta$,
suitable for use as initial data for further hydrodynamic evolution.
This procedure uses strongly coupled holographic dynamics
to map energy density profiles of the initial projectiles,
which may include initial state fluctuations
and have non-vanishing impact parameter,
into hydrodynamic initial data,
without the need to perform full 5D numerical relativity
calculations which are very challenging \cite{Chesler:2015wra}.
As noted above this procedure, based on planar shock results,
should be viewed as the first term in an expansion in (small)
transverse gradients.
It would, of course, be interesting to derive, systematically,
subsequent terms in this expansion.

\begin{figure}
\centering
\includegraphics[scale=0.25]{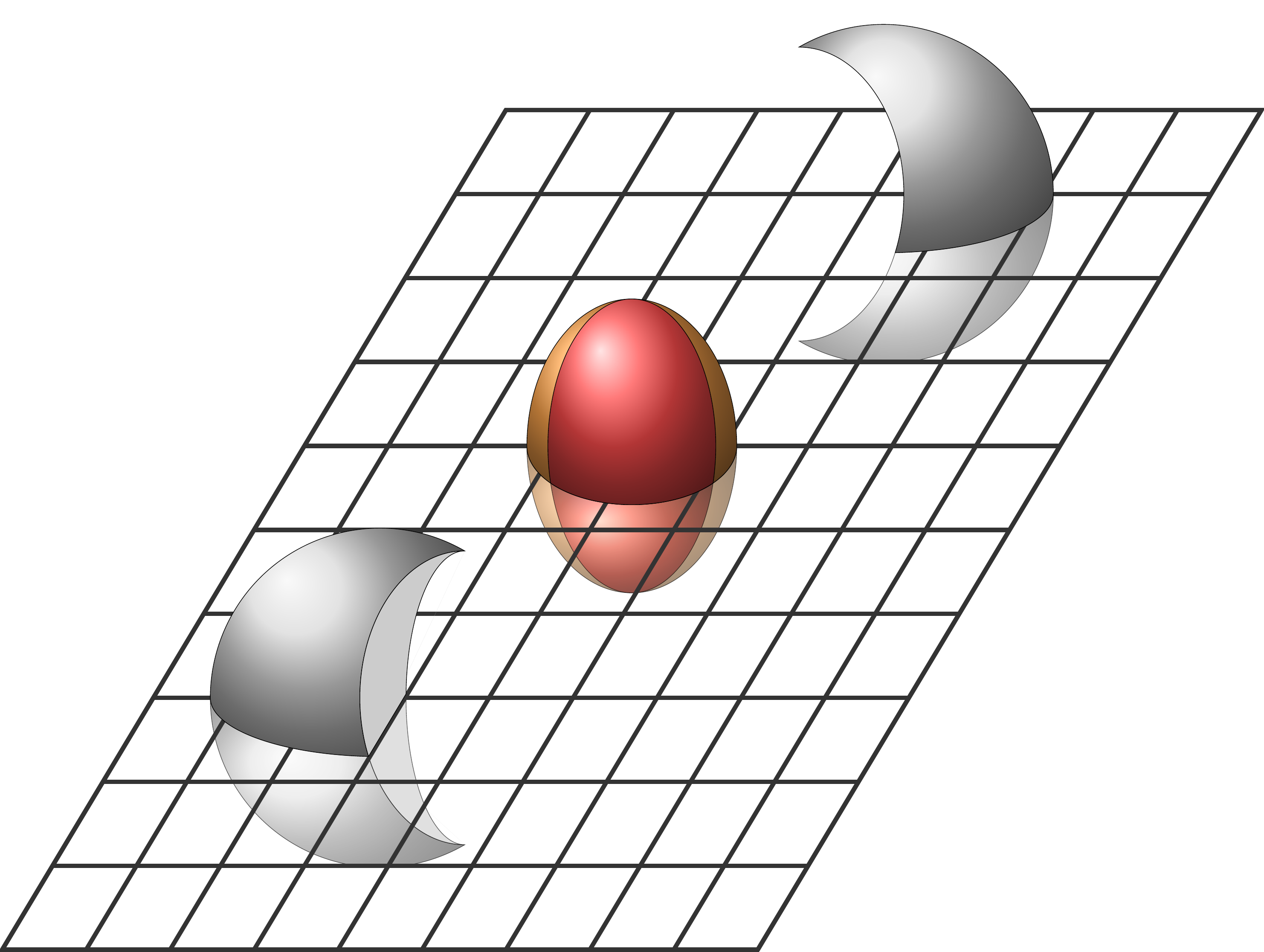}
\caption
    {%
    Sketch of a peripheral heavy ion collision.
    The almond shaped overlap region forms a quark-gluon plasma,
    not the spectator portions (shown in grey).
    The hydrodynamization time increases rapidly as one approaches
    the boundary of the overlap region, whose shape influences
    the value of the experimentally measured elliptic flow parameter $v_2$.
    }
\label{fig:collision}
\end{figure}

Pixels near the periphery of the overlap region of colliding nuclei,
illustrated in Fig.~\ref{fig:collision}, will have decreasing
CM frame transverse energy density $\mu^3$ due to the rapid fall-off
of the transverse energy density of the colliding nuclei
near their periphery.
Given the fact that the hydrodynamization time scales inversely with
$\mu$ (\ref{eq:thydro}), this implies that pixels near the periphery
of the overlap region (shown in orange) will enter the hydrodynamic regime
much later than pixels in the middle of the overlap region.%
\footnote
    {%
    When transforming from the CM frame back to the lab frame,
    the hydrodynamization time $t_{\rm hydro}$ is nearly Lorentz invariant.
    More precisely, as discussed in section \ref{Hydrodynamic flow} and in
    Ref.~\cite{Chesler:2015fpa}, the boundary of the hydrodynamic regime
    is well-described as a Lorentz invariant hyperboloid relative to
    an origin with a modest temporal displacement.
    }
How this impacts an appropriate choice of the initial Cauchy surface
used in hydrodynamic modeling, and the resulting uncertainties in
estimates of, for example, the elliptic flow parameter $v_2$,
is deserving of further study.

The remainder of this paper is organized as follows.
In section \ref{sec:characteristic}
we review the characteristic formulation of
general relativity in asymptotically anti-de Sitter spacetimes
and the initial data for planar shock collisions,
largely following Ref.~\cite{Chesler:2013lia}.
Section \ref{sec:calculation} describes the numerical procedure
and software used to compute shock collisions,
highlighting several issues in greater detail than
in Ref.~\cite{Chesler:2013lia}.
Results are presented in section \ref{sec:results}, followed by
a brief final discussion in section \ref{sec:discussion}.
Readers primarily interested in results should feel free to
turn directly to section \ref{sec:results}.
Additional computational details are presented in the appendix.

\section{Planar shock collisions in asymptotically AdS spacetime}
\label{sec:characteristic}

\subsection{Characteristic formulation}
\label{subsection:Characteristic_formulation}

As shown in Refs.~%
\cite {Chesler:2010bi,
    Chesler:2013lia,
    Heller:2012km,
    Fuini:2015hba,
    Chesler:2015wra},
the characteristic formulation of general relativity,
originally developed by Bondi and Sachs
\cite{Bondi:1960jsa,Sachs:1962wk},
provides a computationally effective method for 
handling the diffeomorphism invariance of general relativity
when studying collisions dynamics in asymptotically
AdS spacetimes.

The characteristic formulation is based on a null slicing of the
geometry in which coordinates are directly tied to a congruence
of null geodesics.
We will use $X \equiv (x,r)$ to denote 5D coordinates,
with $x = (x^0,x^i) \equiv (t,x^i)$ representing ordinary
Minkowski coordinates on the boundary of the AdS spacetime.
Requiring that $t = \rm const.$ surfaces be null hypersurfaces
implies that the one-form $k = \nabla t$ is null,
$0 = k_A \, k^A = g^{AB} k_A k_B$, which means that $g^{tt} = 0$.
Requiring the spatial coordinates $x^i$ to be constant
along the null rays tangent to $k^A$ implies that
$0 = k^A \, \partial_A x^i = g^{AB} (\partial_A t) (\partial_B x^i)$,
which means that $g^{ti} = 0$.
These conditions on the contravariant components of the metric
then imply that $g_{rr} = g_{ri} = 0$.
Hence, under these assumptions
the most general line element may be written in the
generalized infalling (or Eddington-Finkelstein) form,
\begin{align}\label{eqn:EFansatz}
	ds^{2}
	=
	2 dt\left[ \beta(X) \, dr -A(X) \, dt - F_i(X) \, dx^i	\right]
	+ G_{ij}(X) \, dx^i dx^j \,.
\end{align}
It will be convenient to factor the spatial metric $G_{ij}$ into
a scale factor $\Sigma$ times a unimodular matrix $\ghat$,
\begin{equation}
    G_{ij}(X) \equiv \Sigma(X)^2 \> \ghat_{ij}(X) \,,
\end{equation}
with $\det(\ghat) \equiv 1$.
One may fix one further condition, controlling the parameterization
of the null geodesics tangent to $k^A$.
Bondi and Sachs~\cite{Bondi:1960jsa,Sachs:1962wk}
chose to fix the scale factor $\Sigma(X) = r$,
convenient for problems with spherical symmetry.
We instead follow Chesler and Yaffe~\cite{Chesler:2013lia}
and choose to set
\begin{equation}
    \beta(X) = 1 \,.
\end{equation}
This condition leaves a residual reparametrization invariance
in the metric (\ref{eqn:EFansatz}) consisting of radial shifts,
\begin{align}
    r\rightarrow \tilde{r} = r + \delta\lambda(x) \,,
\end{align}
with the shift $\delta \lambda$ depending in an arbitrary fashion
on the boundary coordinates $x$.
Under such a shift, the metric coefficient functions transform as
\begin{subequations}
\begin{align}
	A(x,r) &\rightarrow
	\widetilde A(x,\tilde{r})
	\equiv A(x,\tilde{r}{-}\delta\lambda)
	+ \partial_t \, \delta\lambda(x) \,,
\\
	F_i(x,r) &\rightarrow
	\widetilde F_i(x,\tilde{r})
	\equiv F_i(x,\tilde{r}{-}\delta\lambda)
	+ \partial_i \, \delta\lambda(x) \,,
\\ 
	G_{ij}(x,r) &\rightarrow
	\widetilde G_{ij}(x,\tilde{r})
	\equiv G_{ij}(x,\tilde{r}{-}\delta\lambda) \,.
\end{align}
\end{subequations}
From these transformations of $A$ and $F_i$ it is apparent
that they may be regarded as temporal and spatial components
of a gauge field representing radial shifts.
It is possible to write the
Einstein equations in a manner which is manifestly covariant under
radial shifts.
To do so, it is convenient to define modified
temporal and spatial derivatives,
\begin{align}\label{eqn:d+}
	d_+ \equiv \partial_t + A(X) \, \partial_r \,,\qquad
	d_i &\equiv \partial_i + F_i(X) \, \partial_r \,.
\end{align}
Given these definitions, the Einstein equations,
\begin{align}
	R^{AB} - \tfrac{1}{2} R\, g^{AB} + \Lambda\, g^{AB} = 0 \,,
\end{align}
acquire a nested structure with the schematic form,
\begin{subequations}\label{eq:einsteineqns}
\begin{align}
    \label{eqn:Sigma}
    \left(\partial_r^2+Q_\Sigma[\ghat] \right)\Sigma
    & = 0 \,.
\\[2pt]
    \left( \delta^i_j \, \partial_r^2
	+ P_F[\ghat,\Sigma]_i^j \, \partial_r
	+ Q_F[\ghat,\Sigma]_i^j \right) F_j
    & = S_F[\ghat,\Sigma]_i \,.
    \label{eqn:F}
\\[2pt]
    \left( \partial_r+Q_{d_+\Sigma}[\Sigma] \right)d_+\Sigma
	& = S_{d_+\Sigma}[\ghat,\Sigma,F] \,.
	\label{eqn:d+Sigma}
\\[2pt]
    \left(\delta^k_{(i} \, \delta^l_{j)} \, \partial_r
	+ Q_{d_+\ghat}[\ghat,\Sigma]^{kl}_{ij}\right) d_+\ghat_{kl}	
    & = S_{d_+\ghat}[\ghat,\Sigma,F,d_+\Sigma]_{ij} \,.
	\label{eqn:d+ghat}
\\[2pt]
	\label{eqn:A}
    \partial_r^2 A
    & = S_A[\ghat,\Sigma,F,d_+\Sigma,d_+\ghat] \,.
\\[2pt]
    \left( \delta^j_i \, \partial_r
	+ Q_{d_+F}[\ghat,\Sigma]^j_i \right) d_+F_j
    & = S_{d_+F}[\ghat,\Sigma,F,d_+\Sigma,d_+\ghat,A]_i \,.
\\[2pt]
    \label{eqn:d+d+Sigma}
    d_+\left( d_+\Sigma\right)
    & = S_{d^2_+\Sigma}[\ghat,\Sigma,F,d_+\Sigma,d_+\ghat,A] \,,
\end{align}
\end{subequations}
Each equation is a first or second order linear radial
differential equation for the indicated  metric component(s) or their
modified time derivatives.
The square brackets of each coefficient or source function
indicates on which fields the term depends.
Explicit form of these equations, for the case of planar shocks,
are given in appendix~\ref{subsection:Appendix_EE}.

Given the rescaled spatial metric $\ghat$ on any time slice,
plus suitable boundary conditions,
each radial differential equation may be integrated in turn,
thereby determining both the other metric coefficients and the time
derivative of $\ghat$ on that time slice.
The required boundary conditions may be inferred from
the near-boundary behavior which can be obtained by solving
equations~(\ref{eqn:Sigma}-\ref{eqn:d+d+Sigma}) order by order
in $r$.
One finds~\cite{Chesler:2013lia},
\begin{subequations}\label{eqn:near-boundary-EF}%
\begin{align}
	&A = \tfrac{1}{2}\left(r {+} \lambda\right)^2
	    - \partial_t \lambda
	    + a^{(4)} \, r^{-2}
	    + \mathcal{O}(r^{-3}) \,,
	&F_i &= -\partial_i \lambda
	    + f_i^{(4)} \, r^{-2}
	    + \mathcal{O}(r^{-3}) \,,
\\
	&\Sigma = r  {+} \lambda
	    + \mathcal{O}(r^{-7}) \,,
	&\ghat_{ij} &= \delta_{ij}
	    + \ghat^{(4)}_{ij} \, r^{-4}
	    + \mathcal{O}(r^{-5}) \,,
\\
	&d_+\Sigma = \tfrac{1}{2}\left(r {+} \lambda\right)^2
	    + a^{(4)} \, r^{-2}
	    + \mathcal{O}\left(r^{-3}\right) \,,
	&d_+\ghat_{ij} &= -2 \, \ghat^{(4)}_{ij} \, r^{-3}
	    + \mathcal{O}(r^{-4}) \,.
\end{align}
\end{subequations}
The coefficients $a^{(4)}$, $f_i^{(4)}$ and $\ghat_{ij}^{(4)}$
cannot be determined by a local near-boundary analysis.
Note that $\ghat_{ij}^{(4)}$ is necessarily traceless
(because $\ghat$ has unit determinant).
These coefficients are mapped, via gauge/gravity duality, to
the stress-energy tensor of the dual field theory.
In our infalling coordinates this relation is given by~\cite{Chesler:2013lia}
\begin{align}\label{eqn:stress-energy-metric-EF}
	\frac{2\pi^2}{N_c^2}\left<T_{\mu\nu}\right>
	\equiv 
	\big<\widehat{T}_{\mu\nu}\big>
	= 
	h_{\mu\nu}^{(4)} + \tfrac{1}{4} \, h_{00}^{(4)} \, \eta_{\mu\nu} \,,
\end{align}
with
$
	h_{00}^{(4)} \equiv -2 a^{(4)} 
$,
$
	h_{0i}^{(4)} \equiv -f_i^{(4)}
$,
and
$
	h_{ij}^{(4)} \equiv \ghat_{ij}^{(4)}
$.
Here $N_c$ is the number of colors in the dual field theory, and
$\eta = \mathrm{diag}(-1,+1,+1,+1)$ is the Minkowski metric tensor.
Explicitly,
\begin{align}
    \big<\widehat{T}_{00}\big> = -\tfrac{3}{2} \, a^{(4)} \,,\quad
    \big<\widehat{T}_{0i}\big> = -f_i^{(4)} \,,\quad
    \big<\widehat{T}_{ij}\big> = \ghat_{ij}^{(4)} - \tfrac{1}{2} \, a^{(4)} \, \delta_{ij}
    \,.
    \label{stressenergy}
\end{align}

The radial shift parameter $\lambda(x)$ is completely
undetermined in expansion~(\ref{eqn:near-boundary-EF}) and may
be chosen arbitrarily.
As in previous work 
\cite {Chesler:2010bi,
    Chesler:2013lia,
    Heller:2012km,
    Fuini:2015hba,
    Chesler:2015wra},
we use this freedom to set the radial position $r_h(x)$
of the apparent horizon equal to a fixed value,
\begin{align}\label{eqn:hor_pos}
	r_h(x) = r_h \,.
\end{align}
It is sufficient to solve for the spacetime geometry in
the region between the horizon and the boundary
because information hidden behind the horizon cannot
propagate outward and influence boundary observables.
Thus, the choice~(\ref{eqn:hor_pos}) results in a convenient rectangular
computational domain.

With our metric ansatz~(\ref{eqn:EFansatz}),
demanding a fixed radial position of the apparent horizon
leads to a condition on $d_+\Sigma$~\cite{Chesler:2013lia}.
To derive this condition, one may write the tangents to a radial
infalling null congruence in the form $k_A(X) = \mu(X) \, \nabla_A \phi(X)$
for some scalar functions $\phi$ and $\mu$.
Demanding that the one-form $k$ be null allows one to reexpress the
time derivative of $\phi$ in terms of spatial derivatives.
Requiring that the congruence satisfy the (affinely parameterized)
geodesic equation $k^A k_{B;A} = 0$ allows one to reexpress the
time derivative of the multiplier function $\mu$ in terms of its
spatial derivatives.
Given these time derivatives, one may then compute the expansion
$\theta = \nabla \cdot k$ on the time slice of interest.
Demanding that the expansion vanish on a surface $\phi(X) = \rm const.$
implies that this surface is an apparent horizon.
Applying this procedure to the metric ansatz~(\ref{eqn:EFansatz}) and
specializing to the case $\phi(X) = r$ leads to the desired condition
\cite{Chesler:2013lia},
\begin{align}\label{eqn:horizonPosition}
	d_+ \Sigma\big\vert_{r_h} 
	=
	-\tfrac{1}{2} \, (\partial_r \Sigma) \, F^2
	-\tfrac{1}{3} \, \Sigma \, \nabla\cdot F \,.
\end{align}
This condition must hold at all times if the radial position of the
horizon is to remain fixed at some given value $r_h$.
Consequently, on every time slice the condition
\begin{align}\label{eqn:HorizonStationary}
	\partial_t \, d_+ \Sigma\big\vert_{r_h} 
	=
	\partial_t
	\left[
		-\tfrac{1}{2}(\partial_r \Sigma) F^2
		-\tfrac{1}{3}\Sigma \, \nabla\cdot F
	\right]
\end{align}
is also required to hold.
When combined with the Einstein equation (\ref{eqn:d+d+Sigma}),
this final condition leads to an elliptic differential equation for the value
of the metric function $A$ on the (apparent) horizon.
Explicit forms of the horizon equation (\ref{eqn:horizonPosition})
and the horizon stationarity condition (\ref{eqn:HorizonStationary})
may be found in appendix~\ref{subsection:Appendix_EE}.

\subsection{Solution strategy}
\label{sec:strategy}

To solve the nested form (\ref{eq:einsteineqns}) of the Einstein equations,
one requires appropriate boundary data which picks out the correct
solution for each equation.
The needed boundary conditions are determined by the homogeneous
solutions of each equation and the asymptotic behavior of the desired
solutions.
This information is summarized in table~\ref{tab:boundary_conditions}.
From this table one sees that a choice for the radial shift $\lambda$
along with values of the asymptotic coefficients $a^{(4)}$ and $f_i^{(4)}$
are needed as boundary conditions for the $\Sigma$, $F_i$,
and $d_+\Sigma$ equations and serve to fix the coefficient of a
homogeneous solution to the corresponding differential equation.
The asymptotic coefficients $a^{(4)}$ and $f_i^{(4)}$, proportional
to the boundary energy and momentum density,
are dynamical degrees of freedom (in addition to the metric $\ghat_{ij}$)
and are determined by integrating the stress-energy continuity
equation as discussed below.
The radial shift $\lambda(x)$ will also be treated as a dynamical
degree of freedom, as described below, and adjusted in a manner
which ensures that the apparent horizon remains at a fixed radial position.

Given this boundary data,
together with the value of $\ghat$ on some given time slice,
the radial differential
equations (\ref{eqn:Sigma})--(\ref{eqn:d+ghat}) may each be integrated in turn,
at every spatial location $x^i$,
leading to a determination of $d_+ \ghat_{ij}$ on the time slice.
Two boundary conditions are needed
to integrate the second order equation (\ref{eqn:A})
for the metric function $A$.
As seen in table~\ref{tab:boundary_conditions}, the value of the
radial shift $\lambda$ supplies one condition.
The second boundary condition is supplied by the value of $A$ at the apparent
horizon, which is determined by
solving the horizon stationarity condition (\ref{eqn:HorizonStationary}).

Having determined both $d_+\ghat$ and $A$,
the actual time derivative for the rescaled spatial metric $\ghat$
is then reconstructed as
\begin{align}
	\partial_t \, \ghat_{ij}
	=
	d_+\ghat_{ij} - A \, \partial_r \, \ghat_{ij} \,.
	\label{update_B}
\end{align}
Knowing $d_+\Sigma$ and $A$ (on a given time slice),
the near boundary expansion~(\ref{eqn:near-boundary-EF}) shows that
the time derivative of the the radial shift $\lambda(x)$ may be extracted as
\begin{align}
    \partial_t \lambda = \lim_{r\to\infty}\left(d_+\Sigma - A \right)
    .
\end{align}
Similarly, the asymptotic coefficient $\ghat_{ij}^{(4)}$ 
determining the traceless stress tensor is extracted from the
boundary limit of either $r^4 \, (\ghat_{ij} - \delta_{ij})$ or
$-\half r^3 d_+\ghat_{ij}$.
This information then allows one to determine the time derivatives
of $a^{(4)}$ and $f_i^{(4)}$ using the 
boundary stress-energy continuity equation,
$\nabla^\mu \left<T_{\mu\nu}\right> = 0 $,
which is an automatic consequence of the Einstein equations.
Explicitly,
\begin{align}
    \partial_t \, a^{(4)} = \tfrac{2}{3} \, \partial_i \, f^{(4)}_i \,,\quad
    \partial_t \, f^{(4)}_i = \tfrac{1}{2} \, \partial_i a^{(4)}
			    - \partial_j \, \ghat_{ij}^{(4)} \,.
			    \label{update_af}
\end{align}

\begin{table}
\centering

\begin{tabular}{ | c | c | l | }
    \hline
    field & homogeneous solution(s) & near-boundary behavior \\\hline
    $\Sigma$ & ${}\sim \sigma^{(0)} \, r^1 + \sigma^{(1)} \, r^0$
	    & $\Sigma \sim r + \lambda$\\\hline
    $F_i$ & ${}\sim f_i^{(0)} \, r^2 + f_i^{(4)} \, r^{-2}$
	    & $F_i \sim -\partial_i\lambda +f_i^{(4)}r^{-2}$\\\hline
    $d_+\Sigma$ & ${}\sim a^{(4)} \, r^{-2}$
	    & $d_+\Sigma \sim \frac{1}{2}\left(r{+} \lambda\right)^2
		+ a^{(4)} \, r^{-2}$\\\hline
    $d_+\ghat_{ij}$ & ${}\sim r^{-{3}/{2}}$ 
	    & $d_+\ghat_{ij} \sim -2 \, \ghat_{ij}^{(4)} \, r^{-3}$\\\hline
    $A$ & ${}\sim a^{(1)} \, r^1 + a^{(2)} \, r^0$
	    & $ A \sim \frac{1}{2}\,(r{+} \lambda)^2 -\partial_t \lambda $\\\hline
\end{tabular}
\caption
    {%
    Near-boundary asymptotic behavior of the homogeneous solutions
    to the radial differential equations
    (\ref{eqn:Sigma})--(\ref{eqn:A}) for the indicated fields,
    together with the desired asymptotic behavior of physical solutions.
    The asymptotic coefficients $a^{(4)}$, $f_i^{(4)}$, and $\ghat_{ij}^{(4)}$
    determine respectively the energy density, momentum density,
    and traceless stress tensor of the dual field theory.
    The leading terms in the near-boundary behavior of
    all fields except $\Sigma$ are driven by the inhomogeneous source
    terms in the various equations and do not correspond to homogeneous
    solutions.
    \label{tab:boundary_conditions}
    }
\end{table}

The above procedure, involving integration of a sequence of
linear ordinary differential equations in the radial direction
plus one spatial elliptic equation on the apparent horizon,
determines the time derivatives of the dynamical data
$\{ \ghat_{ij},\, \lambda,\, a^{(4)},\, f_i^{(4)} \}$
given initial values of this data on some time slice.
These time derivatives are then input into a conventional
time integrator, such as fourth order Runge-Kutta,
to advance to the next time slice where the entire
process repeats.

Overall, this characteristic formulation transforms the highly
non-linear coupled Einstein equations into a set of nested linear 
ordinary differential equations and first order time evolution equations.
We solve the radial differential equations, and the horizon
stationarity equation, using spectral methods as described in some
detail in section~\ref{sec:calculation} and
appendix~\ref{subsection:Appendix_PseudospectralMethods}.

\subsection{Planar shocks}

By ``planar shock'' we mean an asymptotically anti-de Sitter solution
of the vacuum Einstein equations whose boundary stress-energy tensor
describes a ``sheet'' of energy density which moves at the speed of light
in some longitudinal direction and is translationally invariant in the
other two transverse spatial dimensions.
For regular solutions, such a sheet of moving energy density will have
some smooth longitudinal profile and non-zero characteristic thickness.

Let $\{ x^i \} \equiv(\bm x_\perp, z)$ denote spatial coordinates separated into
transverse and longitudinal components,
and consider shocks moving in the $\pm z$ direction.
To specialize the general infalling metric ansatz (\ref{eqn:EFansatz})
to the case of planar shock spacetimes,
we impose translation invariance in transverse 
directions plus rotation invariance in the transverse plane, which
implies that all metric components are functions of only of
$r$ and $x_\mp \equiv t \mp z$,
that $F_i$ only has a longitudinal component, 
and that the (rescaled) spatial metric has the form
\cite{Chesler:2013lia},
\begin{equation}
    \ghat = \mathrm{diag}(e^B, e^B, e^{-2B}) \,.
\end{equation}
Consequently,
\begin{equation}
    ds^2
    =
    2 dt\left( dr -A \, dt - F_z \, dz	\right)
    + \Sigma^2 \left( e^B \, d\bm x_\perp^2 + e^{-2B} \, dz^2 \right) .
\label{eq:planarEF}
\end{equation}
The boundary asymptotics (\ref{eqn:near-boundary-EF}) implies that
the ``anisotropy'' function $B$ behaves as
\begin{equation}
    B(x_\mp, r) = b^{(4)}(x_\mp) \, r^{-4} + O(r^{-5}) \,.
\end{equation}
For later computational convenience, let
\begin{equation}
    u\equiv 1/{r}
\end{equation}
denote an inverted radial coordinate, 
so that the spacetime boundary lies at $u=0$.

In general it does not seem possible to find analytic forms of planar
shock solutions using the infalling Eddington-Finkelstein (EF) coordinates
(\ref{eq:planarEF}).
But analytic solutions are available in Fefferman-Graham (FG) coordinates
\cite{Chesler:2015wra, Chesler:2010bi, Janik}.
Using 
$
    \{ \tilde x^\mu, \tilde \rho \}
    \equiv
    \{ \tilde t, \tilde {\bm x}_\perp, \tilde z, \tilde \rho \}
$
as our FG coordinates,
with $\tilde x_\pm \equiv \tilde t \pm \tilde z$ and
$\tilde \rho$ an inverted bulk radial coordinate,
the metric
\begin{align}\label{eqn:ConditionShockEinstein}
    ds^2 = \tilde\rho^{-2}
	\left(
	    -d\tilde x_+ \, d\tilde x_-
	    +d\tilde{\bm{x}}^2_{\bot}
	    +d\tilde{\rho}^2
	\right)
	+ \tilde \rho^2 \, h(\tilde x_\pm) \, d\tilde{x}_{\mp}^2 \,,
\end{align}
is a planar shock solution describing a shock
moving in the $\pm z$ direction with arbitrary
longitudinal energy density profile $h(z)$.
In the calculations described below, we use simple
Gaussian profiles with width $w$ and longitudinally
integrated energy density $\mu^3$,
\begin{equation}
    h(z) \equiv \mu^3 (2\pi w^2)^{-1/2} \, e^{-\frac 12 z^2/w^2} \,.
\label{eq:h}
\end{equation}
The associated boundary stress-energy tensor is just
\begin{equation}
    \widehat T^{00}(\tilde t, \tilde z)
    =
    \widehat T^{zz}(\tilde t, \tilde z)
    =
    \pm \widehat T^{0z}(\tilde t, \tilde z)
    =
    h(\tilde t {-} \tilde z) \,,
\end{equation}
with all other components vanishing.

Focusing, for ease of presentation,
on shocks moving in the $+z$ direction,
the translational symmetries imply that the EF and FG coordinates
will be related by a transformation of the form
\cite{Chesler:2013lia},
\begin{align}
    \tilde{t} =t+u+\alpha(t{-}z,u) \,,\quad
    \tilde{z} =z-\gamma(t{-}z,u) \,,\quad
    \tilde{\rho} =u+\beta(t{-}z,u) \,,
\label{eq:transform}
\end{align}
and
$
    \tilde{\bm{x}}_{\bot}=\bm{x}_\bot
$.

As discussed above, the required initial data for the characteristic evolution
scheme consists of the anisotropy function $B$ plus the boundary data
$\{ a^{(4)},\, f_z^{(4)} \}$ and the radial shift $\lambda$.
Inserting a transformation of the form (\ref{eq:transform}) into the FG
metric (\ref{eqn:ConditionShockEinstein}), a short exercise
\cite{Chesler:2013lia} shows that 
\begin{align}
    B = -\tfrac{1}{3}
    \ln\big[
	    -(\partial_z\alpha)^2
	    +(\partial_z\beta)^2
	    +(1-\partial_z\gamma)^2
	    +(u+\beta)^4
		( 1 - \partial_z\alpha - \partial_z\gamma)^2 \, h
    \big] \,,
\label{eq:B}
\end{align}
while the boundary data is given by
\begin{align}
    a^{(4)} = -\tfrac{2}{3} \, h \,,\qquad
    f^{(4)}_z = h \,,\qquad
    \lambda = -\tfrac{1}{2} \, \partial_u^2\beta\big\vert_{u=0} \,.
\label{eq:bdrydata}
\end{align}

To solve for the transformation functions
$\{ \alpha, \beta, \gamma \}$,
one approach, used in Refs.~\cite{Chesler:2010bi,Chesler:2013lia},
is to insert the transformation (\ref{eq:transform}) into the
FG metric (\ref{eqn:ConditionShockEinstein}) and demand that the result have
the EF form (\ref{eq:planarEF}).%
\footnote
    {%
    An alternative approach, used in Ref.~\cite{Chesler:2015wra}
    for more general metrics,
    is based on observing that the curve defined by fixed values of the
    EF boundary coordinates and all values of $r$,
    $
	X^A(r) = (t_0, x_0^i,r)
    $,
    is a null geodesic of the EF metric (\ref{eqn:EFansatz}) with $r$ an
    affine parameter.
    Therefore the same path in FG coordinates, $\Ytilde(X(r))$, will 
    satisfy the geodesic equation
    $
	\frac{d^2 \Ytilde^A}{dr^2}
	+ \widetilde{\Gamma}(Y)^A_{BC} \,
	    \frac{d \Ytilde^B}{dr} \, \frac{d \Ytilde^C}{dr} = 0
    $
    with $\widetilde \Gamma^A_{BC}$ denoting the FG coordinate
    Christoffel symbols.
    Explicit forms of the resulting equations can be found in
    appendix~\ref{subsection:Appendix_CoordinateTransformation}.
    }
To simplify the resulting equations, it is helpful to redefine
the transformation functions $\alpha$ and $\beta$ via
\begin{align}
    \alpha = -\gamma + \beta + \delta \,,\qquad
    \beta = -\frac{u^2\zeta}{1+u\zeta} \,.
\label{ansatz1}
\end{align}
One finds \cite{Chesler:2013lia}
that the functions $\zeta$ and $\delta$ satisfy a pair of coupled
differential equations,
\begin{subequations}\label{diffeq1}
\begin{align}
\label{diffeq1a}
    \frac{1}{u^2}\frac{\partial}{\partial u}
	\left(u^2 \, \frac{\partial\zeta}{\partial u }\right)
    + \frac{2 u H}{(1+u\zeta)^5} = 0 \,,\qquad
    \frac{\partial\delta}{\partial u}
    - \frac{u^2}{(1+u\zeta)^2} \, \frac{\partial \zeta}{\partial u} = 0 \,,
\end{align}
while $\gamma$ satisfies a decoupled equation,
\begin{align}
\label{diffeq1b}
    \frac{\partial\gamma}{\partial u}
    - \frac{u^2}{(1+u\zeta)^2} \, \frac{\partial\zeta}{\partial u}
    + \frac{u^4}{2(1+u\zeta)^2}\left(\frac{\partial\zeta}{\partial u }\right)^2 
    + \frac{u^4 H}{2(1+u\zeta)^6} = 0 \,,
\end{align}
with $H \equiv h+ \left(t - z +u +\delta - u^2\zeta/(1+u\zeta)\right)$.
The desired solutions have the near-boundary behavior
\begin{equation}
    \zeta \sim \lambda + O(u^3) \,,\quad
    \delta \sim O(u^5) \,,\quad
    \gamma \sim O(u^5) \,.
\label{diffeqbc}
\end{equation}
\end{subequations}
Integrating equations (\ref{diffeq1}) with boundary conditions ensuring
the behavior (\ref{diffeqbc}),
and inserting the resulting transformation functions into
Eqs.~(\ref{eq:B}) and (\ref{eq:bdrydata}),
yields the anisotropy function $B$ and associated boundary data
describing of a single shock.

To construct initial data for colliding shocks, we superpose
counter-propagating single shock data at an initial time $t_0$
when the two shocks are sufficiently widely separated that their
overlap is negligible,
\begin{subequations}\label{eq:combine}
\begin{align}
    B(u,z,t_0) &= B_+(u,t_0{-}z) + B_-(u,t_0{+}z) \label{B_tot} \,,
\\
    a^{(4)}(z,t_0) &= a_+^{(4)}(t_0{-}z) + a_-^{(4)}(t_0{+}z) \,,
\label{a_tot}
\\
    f_z^{(4)}(z,t_0) &= f_{z+}^{(4)}(t_0{-}z) - f_{z-}^{(4)}(t_0{+}z) \,.
\label{lambda_tot}
\end{align}
\end{subequations}
However, unlike for the other functions, the overlap of the radial
shifts $\lambda_{\pm}$ of the left and right moving shocks in the
region close to $z\eq0$ is significant. Since we choose the shocks
on the first time slice to be well separated, we may regard the
geometry in between as deviating negligibly from pure AdS.
This justifies modifying the initial shift function
$\lambda$ in the neighborhood of $z=0$,
without changing the physical data
$\{B(u,z,t_0),\, a^{(4)}(z,t_0),\, f^{(4)}(z,t_0)\}$.
As in Ref.~\cite{Chesler:2013lia}, we adjust the initial
radial shift by setting
\begin{equation}
    \lambda(z,t_0)
    =
    \theta_+(-z) \, \lambda_+(t_0{-}z)+\theta_-(z) \, \lambda_-(t_0{+}z) \,,
\label{eq:sumlambda}
\end{equation}
with 
$
    \theta_\pm(z) \equiv \frac{1}{2}
    \left[1-\mathrm{erf}(-z/(\sqrt{2}w_\pm)) \right]
$
a smoothed step function.

In practice, we slightly modify the above superposition procedure.
Following Refs. \cite{Chesler:2013lia, Chesler:2010bi}, we replace Eq.~(\ref{a_tot}) with
\begin{equation}
    a^{(4)}(z,t_0) = a_+^{(4)}(t_0{-}z) + a_-^{(4)}(t_0{+}z)
    - \tfrac 23 \, \epsilon_0 \,.
\label{eq:BED}
\end{equation}
From the form (\ref{stressenergy}) of the
stress-energy tensor, one sees that $\epsilon_0$
is a constant additive shift in $\widehat T^{00}$.
In other words, $\epsilon_0$ is an (artificial) uniform background
energy density.
Adding a small background energy density helps alleviate
numerical problems, as discussed below, and physically means
that the colliding shocks will be propagating through a 
background thermal medium.
If the background energy density $\epsilon_0$ is sufficiently small
compared to the energy densities in the colliding shocks,
then the background will effectively be very cold (compared to
the energy scale $\mu$ of the shocks) and there will be little
dissipation to the medium.
This modification is done purely for numerical convenience and
we will be interested in results extrapolated to vanishing
background energy density.

\section{Computational methods and software construction}
\label{sec:calculation}

The aim of this section is to describe
the construction of a planar shockwave collision code
in sufficient detail so that an interested reader could
create their own version with relatively modest effort.
Those primarily interested in results should skip to the next section.

\subsection{Transformation to infalling coordinates}
\label{sec:transform}

As explained in Ref.~\cite{Chesler:2013lia} and the previous section,
the transformation from Fefferman-Graham to infalling
coordinates may be computed by first solving for the congruence of
infalling geodesics in FG coordinates.
Or, in the special case of planar shock geometries,
one can directly solve the simplified transformation equations
(\ref{diffeq1}).
We implemented both approaches, and found them to have comparable
numerical efficiency.
Here, we focus on the direct approach of solving Eqs.~(\ref{diffeq1})
for the case of a right moving shock.
Henceforth, for convenience, we also set $\mu \eq 1$.
Appropriate factors of $\mu$ can always be reinserted via
dimensional analysis.

We solve the coordinate transformation equations (\ref{diffeq1})
in the rectangular region $u \in [0,u_{\rm end}]$,
$z \in [-L_z/2,L_z/2]$
using Newton-Raphson iteration
(i.e., linearizing each equation in the deviation of the solution
from the current approximation), and solving the resulting linear
equations using spectral methods with domain decomposition.%
\footnote
    {%
    A good introduction to spectral methods may be found in,
    for example, Ref.~\cite{Boyd:Spectral}.
    }

Periodic boundary conditions are imposed in the longitudinal direction
and functions of $z$ are approximated as truncated Fourier series.
This is exactly equivalent to characterizing any function $f(z)$
by a list of its values, $\{ f_l \equiv f(z_l) \}$,
on an evenly spaced Fourier grid composed of $N_z$ points,
\begin{equation}
    z_l \equiv L_z (-\half + l/N_z) \,,
\end{equation}
for $k = 0, {\cdots}, N_z{-}1$.
Derivatives with respect to $z$ turn into the application
of a Fourier grid differentiation matrix $D_z = \| (D_z)_{kl}\|$
applied to the vector of function values,
\begin{equation}
    f'(z_k)
    =
    \sum_l (D_z)_{kl} \, f_l \,.
\end{equation}
Explicit expressions for the Fourier grid differentiation matrix components
$(D_z)_{kl}$ are given in 
appendix \ref{subsection:Appendix_PseudospectralMethods}.
A rather fine longitudinal grid is required to accurately represent
thin shocks within a large longitudinal box.
We used Fourier grids with $N_z \eq 960$
for $L_z \eq 12$ and shock widths down to $0.075$.

To represent the dependence of functions on the radial coordinate $u$
we first decompose the domain $[0,u_{\rm end}]$ into $M$ equally sized
subdomains, and then use a Chebyshev-Gauss-Lobatto grid with $N_u$ points
within each subdomain.
This amounts to using a radial grid composed of the points
\begin{equation}
    u_{jk}
    \equiv
    \frac {u_{\rm end}}{2M}
    \left( 2j-1 - \cos\frac {\pi k}{N_u{-}1} \right) ,
\label{eq:ujk}
\end{equation}
for $j = 1,{\cdots},M$ and $k = 0,{\cdots},N_u{-}1$.
The radial dependence of some function $g(u)$ is represented by
the list of $M \times N_u$ function values on this grid,
$\{ g_{jk} \equiv g(u_{jk}) \}$,
and derivatives with respect to $u$ turn into the application
of a (block diagonal) Chebyshev differentiation matrix
$D_u$ applied to this list of function values,
\begin{equation}
    g'(u_{jk})
    =
    \sum_{l} (D_u)_{kl} \, g_{jl} \,.
\end{equation}
Explicit expressions for the components of the Chebyshev differentiation
matrix $D_u$ are given in 
Eq.~(\ref{diffmat}).
As discussed in Ref.~\cite{Chesler:2013lia},
using domain decomposition (i.e., $M > 1$)
helps to avoid excessive precision loss in the numerical evaluation
of equations near the $u = 0$ boundary, and allows the use of a
larger time step without running afoul of CFL instabilities.
To integrate radial equations down to $u_{\rm end} \eq 2$,
we used radial grids with up to $M \eq 22$ domains and $N_u \eq 12$
points within each subdomain.

The product of these 1D grids defines our 2D spectral grid.
Any function $f(u,z)$ becomes a set of
$N_{\rm tot} \equiv M \times N_u \times N_z$ values on these grid points,
\begin{equation}
    \{ f_{jkl} \equiv f(u_{jk},z_l) \} \,.
\end{equation}
Fortunately,
the differential equations (\ref{diffeq1}) are completely local in $z$.
So these equations, evaluated on the 2D grid with derivatives replaced
by the corresponding differentiation matrices,
do not become a single set of
$2 N_{\rm tot}$ (for Eq.~\ref{diffeq1a})
or $N_{\rm tot}$ (for Eq.~\ref{diffeq1b})
coupled algebraic relations.
Rather they yield $N_z$ decoupled systems, each involving
$2M N_u$ (for Eq.~\ref{diffeq1a})
or $M N_u$ (for Eq.~\ref{diffeq1b}) variables.

For each set of equations, linearization around some initial,
or current, guess for a solution leads to a set of
linear equations of the generic form $\mathcal M \, f = -S$,
where $f$ is the unknown vector of function deviations from the current guess,
$S$ is the vector of residuals,
and $\mathcal M$ is the spectral approximation to the linear operator
which results from the linearization of the differential equation(s)
at some given value of $z$.

At this point, these linear equations are singular.
First, $u\eq0$ is a regular singular point of the differential
equations (\ref{diffeq1a}) and (\ref{diffeq1b}); one cannot
simply evaluate, numerically, these equations at $u\eq0$.
Moreover, solutions to these differential equations are,
of course, non-unique.
One must complement the differential equations with suitable boundary
conditions to specify a unique solution.
With spectral methods, fixing one of these problems fixes the other.
Prior to linearization,
one simply replaces the (ill-defined) evaluation of the equations
at $u \eq 0$ by constraints encoding required boundary conditions.

Examining equations (\ref{diffeq1a}) and (\ref{diffeq1b}), one
sees that the most general near-boundary behavior is
\begin{equation}
    \zeta \sim \zeta_{-1} \, u^{-1} + \lambda + O(u^3) \,,\quad
    \gamma \sim \gamma_0 + O(u^5) \,,\quad
    \delta \sim \delta_0 + O(u^5) \,,
\end{equation}
for arbitrary values of the coefficients $\zeta_{-1}$, $\lambda$,
$\gamma_0$ and $\delta_0$.
We want to set the leading coefficients
$\zeta_{-1}$, $\gamma_0$ and $\delta_0$ to zero.
To implement this Dirichlet condition for $\gamma$ and $\delta$ in a manner
which avoids unnecessary precision loss when computing derivatives of
these functions at the boundary,
it is convenient first to redefine 
\begin{equation}
    \gamma(z,u) \equiv u^3 \, \tilde \gamma(z,u) \,,\quad
    \delta(z,u) \equiv u^3 \, \tilde \delta(z,u) \,,
\end{equation}
and then reexpress equations (\ref{diffeq1}) in terms of
$\tilde \gamma$ and $\tilde \delta$.
Unwanted solutions with non-zero boundary values for $\gamma$ or $\delta$
are then simply not representable when using our spectral representation for
$\tilde \gamma$ or $\tilde\delta$.
Similarly, using our spectral representation for $\zeta$ automatically
eliminates unwanted solutions where $\zeta$ has singular $1/u$ behavior.

The continuum differential equations imply
that $\tilde\gamma$ and $\tilde\delta$ both vanish, and have vanishing
first derivatives, at the boundary.
To deal with the $u\eq0$ regular singular point in the
discretized equations for $\tilde\gamma$ and $\tilde\delta$ it is
sufficient to replace the equations at $u\eq0$
with constraints setting $\tilde\gamma$ and $\tilde\delta$ to zero.
If we wished to fix the radial shift $\lambda$ by simply
specifying its value, we could similarly redefine 
$\zeta = \lambda + u \, \tilde \zeta$ and require
$\tilde \zeta$ to vanish at the boundary.
However, we found it more convenient to fix $\lambda$
indirectly by demanding that $\zeta$ vanish at our chosen
value of $u_{\rm end}$.
Referring to Eqs.~(\ref{eq:transform}) and (\ref{ansatz1}),
one sees that this condition will make the $u = u_{\rm end}$ surface
coincide with a surface of constant FG radial coordinate,
$\tilde \rho = u_{\rm end}$.
In other words, with this condition the FG computational domain
$\tilde\rho \in [0,\tilde\rho_{\rm end}]$ 
is the same as the EF domain $u \in [0,u_{\rm end}]$.

The net effect of the above procedure,
in the discretized equations for
$\zeta$, $\tilde\delta$ and $\tilde\gamma$
at longitudinal position $z_l$, is to replace the
the (degenerate) equations at $u\eq0$ by
the respective constraints%
\footnote
    {%
    Although not required, we also replaced a second row in the
    linearized equation for $\zeta$ by the condition that
    the first derivative of $\zeta$ vanish on the boundary,
    $
    \sum_{j}(D_u)_{0j}\,\zeta_{1,j,l}=0 
    $.
    The continuum equations automatically imply this behavior,
    but imposing it explicitly in the discretized equations
    helped to minimize precision loss associated with unwanted 
    solutions that diverge on the boundary.
    }
\begin{equation}
    \zeta_{M,N_u-1,l} = 0 \,,\quad
    \tilde\delta_{1,0,l} = 0 \,,\quad
    \tilde\gamma_{1,0,l} = 0 \,.
\end{equation}

In addition to applying boundary conditions at $u \eq 0$, when using
domain decomposition one must also impose continuity conditions at
subdomain boundaries.
Our set (\ref{eq:ujk}) of radial grid points redundantly duplicates
the interior endpoints of each subdomain,
$
    u_{j,N_u-1} = u_{j+1,0}
$
for $j = 1,{\cdots},M{-}1$,
and hence two different rows of the linear equation $\mathcal M f = -S$
represent the differential equation evaluated at the same physical point.
One could deal with this by eliminating the duplication of subdomain endpoints
and suitably redefining the differentiation matrix $D_u$.
But it is even easier to fix the problem by simply replacing one of the
rows representing an interior subdomain endpoint with a constraint equation
enforcing the equality of duplicated function values at this point,
$f_{j,N_u-1,l} - f_{j+1,0,l} = 0$.%
\footnote
    {%
    There is a subtlety involving the choice of which row to replace as,
    relative to a given interior subdomain endpoint, one row approximates $u$
    derivatives using information on one side of the endpoint, while the
    other row approximates $u$ derivatives using information on the other side.
    Since the behavior of the transformation functions is fixed, and known,
    at the $u \eq 0$ boundary,
    one should regard the transformation equations (\ref{diffeq1})
    as describing the propagation of information from the boundary into the bulk.
    Consequently, one should retain the row corresponding to
    $u_{j,N_u-1}$ and replace the row corresponding to $u_{j+1,0}$.
    }

After these row replacements, the modified linear system is reasonably
well conditioned and, with a sufficiently good initial guess,
Newton iteration rapidly converges quadratically.
To generate an initial guess, it is natural to work sequentially in $z$.
If the shock is propagating in the $+z$ direction with the profile function
$h(z)$ having its maximum at $z \eq 0$,
then at the furthest point behind the shock,
$z_0 = -L_z/2$, the geometry deviates negligibly from pure AdS and
$\zeta = \tilde\gamma = \tilde\delta = 0$ is a fine initial guess.
Thereafter, we use the converged solution at each $z_i$ as an initial
guess for the solution at $z_{i+1}$.
This provides a good initial guess provided the longitudinal grid spacing
is sufficiently fine.

The above procedure for solving the transformation equations (\ref{diffeq1})
using spectral methods works well as long as the radial depth $u_{\rm end}$ to
which one integrates is not too large.
The key advantage of this approach is that the precision of the obtained
solutions do not degrade near the boundary, even through $u\eq0$ is a singular
point of the differential equations.
That is to say, spectral methods are excellent for finding well-behaved
solutions of equations having regular singular points.
However,
as $u_{\rm end}$ increases the linear operators one inverts in
this Newton iteration scheme become increasingly ill-conditioned.
Unfortunately, the depth to which one must integrate in order to
locate the apparent horizon (discussed next) after superposing shocks
grows with increasing separation of the initial shocks.
We used two strategies to cope with this difficulty.

First, following Refs.~\cite{Chesler:2013lia, Chesler:2010bi}, we
added a small artificial background
energy density $\epsilon_0$ when superposing shocks as described above.
Increasing the background energy density decreases the depth at which
an apparent horizon forms.
Second, after using the above approach to find the transformation functions
for $u < u_{\rm end}$,
we integrate further into the bulk by switching to an adaptive 4th order
Runge-Kutta integrator,
with the spectral solution at $u_{\rm end}$ providing initial data.
(A description of this standard integrator is given in appendix~\ref{RK4}.)
For simplicity, we choose to integrate to a fixed value
$u = u_{\rm max}$,
instead of a fixed value of $\tilde\rho$.

For our chosen range of shock parameters,
with widths down to $w \eq 0.075$, using a spectral grid
down to $u_{\rm end} \eq 2$ worked well.
With a longitudinal box size $L_z \eq 12$ and background energy densities
in the range of 1--5\% of the peak energy density,
it turned out that only a modest further integration 
with the adaptive integrator
down to $u_{\rm max} = 2.11$ was sufficient to reach the apparent horizon
throughout the longitudinal box.%
\footnote
    {%
    For the parameters which we chose,
    displayed in  Table~\ref{tab:cases} and
    discussed below in section~\ref{sec:results},
    it turned out that using an adaptive integrator
    to probe deeper into the bulk was not essential,
    as the apparent horizon was found to lie within
    the domain of integration reached with spectral methods.
    However,
    as we used a relaxation algorithm to find the horizon, it was
    convenient to have additional surplus depth available,
    especially for small values of $\epsilon_0$,
    since on some early iteration steps the current guess for the
    apparent horizon would lie deeper than the final value,
    possibly beyond the spectral solution endpoint.
    }
Having transformed a right-moving single shock solution to
infalling coordinates, and extracted the resulting initial
data $\{B_+,a_+^{(4)},f_{z+}^{(4)},\lambda_+\}$ for evolution
using Eqs.~(\ref{eq:B})--(\ref{eq:bdrydata}),
a simple reflection generates
corresponding data for a left-moving shock,
\begin{subequations}
\begin{align}
    B_-(u,z) &= B_+(u,-z) \,,
    & a_-^{(4)}(z) &= a_+^{(4)}(-z) \,,
\\
    \lambda_-(z) &= \lambda_+(-z) \,,
    & f_{z-}^{(4)}(z) &= -f_{z+}^{(4)}(-z) \,.
\end{align}
\end{subequations}

We construct initial data
for counter-propagating shocks by combining single shock solutions
as described earlier in Eqs.~(\ref{eq:combine})--(\ref{eq:BED}).
We chose the initial time $t_0$ for this superposition so that
the initial separation between the shocks, $\Delta z_0 = -2t_0$,
is large compared to the shock widths.
We used $\Delta z_0 = 4$ for symmetric collisions of broad shocks,
$\Delta z_0 = 2$ for symmetric collisions of thin shocks, and
$\Delta z_0 = 3$ for asymmetric collisions of shocks.

For thin shockwave collisions with small background energy density,
avoiding numerical instabilities associated with short wavelength
perturbations is challenging.
As discussed in Ref.~\cite{Chesler:2013lia}, it is helpful
to damp discretization induced perturbations using appropriate
filtering.
We constructed and applied smoothing filters to the initial data
in both longitudinal and radial directions.
Details of these filters are presented in
appendix~\ref{Near boundary filter}.

\subsection{Horizon finding}
\label{sec:horizon}

After transforming chosen single shock solutions to infalling
coordinates, as just discussed, and combining two counter-propagating
shocks as shown in Eqs.~(\ref{eq:combine})--(\ref{eq:BED})),
the final step in the construction of initial data is locating
the apparent horizon which serves as an IR cutoff in the bulk.%
\footnote
    {%
    One subtlety is that the transformation to infalling coordinates
    is only computed to some finite depth $u_{\rm max}$.
    For a given configuration of initial shocks and chosen
    value of the background energy density $\epsilon_0$,
    it is a matter of trial and error to find a value of
    $u_{\rm max}$ for the transformation which is sufficiently deep
    so that the apparent horizon lies above this depth,
    for all values of $z$ within the computational domain.
    The required value of $u_{\rm max}$ increases with the
    size of the longitudinal domain and separation of the
    initial shocks.
    }

In our planar shock geometries, the apparent horizon condition
(\ref{eqn:horizonPosition}) becomes
\begin{equation}
    0 = 
    d_+ \Sigma 
    +
    \frac {e^{2B}}{6 \Sigma^2}
    \left(
	3 F^2 \, \partial_r \Sigma
	+ 2 \Sigma \, \partial_z F
	+ 4 F \, \Sigma \, \partial_z B + 2 F \, \partial_z \Sigma
    \right)
    \Big|_{r=r_h} .
\label{findh}
\end{equation} 
A radial shift, $r = \bar r + \delta\lambda$,
corresponds in our inverted radial coordinates to
\begin{equation}
    u = \frac{\bar{u}}{1+\bar{u} \, \delta \lambda} \,.
\label{eq:ushift}
\end{equation}
If $\bar u \in [0,u_{\rm max}]$ represents the radial coordinate
used in the transformation to infalling coordinates,
then we wish to determine the value of a further shift
$\delta\lambda(z)$ such that condition (\ref{findh}) holds
at some value of $u_h \equiv 1/r_h$ which may, for convenience,
be chosen to equal the same value $u_{\rm max}$ from the
coordinate transformation.
With this choice, $\delta\lambda$ must be negative for the
sought-after apparent horizon to lie within the coordinate
transformation domain.

Equation (\ref{findh}) is a nonlinear but
ordinary differential equation for the shift function
$\delta\lambda(z)$.
To solve it,
we use spectral methods (with the same Fourier grid in $z$)
combined with a root finding routine.
Linearizing equation (\ref{findh}) in $\delta \lambda$
allows us to apply Newton iteration.
Each iteration step starts with a trial value of
the radial shift, $\delta \lambda^{(m)}$ in iteration $m$,
and computes the residual
(i.e., the right-hand side of Eq.~(\ref{findh}))
and its variation with respect to $\delta\lambda$,
and solves the linearized equation to find an improved
value $\delta\lambda^{(m+1)}$ of the shift.

To evaluate the residual and its variation,
we first integrate Eqs.~(\ref{eqn:Sigma})--(\ref{eqn:d+Sigma}),
using the current value of $B(z,u)$ and $\lambda(z)$,
to find the auxiliary functions $\Sigma$, $F$ and $d_+\Sigma$.%
\footnote
    {%
    Explicit forms of these equations are shown in
    appendix~\ref{subsection:Appendix_EE}.
    After the first integration of these equations, one could
    thereafter use off-grid spectral interpolation to evaluate
    the radially-shifted auxiliary functions on the spectral grid.
    But it is just as easy to reintegrate 
    Eqs.~(\ref{eqn:Sigma})--(\ref{eqn:d+Sigma}) on every Newton iteration step.
    }
After each step we convert the spectral representation of  $B(z,u)$
to a new radial grid 
with grid points shifted according to Eq.~(\ref{eq:ushift}).
To do so,
we perform off-grid interpolation using a sum of Chebyshev
cardinal functions \cite{Boyd:Spectral} with coefficients given by
the on-grid values of $B(z,u)$.

For  our settings of longitudinal box size and shock parameters,
we found it advantageous to choose the initial guess
$\delta \lambda^{(0)}$  to be $0.1$.
It was also helpful to start with a relatively large background
energy density $\epsilon_0$ of about $10\%$ of the peak shock energy
density, and then gradually decrease $\epsilon_0$ during each iteration step
until it reached the desired final value before Newton iteration
convergence.

During time evolution, described next,
solving the horizon stationarity condition (\ref{eqn:HorizonStationary})
on each time step yields the time derivative of the radial shift 
thereby providing the information needed to integrate
$\lambda$ forward in time.
(The explicit form of Eq.~(\ref{eqn:HorizonStationary}) for our
planar shock geometries is given in Eq.~(\ref{stationary}).)
To prevent discretization errors from driving long term drift
away from the desired horizon condition (\ref{findh}),
we also directly recomputed the apparent horizon position every
10--100 time steps using the above iterative procedure.

\subsection{Time evolution}

As described above in section (\ref{sec:strategy}),
the data on any time slice needed to integrate forward in time
consists of $\{ B(z,u),\, a^{(4)}(z),\, f^{(4)}(z),\, \lambda(z) \}$.
To compute the time derivative of this data,
we successively solve Eqs.~(\ref{eqn:Sigma})--(\ref{eqn:A})
as discussed earlier.
Explicit forms of these equations are given in
Eqs.~(\ref{eq1})--(\ref{eq5}) of appendix~\ref{subsection:Appendix_EE}.
We use the same multi-domain spectral methods described above
in section \ref{sec:transform}.
These methods presume that functions being represented by
their values on the spectral grid are well behaved throughout
the computational domain.%
\footnote
    {%
    See, for example, Ref.~\cite{Boyd:Spectral} for a good discussion of the
    connection between analyticity properties and
    convergence of spectral representations.
    }
Our functions $\Sigma$ and $A$ have divergent near-boundary
behavior, as shown in Table \ref{tab:boundary_conditions},
so for computational purposes we use
redefined functions in which the leading near-boundary
behavior is subtracted.
For most functions,
we also choose redefinitions such that the
new functions either have known non-zero boundary values
or vanish linearly at the boundary.
Specifically, we use the following redefinitions,
\begin{subequations}\label{eq:redefs}%
\begin{align}
    B(u,z,t)
    &=
    \Big(\frac{u}{1+u\lambda} \Big)^3 \> b(u,z,t) \,,
\label{redefinition_1}
\\[2pt]
    \Sigma(u,z,t)
    &=
    \Big(\frac{u}{1+u\lambda} \Big)^{-1}
    + \Big(\frac{u}{1+u\lambda} \Big)^4 \> \sigma(u,z,t) \,,
\label{redefinition_2}
\\[2pt]
    F_z(u,z,t)
    &=
    -\partial_z \lambda
    + \Big(\frac{u}{1+u\lambda} \Big)^2 \> f(u,z,t) \,,
\\[2pt]
    d_+\Sigma(u,z,t)
    &=
    \tfrac{1}{2} \Big(\frac{u}{1+u\lambda}\Big)^{-2}
    +\Big(\frac{u}{1+u\lambda} \Big)^2 \> d_+\sigma(u,z,t) \,,
\label{eq:d+sig}
\\[2pt]
    d_+B(u,z,t)
    &=
    \Big(\frac{u}{1+u\lambda} \Big)^2 \> d_+b(u,z,t) \,,
\label{eq:d+b}
\\[2pt]
    A(u,z,t)
    &=
    \tfrac{1}{2}\Big(\frac{u}{1+u\lambda}\Big)^{-2}
    + a(u,z,t),
\end{align}
\end{subequations}
We use factors of $u/(1+u\lambda) = (r+\lambda)^{-1}$ in these
redefinitions, instead of pure powers of $u$,
so that the new functions transform simply under radial shifts.
This is natural as it preserves manifest radial shift
covariance in the equations for the new functions,
but is not essential.
In relations (\ref{eq:d+sig}) and (\ref{eq:d+b}),
and henceforth,
$d_+\sigma$ and $d_+b$ are simply names for 
redefined functions encoding $d_+\Sigma$ and $d_+B$, respectively,
and are not themselves modified $d_+$
time derivatives applied to $\sigma$ or $b$.

Referring to Table~\ref{tab:boundary_conditions} and
Eq.~(\ref{eqn:near-boundary-EF}),
one sees that the new functions $b$ and $d_+b$
vanish linearly as $u \to 0$, while $f$ and $d_+\sigma$ have
non-zero boundary values of $f_z^{(4)}$ and $a^{(4)}$,
respectively.
The new function $a$ has a boundary value of $-\partial_t \lambda$
which is an output, not an input, of the radial integration determining~$a$.

Arranging to have constant or linear near-boundary behavior 
of redefined functions minimizes the precision loss which
can occur when evaluating derivatives very near the boundary.
In particular, extracting the third power of $u/(1+ u\lambda)$ in
the definition (\ref{redefinition_1}) of $b$
is essential for the  numerical stability.

After inserting the redefinitions (\ref{eq:redefs}) into
the relevant radial equations (\ref{eq1})--(\ref{eq5}),
it is crucial to simplify the resulting
equations, prior to numerical implementation,
in such a way that cancellations of terms with the most
divergent near-boundary behavior are performed exactly,
analytically.
When each radial differential equation is written in canonical
form (with a unit coefficient of the highest order $u$-derivative),
no term in the inhomogeneous source term of the equation should be
more singular than $1/u$ for first order and $1/u^2$ for second order equations,
otherwise unnecessary precision loss will occur during the numerical
evaluation of the equation.%
\footnote
    {%
    Such analytic simplification, eliminating what would otherwise be
    huge cancellations near the boundary, is essential when performing
    calculations using machine precision (64 bit) arithmetic.
    If one instead uses arbitrary precision arithmetic
    (in, for example, Mathematica),
    one might think such careful simplification prior to programming
    is unnecessary.
    However, failure to properly simplify expressions will then
    require the use of extraordinarily high precision arithmetic
    with concomitant poor performance.
    }

In solving the successive radial equations
Eqs.~(\ref{eqn:Sigma})--(\ref{eqn:A})
[or (\ref{eq1})--(\ref{eq5})],
we implement the following boundary conditions at $u\eq0$
using the row replacement technique described in
section~\ref{sec:transform},
\begin{align}
	\sigma(0,z) = 0
	\,,\quad
\label{bc_start}
	f(0,z) = f^{(4)}(z)
	\,,\quad
	d_+\sigma(0,z) = a^{(4)}(z)
	\,,\quad
	d_+b(0,z) = 0 \,.
\end{align}
Equation (\ref{eqn:Sigma}) [or (\ref{eq1})] for $\Sigma$ is a
second order differential equation, but after conversion to
an equation for $\sigma$ both homogeneous solutions are
divergent at the boundary and lie outside our
spectral representation function space.
Using row replacement to encode $\sigma(0,z) \eq 0$ 
(which is an automatic consequence of the differential equation
for $\sigma$) is the easiest way to handle the singular
boundary point on the spectral grid.
For the $F$ equation (\ref{eqn:F}) [or (\ref{eq2})],
after conversion to an equation for $f$
only one boundary condition fixing the
coefficient $f_z^{(4)}$ of the normalizable homogeneous
solution is needed as the spectral representation for $f$
automatically precludes any non-normalizable homogeneous solution.
Likewise for the $d_+\Sigma$ equation (\ref{eqn:d+Sigma})
[or (\ref{eq3})],
a single boundary condition fixing the
coefficient $a^{(4)}$ of the normalizable homogeneous solution
is needed.
For the $d_+B$ equation (\ref{eqn:d+ghat})
[or (\ref{eq4})],
after conversion to an equation for $d_+b$ the one homogeneous
solution is again outside the spectral representation function
space, and encoding $d_+b(0,z) \eq 0$ via row replacement
is again the easiest way to handle the singular boundary point.

Finally, for the $A$ equation (\ref{eqn:A})
[or (\ref{eq5})],
after conversion to an equation for the new function $a$
the non-normalizable homogeneous solution is automatically
excluded by the spectral representation for $a$.
One boundary condition is needed
to fix the coefficient of the normalizable homogeneous solution.
Referring to table \ref{tab:boundary_conditions},
specifying the boundary value of $a$ is the same as
fixing the time derivative of the radial shift,
$a(0,z) = -\partial_t \lambda(z)$.
But we do not wish to input some arbitrary choice for this
time derivative.
Instead, prior to solving the $A$ equation (\ref{eqn:A})
we first solve the horizon stationarity condition
(\ref{eqn:HorizonStationary}) which determines the
value of $A$ on the horizon.
This is an inhomogeneous differential equation 
involving $A$ and its first and second order longitudinal
derivatives, evaluated on the horizon.
The explicit form is given in (\ref{stationary}) of
appendix~\ref{subsection:Appendix_EE}.
Then, to solve the radial equation (\ref{eqn:A})
[or (\ref{eq5})], converted to an equation for $a$,
we replace the $u \eq 0$ row in the spectral discretization
of this equation with a row fixing the value of $a$ at the horizon,
i.e., equating $a(u_{\rm max},z)$ to the value determined by
the horizon stationarity condition.

To recap, every time step begins with the sequential solution of
equations (\ref{eq1})--(\ref{eq5}), plus the horizon stationarity
condition,
using the same spectral methods and Chebyshev grid employed
in the preparation of initial data.
This yields $\Sigma$, $F$, $d_+\Sigma$, $d_+B$ and $A$,
from which the ordinary time derivatives of 
$B$, $a^{(4)}$, $f_z^{(4)}$ and $\lambda$ are extracted using
relations (\ref{update_B})--(\ref{update_af}).
This is the information needed to integrate forward
in time.

To perform time integration we use a discrete 
approximation with non-zero time step $\delta t$.
We specifically choose the well known
fourth order Runge-Kutta method (RK4),
which uses four ``substeps'' each involving the evaluation
of time derivatives, performed as described above for each
point on our longitudinal grid.
(The relevant RK4 formulas are shown in appendix \ref{RK4}.)

A time step $\delta t = 0.002$ was used in all integrations,
which was sufficient to deliver stable evolution for
all shock widths considered.
For broader shocks ($w>0.3$) a lower order time integrator
would have sufficed, but for shocks with width $w < 0.1$
we found using at least RK4 to be essential,
with our time step, to achieve accurate results.
After each time step of the evolution we
filter the final results for the propagating data
$\{B,a^{(4)}, f^{(4)}, \lambda\}$ in the longitudinal direction
using a low-pass filter,
as detailed in \ref{Filtering},
which damps the upper third of the spectral bandwidth.
The filtering is applied to the final RK4  outputs,
not during the RK4 substeps.
This damps short wavelength discretization-dependent fluctuations;
such filtering should be viewed as a part of the spectral
discretization prescription.
We do not apply filtering to any interim results while solving
the radial equations (\ref{eqn:Sigma})--(\ref{eqn:A}).

\section{Results} 
\label{sec:results}

\subsection {Calculated collisions}

Using the above described techniques and associated software,
planar shock collisions were computed for various combinations
of incoming shock widths.
All initial shocks had Gaussian profiles (\ref{eq:h})
and identical
transverse energy density $\mu^3$.
In units in which
$\mu \equiv 1$,
shock widths ranged between 0.075 and 0.35.
For technical reasons involving the damping of numerical artifacts,
as discussed above, an artificial background energy
density was added whose size ranged from $5.5\%$ down to $1.2\%$ of
the peak energy density of the narrower shock.
Periodic boundary conditions
were applied in the longitudinal direction, with this dimension
then discretized with a uniformly spaced (Fourier) grid
having of up to $N_z = 720$ points.
The longitudinal period $L_z$ was set to 10, 11, or 12
for collisions of narrow, asymmetric, or broad shocks, respectively.
In the radial direction, domain decomposition with
$M=22$ subdomains of uniform size in the inverted radial
coordinate $u = 1/r$ was used, with a Chebyshev-Gauss-Lobatto grid
of 
$N_u = 13$ points within each subdomain.
Time evolution used RK4 time-stepping with a step size
$\delta t =0.002$
and total time duration ranging from $t=5/\mu$  to $t=6/\mu$.
Table \ref{tab:cases} lists the parameters of specific
calculations.

\begin{table}
\begin{center}
\begin{tabular}{ccccccccc}
run & $w_+$ & $w_-$ & $N_z$ & $\epsilon_0$ 
\\
\hline
1 & 0.35 & 0.35 & 720 & $\{ 0.055,\, 0.066 \}$ 
\\
2 & 0.25 & 0.25 &480 & $\{ 0.039,\, 0.045 \}$ 
\\
3 & 0.1 & 0.25 &660 & $\{ 0.015,\, 0.017 \}$ 
\\
4 & 0.1 & 0.1 & 600 & $\{ 0.015,\, 0.017 \}$ 
\\
5 & 0.075 & 0.35 &660 & $\{ 0.012,\, 0.015 \}$ 
\\
6 & 0.075 & 0.25 &660 & $\{ 0.012,\, 0.015 \}$ 
\\
7 & 0.075 & 0.075 &600 & $\{ 0.012,\, 0.015 \}$ 
\end{tabular}
\end{center}
\caption
    {%
    Physical and computational parameters
    of specific computed collisions.
    Shown are the incoming shock widths $w_\pm$,
    number of longitudinal grid points $N_z$, and
    background energy densities $\epsilon_0$.
    Shock widths $w_\pm$ 
    are measured in units of $\mu^{-1}$.
    The background energy density $\epsilon_0$ is in units of the
    peak energy density of the narrower shock,
    or $\mu^3 w_+^{-1}/\sqrt{2\pi}$.
    Computed results at the two listed values of $\epsilon_0$
    were used to extrapolate to vanishing background energy density.
    \label{tab:cases}
    }
\end{table}

\begin{figure}[htb]
\centering
  \begin{center}
  \hspace*{-5pt}%
  \includegraphics[scale=0.30]{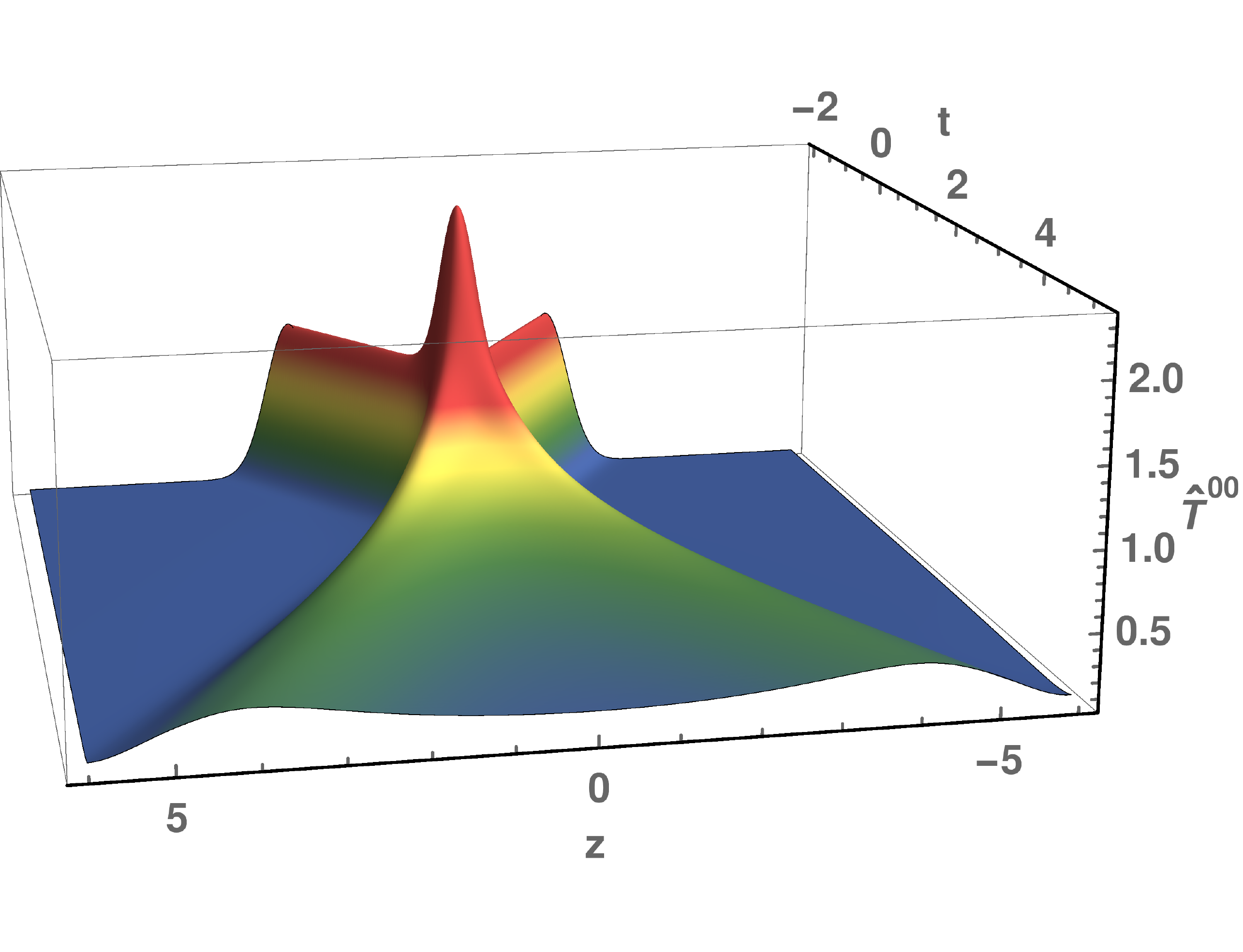}\hfill
  \includegraphics[scale=0.30]{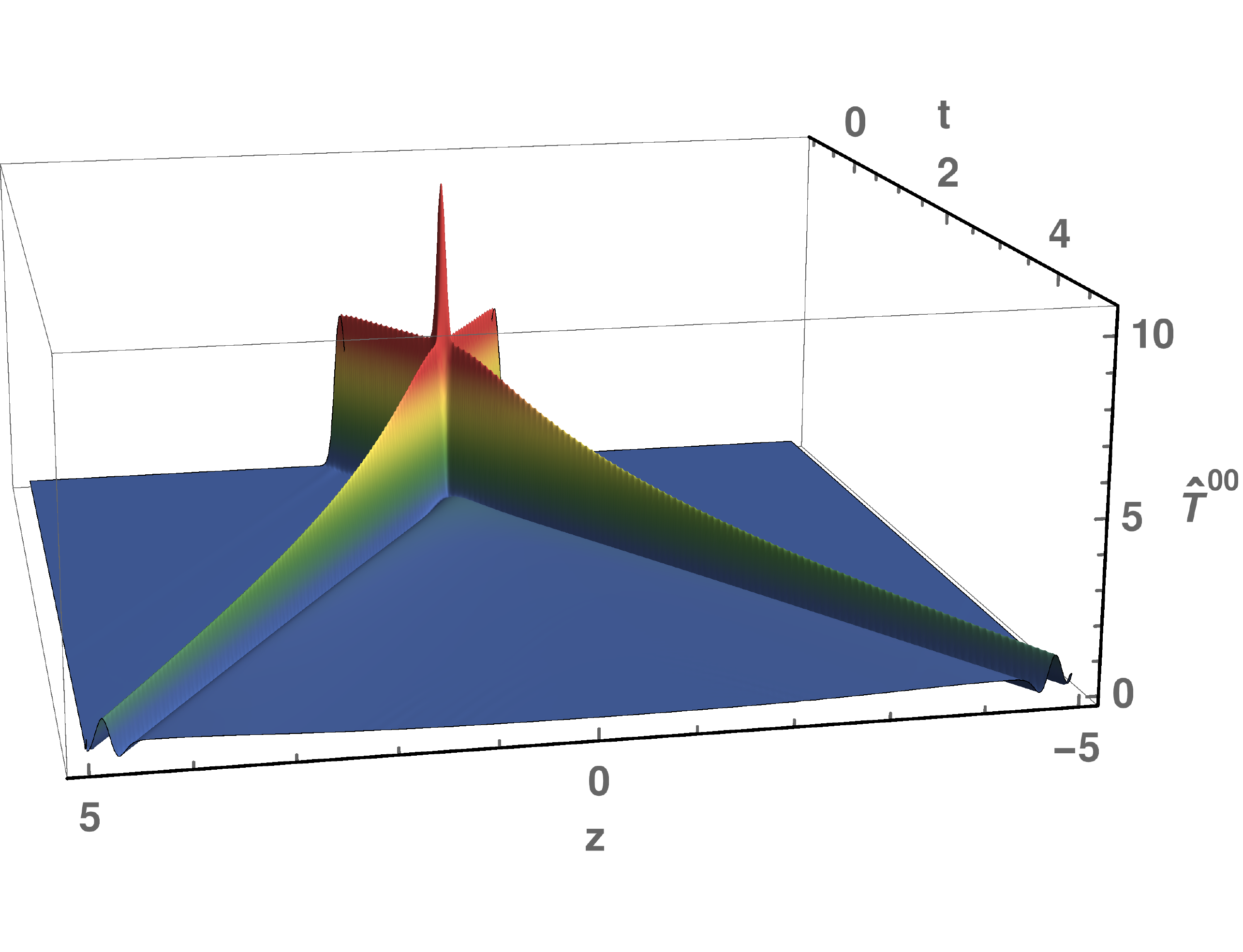}%
  \hspace*{-20pt}
  \\[-25pt]
  \includegraphics[scale=0.3]{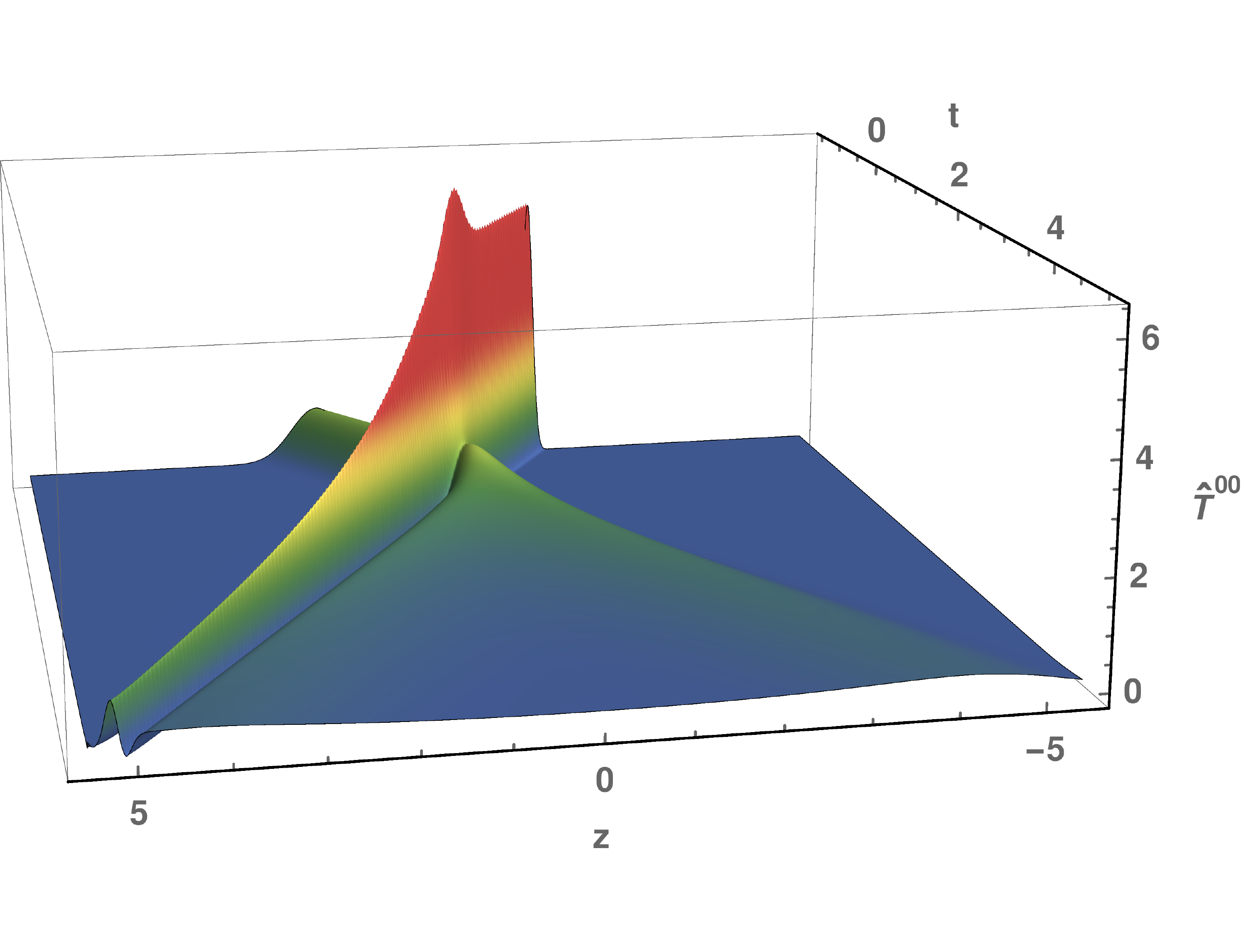}
  \vspace*{-25pt}
  \end{center}
  \caption   {%
    The energy density $\widehat T^{00}(t,z)$ plotted as
    a function of time $t$ and longitudinal position $z$ for
    symmetric collisions with shock width
    $w_\pm = 0.35/\mu$ (upper left) and
    $w_\pm = 0.075/\mu$ (upper right),
    and the corresponding asymmetric collision (bottom)
    involving shocks of widths
    $(w_+,w_-) = (0.075/\mu,0.35/\mu)$.
    All shocks have equal transverse energy density $\mu^3$.
    \label{fig:output}
    }
\end{figure}

\begin{figure}[tp]
\centering
\includegraphics[scale=0.9]{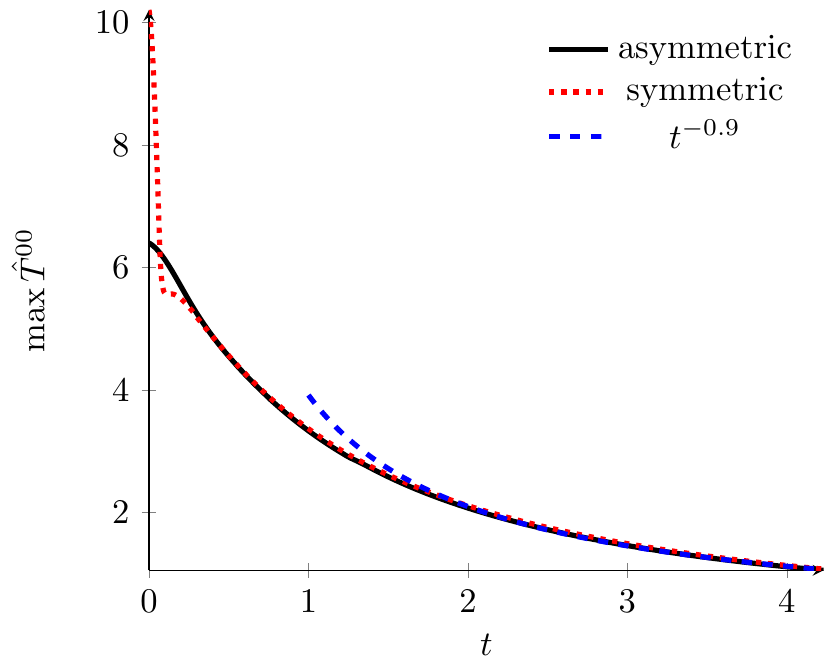}
\caption
    {%
    Comparison of time dependent amplitudes of the
    energy density maxima (associated with the thinnest shock)
    in a symmetric collision of two narrow shocks of width $w=0.075$
    (solid blue line)
    and an asymmetric collision of shocks
    having widths $w=0.075$ and $w=0.25$
    (dashed red line),
    with all incoming shocks having the same transverse energy density.
    Also shown is a $t^{-0.9}$ asymptotic form (long dashed blue line).
    Except for short time transients ($t \lesssim 0.3$),
    the maxima in the symmetric and
    in the asymmetric case behave identically.
    For $t \gtrsim 1.5$ the amplitude decrease is well described
    as $t^{-0.9}$,
    as previously found in Ref.~\cite{Chesler:2013lia}.
    \label{fig:maxima}
    }
\end{figure}

Figure \ref{fig:output} shows the energy density
$\widehat T^{00}(t,z)$, in units of $\mu^4$,
for three representative collisions.
The top row shows symmetric collisions of shocks
with widths $w_\pm = 0.35$ (upper left) and $w_\pm = 0.075$
(upper right), while the lower row displays results
from the corresponding asymmetric collision with 
$(w_+,w_-)=(0.075,0.35)$.

Local maxima in the
energy density are present on the forward lightcone,
as clearly seen in Fig.~\ref{fig:output}.
These local maxima lie outside the hydrodynamic region
(discussed below).
In asymmetric collisions,
the width of a given postcollision local maxima largely
reflects the width of the corresponding incoming projectile.
As shown in Fig.~\ref{fig:maxima},
the amplitude of these local maxima decay with the
same power-law time dependence seen in symmetric collisions.

\subsection {Hydrodynamic flow}
\label{Hydrodynamic flow}

At every spacetime event inside the forward lightcone of a collision,
the timelike eigenvector and corresponding eigenvalue of the
holographically computed stress-energy tensor determine the
fluid 4-velocity $u^\mu$ and proper energy density $\epsilon$,%
\footnote
    {%
    A real timelike eigenvector (\ref{eq:umu})
    can fail to exist in spacetime regions where hydrodynamics
    is not applicable \cite{Chesler:2013lia,Arnold:2014jva}.
    As we are interested in behavior in the
    hydrodynamic region, this is not a concern.
    }
\begin{equation}
    {\widehat T^\mu}{}_\nu \, u^\nu = -\epsilon \, u^\mu \,,
\label{eq:umu}
\end{equation}
with normalization $u^\mu u_\mu = -1$ and $u^0>0$.
Given the flow velocity and energy density,
we use the first order hydrodynamic constitutive relation
to construct the hydrodynamic approximation to the stress-energy
tensor,
\begin{equation}
    \widehat T^{\mu\nu}_{\rm hydro}
    = p \, g^{\mu\nu} + (\epsilon{+}p) \, u^\mu u^\nu + \Pi^{\mu \nu} \,,
    \label{hydro_approx}
\end{equation}
where the viscous stress (to first order in gradients)
is given by
\begin{equation}
    \Pi_{\mu\nu}
    =
    -\eta \,
    \left[
    \partial_{(\mu}u_{\nu)}
    + u_{(\mu} u^\rho \partial_\rho u_{\nu)}
    - \tfrac 13 \,  \partial_\alpha u^\alpha
    (\eta_{\mu\nu} + u_\mu u_\nu)
    \right]
    + \mathcal{O}(\partial^2) \,.
\end{equation}
For the conformal fluid of $\Nfour$ Yang-Mills theory,
the pressure
$
    p = \epsilon/3
$
and the shear viscosity
$
    \eta = (\epsilon/3)^{3/4} / \sqrt{2}
$.%
\footnote
    {%
    This value for $\eta$ has been rescaled by the same factor
    of $2\pi^2/\Nc^2$ used in the definition of the rescaled
    stress-energy tensor (\ref{eq:That}).
    }

Following Ref.~\cite{Chesler:2015fpa}, we define the spacetime
region $\mathcal{R}$ in which hydrodynamics provides a good description
as the largest connected region within the future lightcone in
which the normalized residual,
\begin{equation}
    \Delta
    \equiv
    \frac{1}{p} \, \sqrt{\delta T^{\mu \nu} \, \delta T_{\mu \nu}} \,,
\qquad
    \delta T^{\mu \nu}
    \equiv
    T^{\mu \nu}-T^{\mu \nu}_{\text{hydro}},
    \label{Delta}
\end{equation}
measuring the difference between the holographically computed
stress energy tensor and its hydrodynamic approximation,
is smaller than $0.15$.

\begin{figure}
\includegraphics[width=2.7in,height=2.2in]{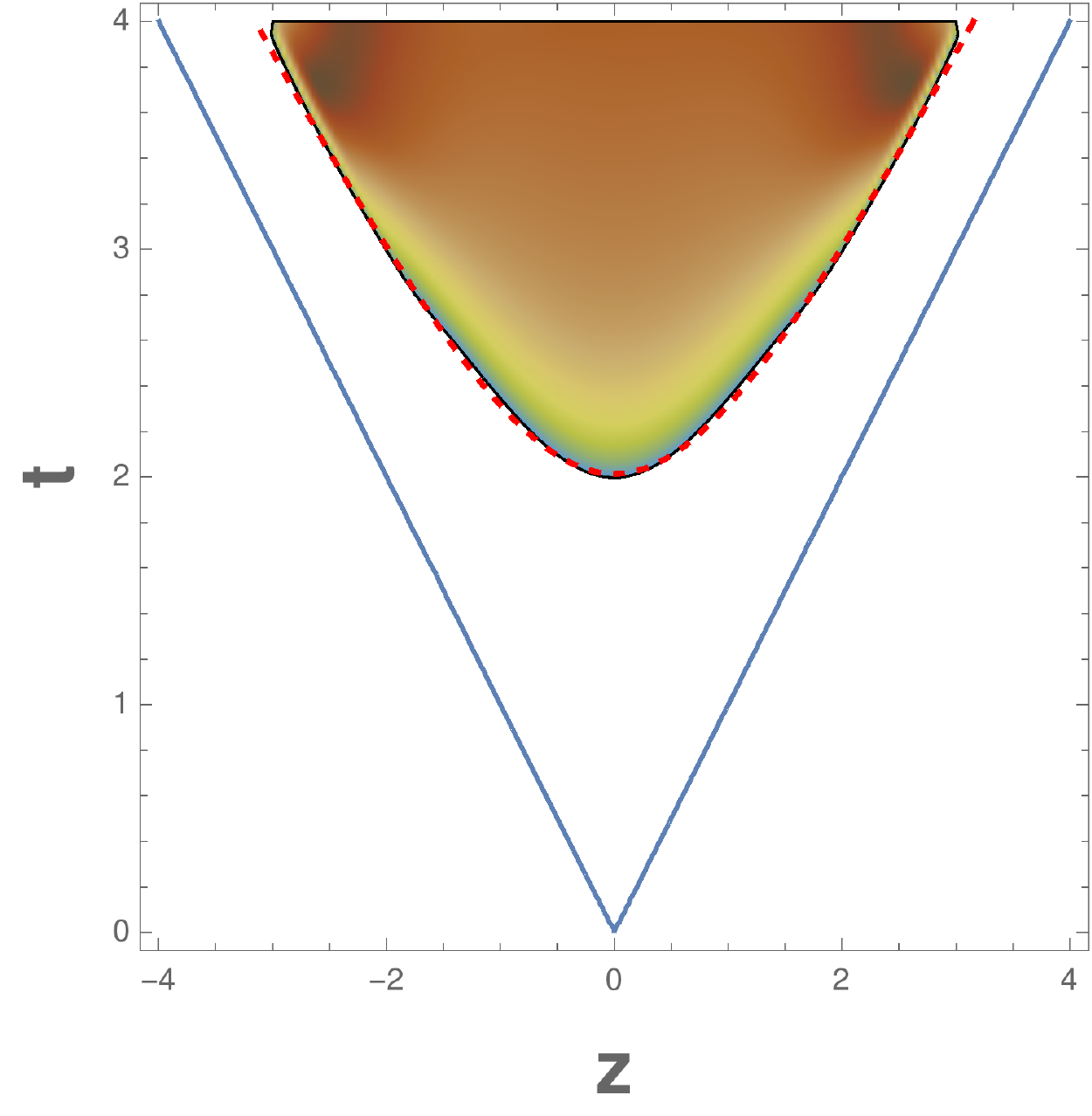}
\includegraphics[width=2.7in,height=2.2in]{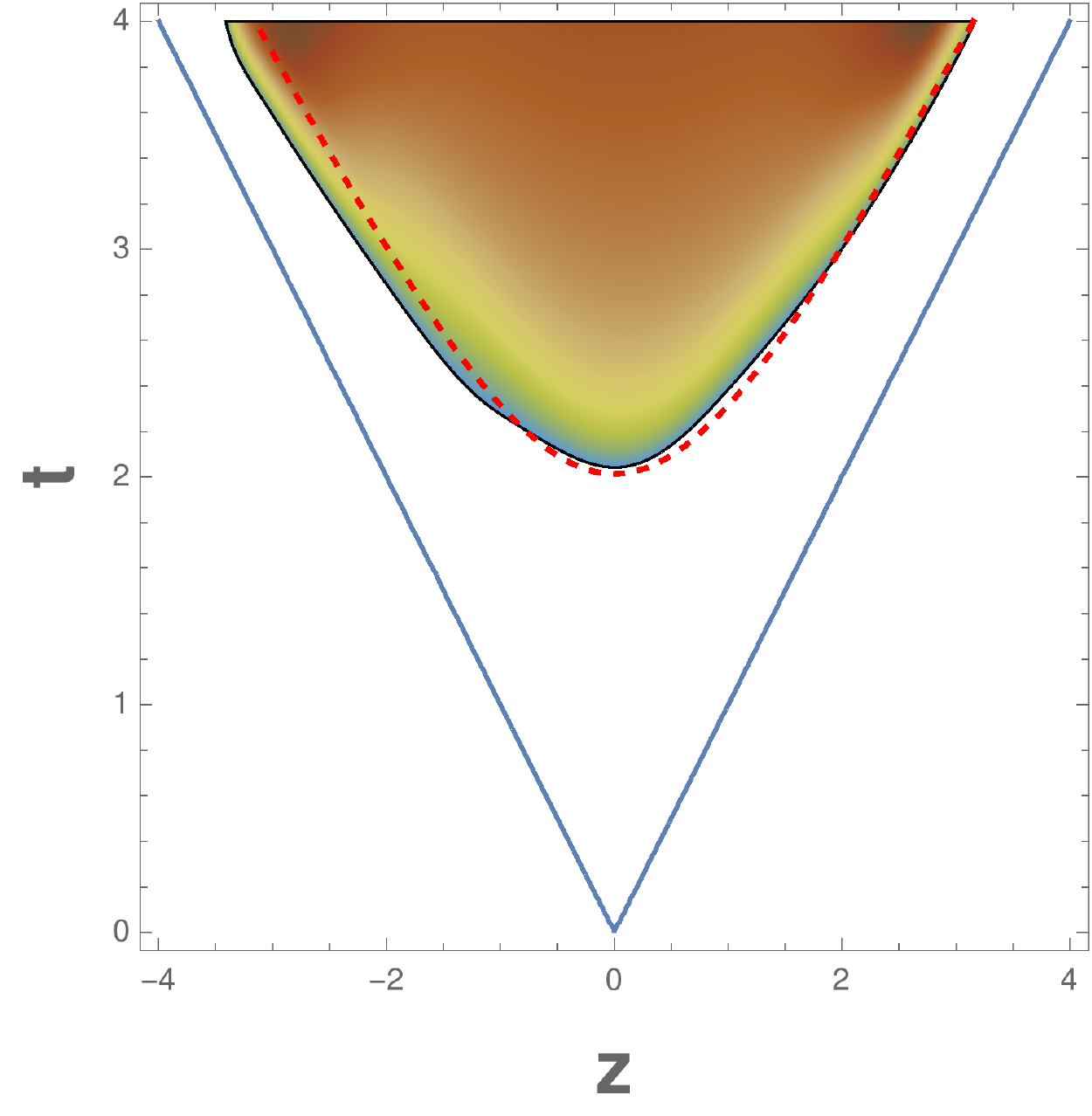}
\hfill
\raisebox{1em}{\includegraphics[scale=0.6]{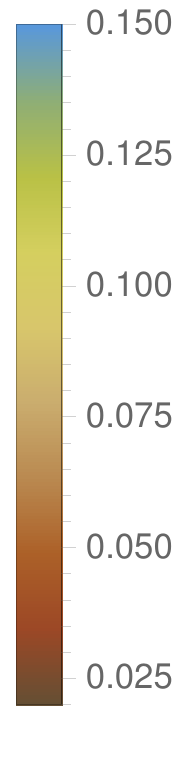}}
\caption
    {%
    The largest connected spacetime region $\mathcal{R}$
    in which the hydrodynamic residual $\Delta < 0.15$ 
    for a collision of narrow symmetric shocks,
    $w_\pm\eq 0.075$, on the left,
    and asymmetric shocks, $(w_+,w_-)=(0.1,0.25)$, on the right.
    The red dotted line shows the
    hyperbola $(t-0.5)^2-z^2 =\tau^2$, with $\tau =1.5$. For both
    asymmetric and symmetric collisions the region $\mathcal{R}$
    starts at $t_{\text{hydro}}\approx 2$.
    }
    \label{R_hydro}
\end{figure}  

For all collisions studied, symmetric and asymmetric,
with various combinations of incoming shock widths ranging
from 0.35 down to 0.075,
we found that the boundaries of the region $\mathcal{R}$
differ very little from one another, as  illustrated in Fig.~\ref{R_hydro}.%
\footnote
    {%
    By suitably adjusting the filtering of discretization induced artifacts,
    as discussed in the appendix~\ref{Filtering},
    we could decrease the background energy density in our computations
    of asymmetric collisions to about 1\% of the peak value
    of the energy density of the narrower shock.
    For asymmetric collisions
    it turned out to be quite challenging to achieve high precision
    and numerical stability with significantly smaller background energy densities.
    In this and subsequent figures,
    we perform a linear extrapolation to vanishing background energy density
    using calculated results at the non-zero background energy densities
    shown in table \ref{tab:cases}.
    At sufficiently late times, this linear extrapolation 
    ceases to be a reliable approximation to the limit of vanishing background
    energy density.
    A simple linear extrapolation, with our values of $\epsilon_0$,
    is adequate in the $t \leq 4$ interval displayed in Fig.~\ref{R_hydro},
    which coincides with the time interval
    shown in Ref.~\cite{Chesler:2015fpa} of
    the hydrodynamic region~$\mathcal R$.
    }
At $z=0$ we find that time at which hydrodynamics first becomes valid
(i.e., $\Delta <0.15$) to be essentially the same for asymmetric and
symmetric collisions and given by
\begin{equation}
    t_{\text{hydro}}\approx 2\,.
\end{equation}
In the symmetric case this confirms earlier results found in
Refs.~\cite{Chesler:2013lia,Chesler:2015fpa}.

\begin{figure}
\includegraphics[scale=0.80]{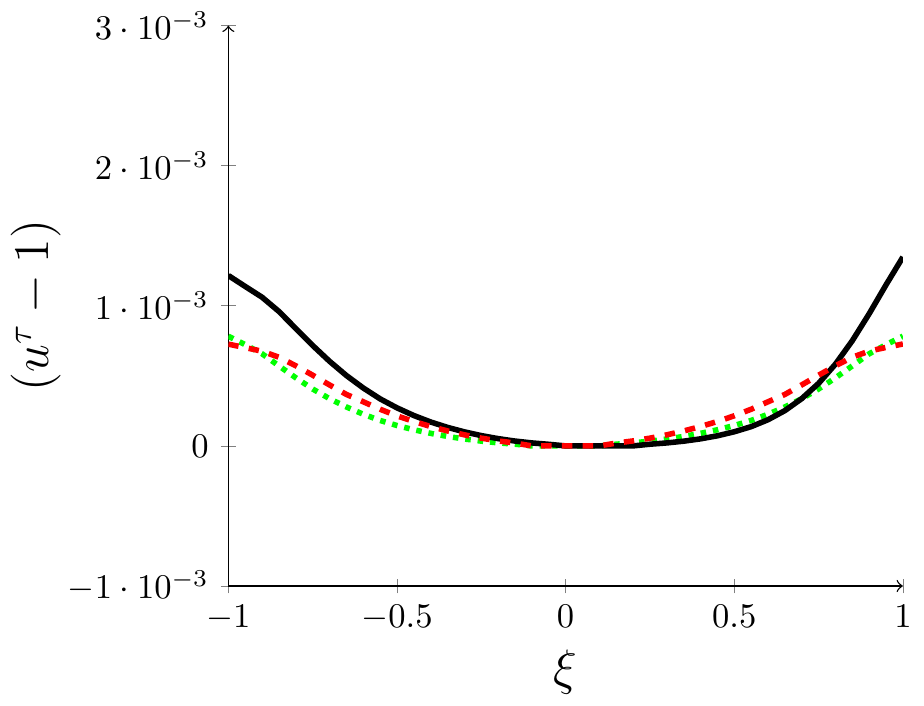}
\hfill
\includegraphics[scale=0.80]{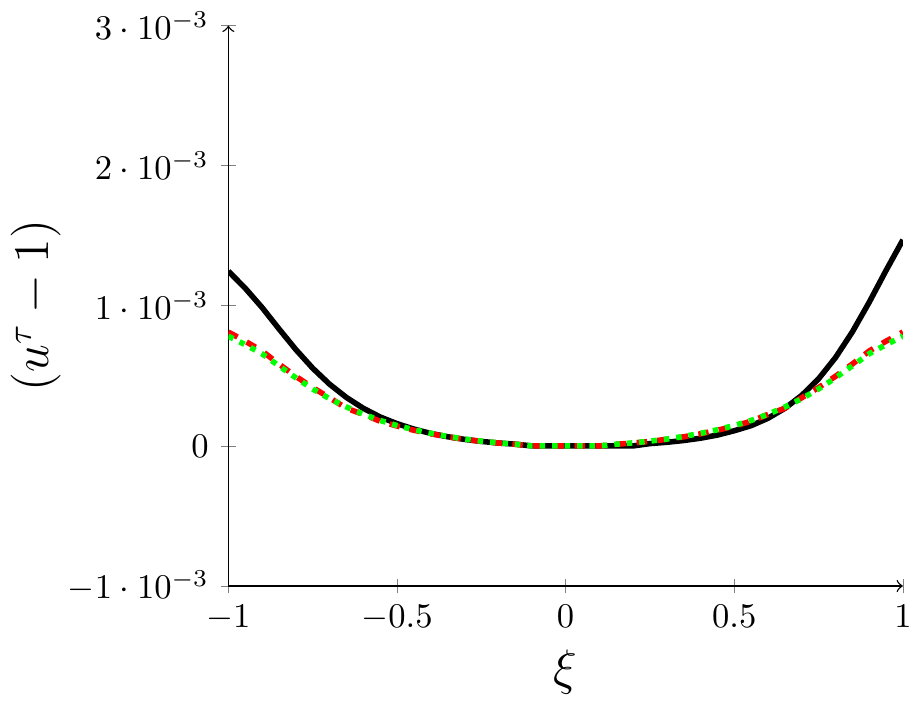}
\caption
    {%
    Left:
    the difference of the proper time
    component of the fluid velocity from unity,
    $u^\tau-1$, plotted as a function of rapidity
    at proper time $\tau = 3$ for the asymmetric collision
    $(w_+,w_-)=(0.075,0.35)$ (black line) and the symmetric collisions
    $w_\pm=0.075$ (red dashed line)
    and $w_\pm=0.35$ (green dotted line).
    Right:
    the analogous comparison for the asymmetric collision
    $(w_+,w_-)=(0.075,0.25)$ (black line),
    and corresponding symmetric collisions
    $w_\pm=0.075$ (red dashed line) and
    $w_\pm=0.25$ (green dotted line).
    As in Ref.~\cite{Chesler:2015fpa}
    we find that $u^\tau \approx 1$ with a deviation of a few parts
    in $10^{-3}$, showing that the fluid velocity
    is quite well described by boost invariant flow.
    }
\label{fig:flow}
\end{figure}

For symmetric collisions, we reproduced the key results of
Ref.~\cite{Chesler:2015fpa}:
boost invariant flow (\ref{eq:boostinvflow}) within the hydrodynamic
region to within a precision of $\mathcal{O}(10^{-3})$,
Gaussian rapidity dependence of the proper energy density
(\ref{eq:Gaussian}) at fixed proper time,
with the amplitude and width of this Gaussian well described
by the analytic forms (\ref{eq:Aandw}) at $\tau_{\rm init} = 3$.

\begin{figure}
\includegraphics[scale=0.77]{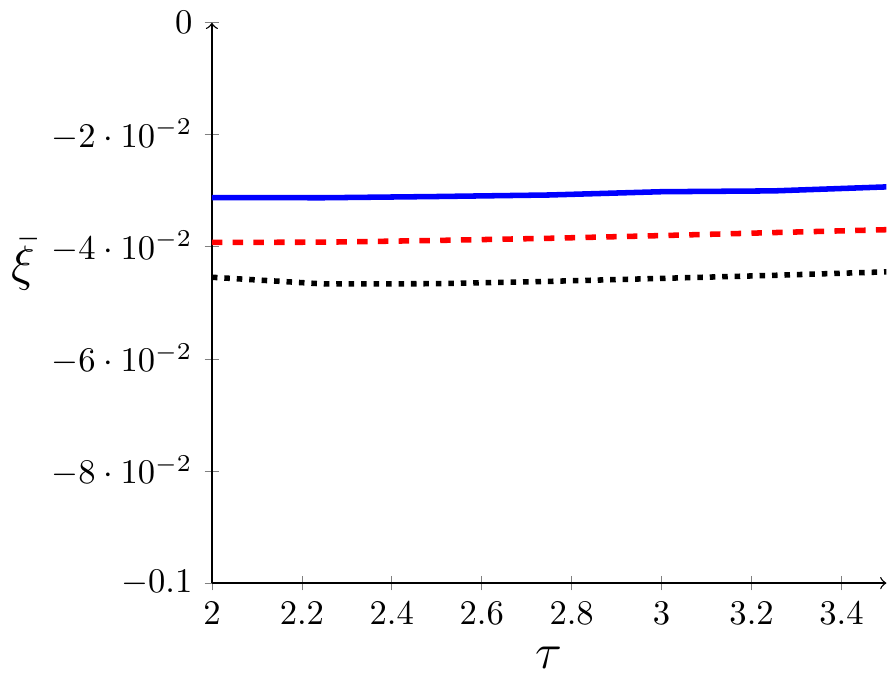}
\hfill
\includegraphics[scale=0.79]{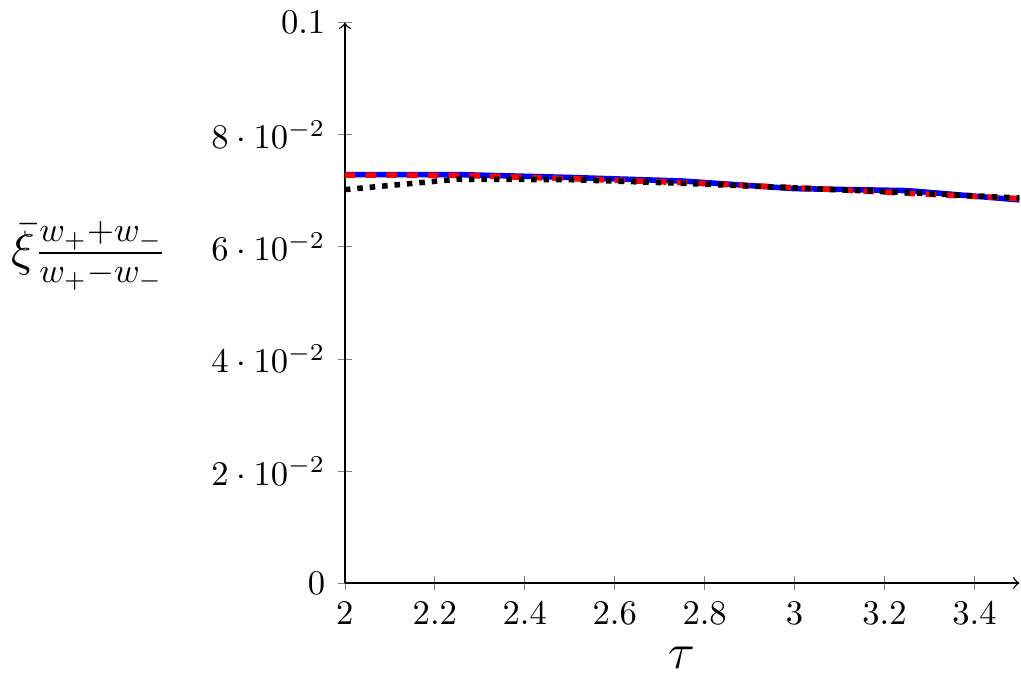}
\caption
    {     
    Left: the rapidity shift $\bar\xi(w_+,w_-;\tau)$
    of the proper energy density distribution,
    as a function of proper time $\tau$,
    for asymmetric collisions with shock widths
    $(w_+,w_-) = (0.075,0.25)$ (dashed red line),
    $(w_+,w_-) = (0.1,0.25)$ (solid blue line),
    and $(w_+,w_-) = (0.075,0.35)$ (dotted black line).
    Right:
    the coefficient function
    $
    \Xi(\tau) \equiv
    \bar\xi(w_+,w_-;\tau)
    \big(\frac{w_++w_-}{w_--w_+}\big)
    $
    for the same three cases.
    }
\label{Xi_bar}
\end{figure}

Turning to asymmetric collisions of shocks with differing widths,
we again find that flow within the hydrodynamic region $\mathcal R$
is very close to ideal boost invariant flow (\ref{eq:boostinvflow}),
as illustrated in Fig.~\ref{fig:flow} for rapidity $\xi \in [-1,1]$.
Moreover,
the rapidity distribution of the proper energy density
on a surface of constant proper time
$\tau \gtrsim \tau_{\rm hydro}$
continues to be  well approximated by a Gaussian but now with
a peak which is shifted away from vanishing rapidity:
\begin{equation}
    \epsilon(\xi, \tau)
    =
    A(w_+,w_-;\tau) \,
    e^{-\frac 12 \, (\xi-\bar\xi(w_+,w_-;\tau))^2/\sigma(w_+,w_-;\tau)^2}
    \,.
\label{eq:Gaussian2}
\end{equation}
Our results for the rapidity shift $\bar\xi(w_+,w_-;\tau)$
are shown in Fig.~(\ref{Xi_bar})
for three examples of asymmetric collisions.
To a good approximation, the width dependence
of the rapidity shift has a simple factorized form for
$\tau \gtrsim 2$,
\begin{equation}
    \bar\xi(w_+,w_-;\tau)
    \approx
    \Xi \> \frac{w_+ - w_-}{w_+ +w_-} \,,
\label{eq:xibarform}
\end{equation}
with a coefficient $\Xi\approx 0.07$ that
is essentially constant for $\tau>2$.

\begin{figure}
\centering
  \begin{tabular}{@{}cc@{}}
      \includegraphics[scale=0.9]{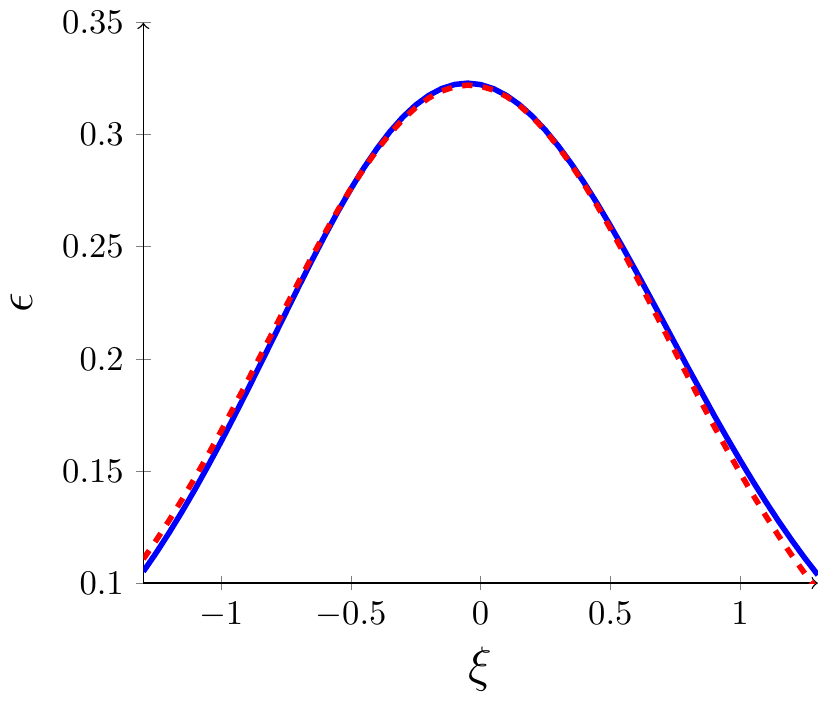} &
    \includegraphics[scale=0.9]{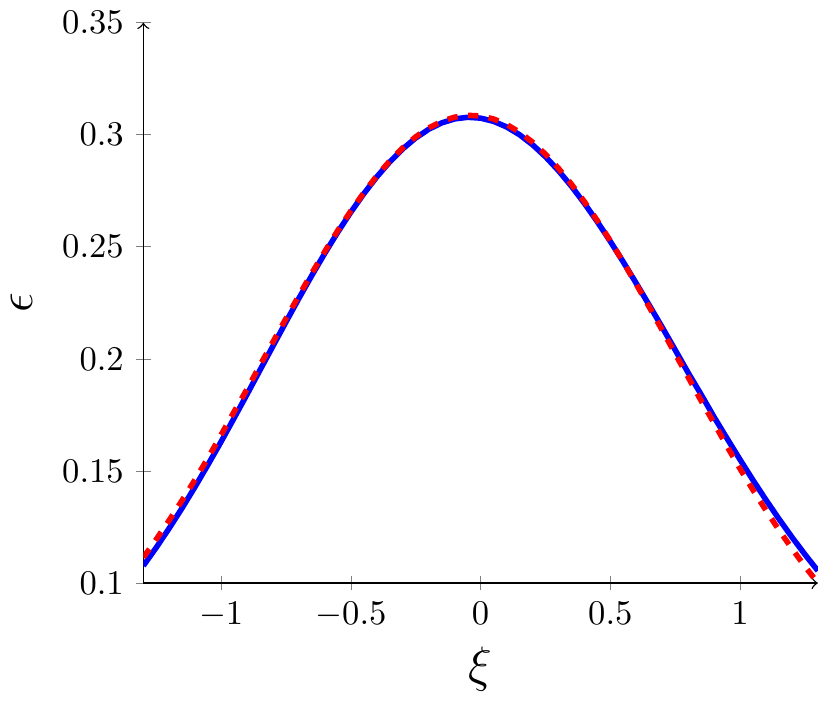}\\
    \includegraphics[scale=0.9]{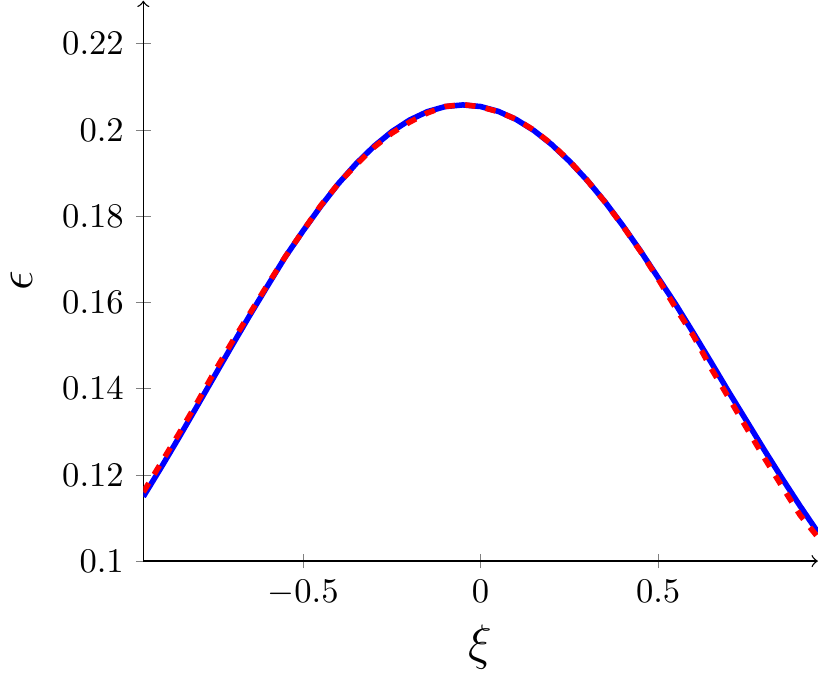} &
    \includegraphics[scale=0.9]{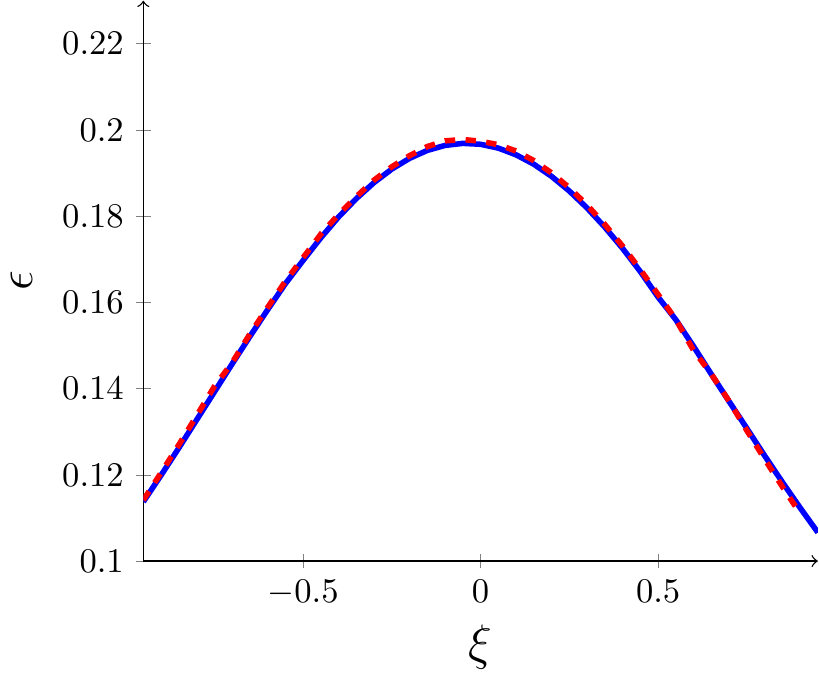} \\
  \end{tabular}
\caption
    {%
    The proper energy density $\epsilon$ as a function of rapidity
    $\xi$ at constant proper time $\tau=2$ (first row) and
    $\tau= 3$ (second row) for asymmetric collisions with
    $(w_+,w_-)=(0.075,0.35)$ (left) and $(w_+,w_-)=(0.075,0.25)$ (right)
    displayed as the solid blue curves. 
    On each plot, the red dashed curve shows
    the geometric mean of the corresponding symmetric distributions
    shifted by $\bar \xi$ as given in Eq.~(\ref{eq:xibarform}).
    Only at $|\xi| \gtrsim 1$ is a slight deviation between the two visible.
    }
\label{plots}
\end{figure}

We find that the rapidity distribution of of the proper energy
density for the asymmetric collisions is well approximated by the
shifted geometric mean of the corresponding symmetric collision
results,
\begin{equation}
    \epsilon(\xi,\tau;w_+,w_-)
    \approx
    \left[
	\epsilon(\xi-\bar\xi(w_+,w_-;\tau),\tau;w_+,w_+)
	\>
	\epsilon(\xi-\bar\xi(w_+,w_-;\tau),\tau;w_-,w_-)
    \right]^{1/2} \,.
\label{eq:simple}
\end{equation}
The efficacy of this relation is illustrated in
Fig.~\ref{plots}, which  shows the proper energy density as a function
of the  rapidity $\xi$ at proper times
$\tau =2$ (top) and $3$ (bottom)
for the case of
$(w_+,w_-) = (0.075, 0.35)$ (left) and
$(w_+,w_-) = (0.075, 0.25)$ (right).
In each plot the solid blue line shows the 
asymmetric collision result while
the red dashed curve shows the shifted geometric
mean of the corresponding symmetric collision results.
For $|\xi|< 1$ this model fits almost perfectly, while
for $|\xi|>1$ small deviations from this simple description begin to show.

To motivate a more elaborate model which captures these deviations from
the simple model (\ref{eq:simple}), let
\begin{equation}
    \langle X \rangle_p
    \equiv
    \left[
    \tfrac 12 X(w_+)^p + \tfrac 12 X(w_-)^p
    \right]^{1/p}
    \label{eq:pmean}
\end{equation}
denote the generalized mean with power $p$ of some quantity $X$ which is
observable in symmetric collisions of shocks with widths $w_+$ and $w_-$,
and then define $p[X]$ as the power for which the generalized mean
of symmetric collision results gives the result $X(w_+,w_-)$ of this
observable in an asymmetric collision with shock widths $(w_+,w_-)$.
In other words, $p[X]$ is the solution to the equation
\begin{equation}
    \langle X \rangle_{p[X]} = X(w_+,w_-) \,.
\label{def1}
\end{equation}
Recall that the geometric mean is
the $p \to 0$ limit of the generalized mean (\ref{eq:pmean}).

\begin{figure}
\includegraphics[scale=0.8]{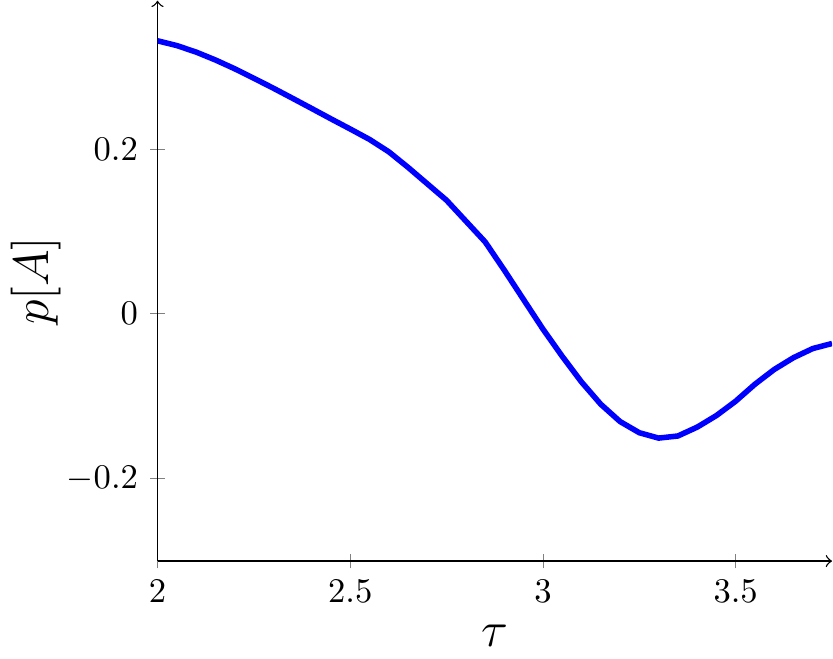}
\hfil
\includegraphics[scale=0.8]{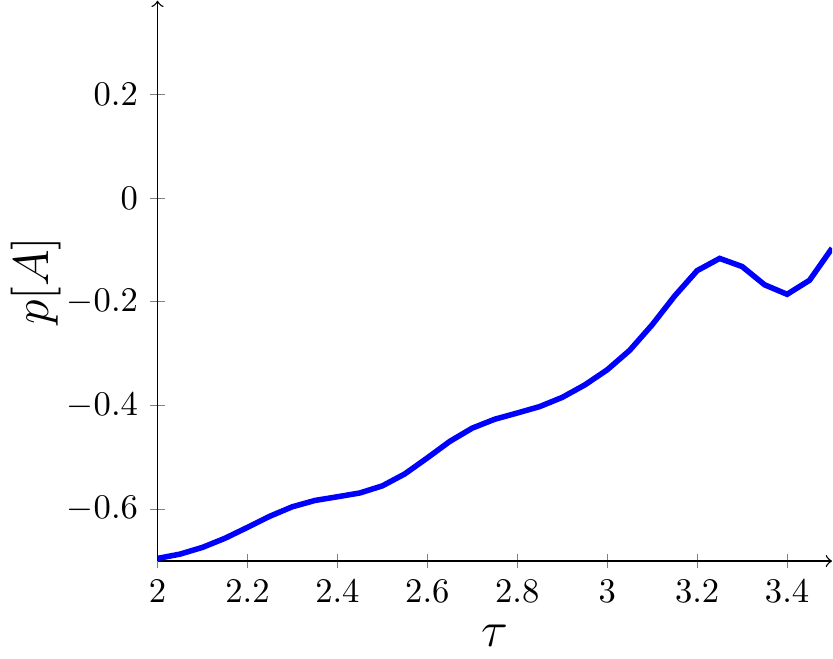}
\caption
    {%
    The exponent $p[A(\tau)]$,
    defined as the solution to relation (\ref{def1})
    for the rapidity distribution amplitude $A$,
    as a function of proper time $\tau$,
    for collisions with $(w_+,w_-)=(0.075,0.35)$ (left)
    and $(w_+,w_-)=(0.1,0.25)$ (right).
    }
\label{plots3}
\end{figure}

\begin{figure}
\includegraphics[scale=0.8]{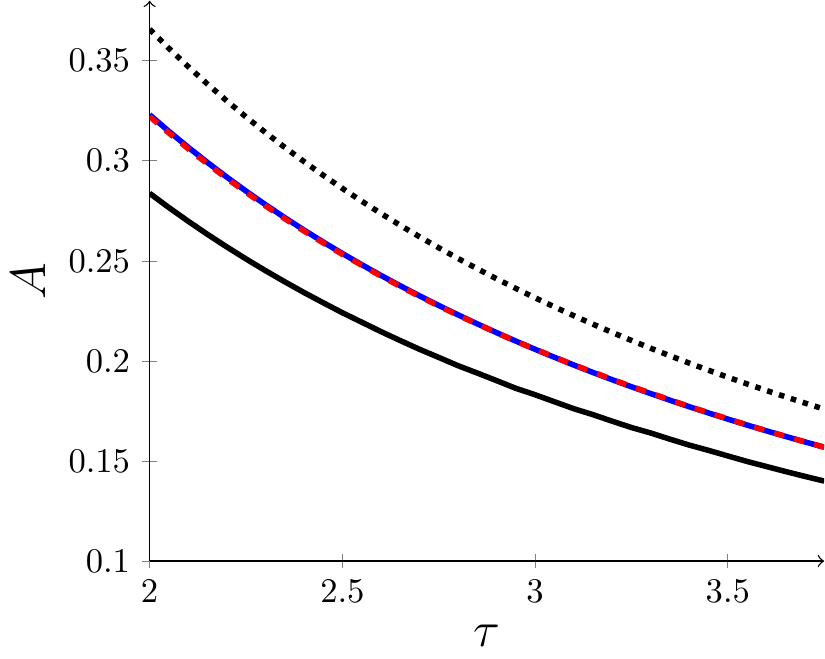}
\hfil
\includegraphics[scale=0.8]{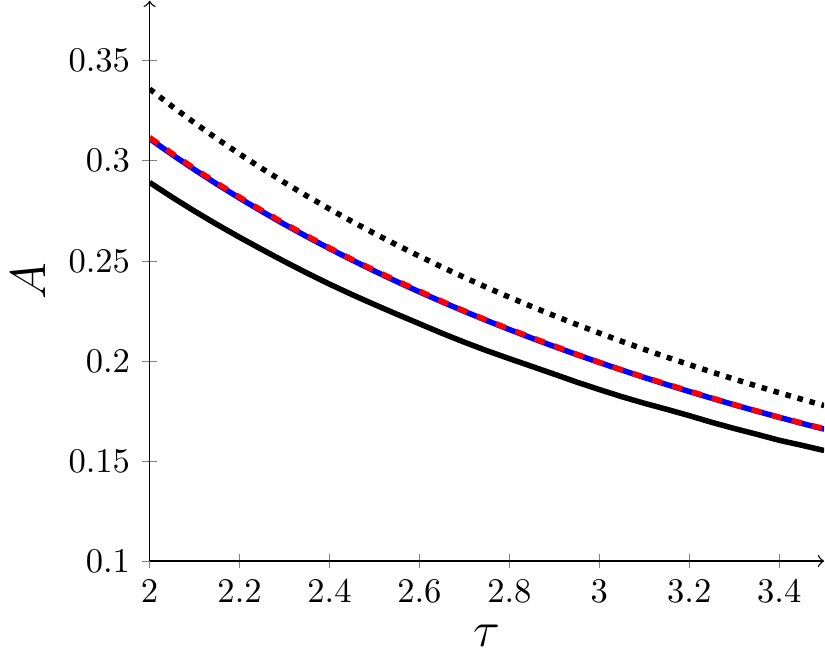}
\caption
    {%
    The amplitude $A$ (or maximum of the energy density rapidity distribution)
    for asymmetric collisions with
    $(w_+,w_-)=(0.075,0.35)$ (left) and $(w_+,w_-)=(0.1,0.25)$ (right),
    shown as the (middle) blue line.
    In each plot, the upper (dotted) line and lower (solid) line show
    the amplitude for the corresponding symmetric collision with wider
    or narrower width, respectively.
    In each plot, the red dashed line, overlaying the middle blue curve,
    shows the geometric mean of the respective symmetric collision results.
    }
\label{maxima_plot}
\end{figure}

Fig.~\ref{plots3} displays the resulting power
$p[A(\tau)]$ for the amplitude $A$ of the distributions in rapidity
of the proper energy density,
as a function of proper time $\tau$,
resulting from collisions with widths
$(w_+,w_-) = (0.075,0.35)$ on the left and $(w_+,w_-) = (0.1,0.25)$ on the right.
One sees that $p[A]$ is quite small, appearing to approach $0$ at late times.
Fig.~\ref{maxima_plot} directly compares the amplitude $A(\tau)$ for asymmetric
collisions with the geometric mean of the corresponding symmetric collision results.
For times $\tau > 2$, the difference is negligible.

To construct an improved model, let
\begin{equation}
    g_\pm(\xi,\tau) \equiv  e^{-\frac 12 \, \xi^2/\sigma(w_\pm;\tau)^2}
\end{equation}
denote the Gaussian of a symmetric collision rapidity distribution
(without the corresponding amplitude).
Then replace the geometric mean of the simple model (\ref{eq:simple})
by a biased mean of symmetric collision Gaussians,
\begin{align}
    \epsilon(w_+,w_-;\zeta,\tau)
    & \approx
    \sqrt{A(w_+;\tau)A(w_-;\tau)}
    \;
    g_+(\xi{-}\bar\xi,\tau)^{1/2-a(w_+,w_-;\xi-\bar{\xi},\tau)}
\nonumber\\ & \hspace{1.5in} \times
    g_-(\xi{-}\bar\xi,\tau)^{1/2+a(w_+,w_-;\xi-\bar{\xi},\tau)} \,,
\label{weighted_Gaussian}
\end{align}
where, once again,
$A(w_\pm,\tau)$ is the amplitude of the rapidity distribution for 
symmetric collisions of width $w_\pm$,
and the rapidity shift $\bar \xi$ is given in  Eq.~(\ref{eq:xibarform}).
If the bias $a(w_+,w_-;\xi,\tau)$ vanishes, then this form reduces
to the previous simple model (\ref{eq:simple}).

\begin{figure}
\begin{center}
\includegraphics[scale=0.75]{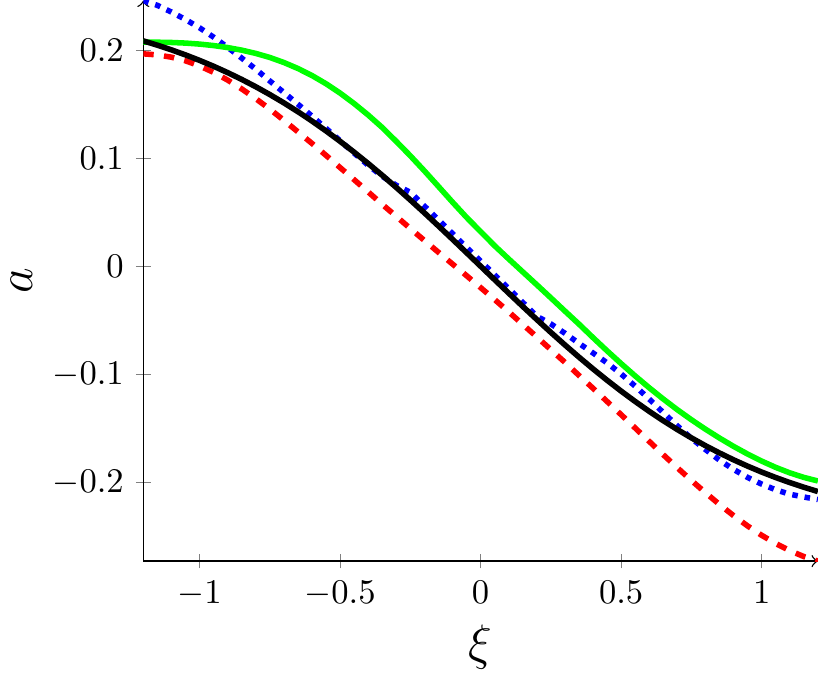}
\caption
    {
    The bias function $a$ as a
    function of rapidity $\xi$, evaluated at $\tau = 2$, for the cases
    $(w_+,w_-)=(0.075,0.35)$ (green line),
    $(w_+,w_-)=(0.075,0.25)$ (red dashed line), and
    $(w_+,w_-)=(0.1,0.25)$ (blue dotted line).
    The black line corresponds to the fitting function $
    f(\xi)=-\tanh(\xi)/4$.
    }
\label{weighted mean}
\end{center}
\end{figure}

\begin{figure}
    \centering
    \includegraphics[scale=0.9]{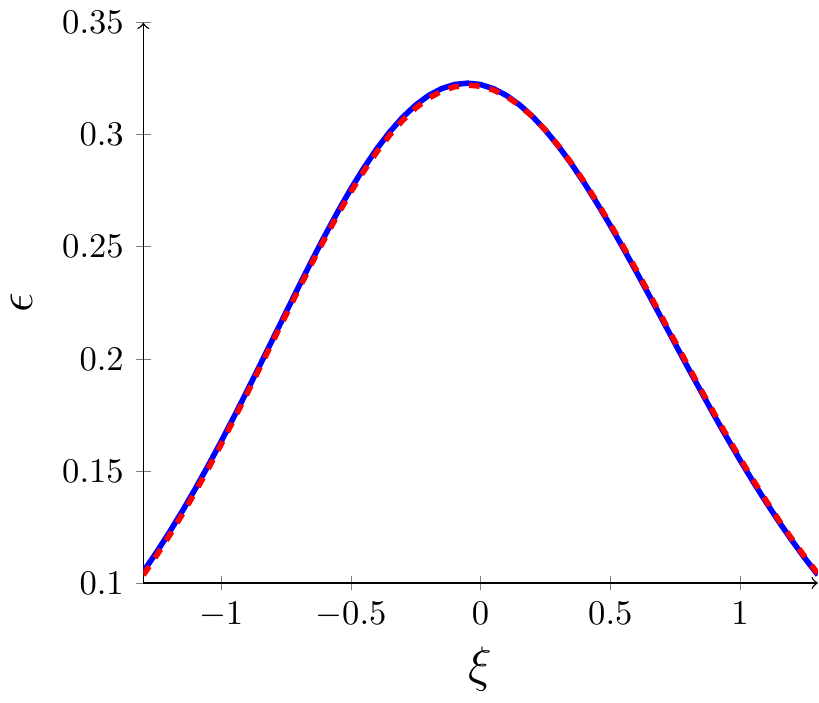} \hfill
    \includegraphics[scale=0.9]{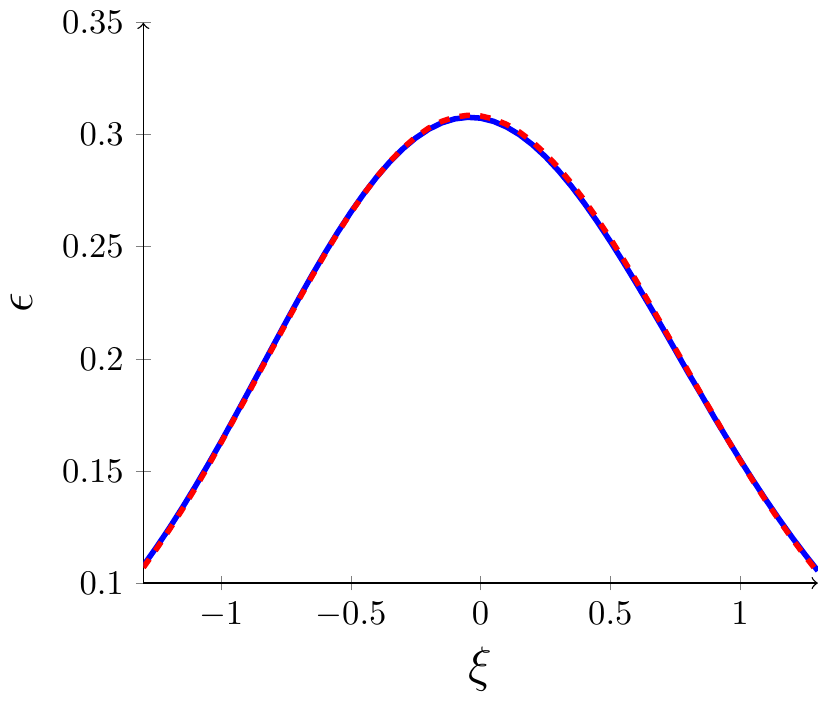}
    \\
    \includegraphics[scale=0.9]{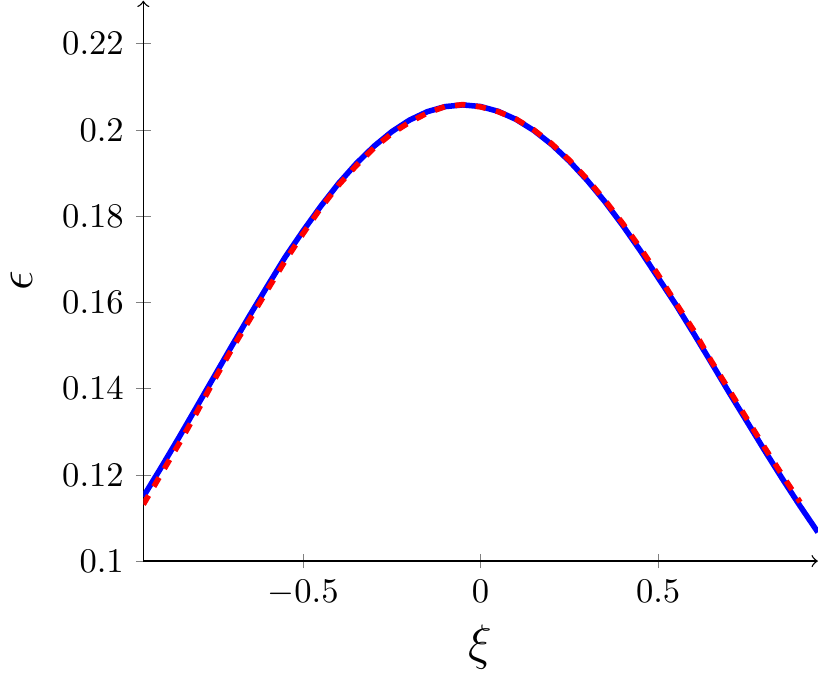} \hfill
    \includegraphics[scale=0.9]{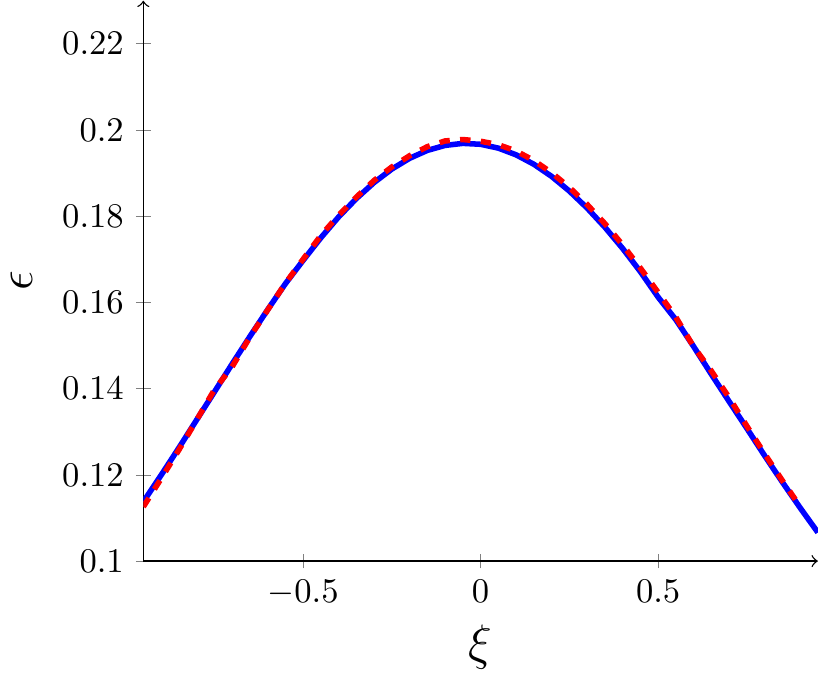}
  \caption
    {%
    The proper energy density $\epsilon$ as a function of rapidity
    $\xi$ at constant proper time $\tau=2$ (first row) and
    $\tau = 3$ (second row) for asymmetric collisions with
    $(w_+,w_-)=(0.075,0.35)$ (left) and $(w_+,w_-)=(0.075,0.25)$ (right),
    displayed by the solid blue curve.
    The overlaid red dashed curve shows
    the result obtained from the improved model (\ref{weighted_Gaussian}),
    with bias function
    $
	a(\xi) = -\tfrac 14 \, \tanh \xi
    $,
    and the
    respective Gaussian distributions for the corresponding symmetric collisions.
    }
\label{fig_impr_model}
\end{figure}

Fitting the improved model (\ref{weighted_Gaussian})
to our numerical results, we find that the resulting bias function 
$a(w_+,w_-;\xi,\tau)$ is remarkably
insensitive to the widths $(w_+,w_-)$ and is also constant
for $\tau>2$ to quite good accuracy.
Our results for $a$
are displayed in Fig.~\ref{weighted mean} for the
the cases $(w_+,w_-)=(0.075,0.35)$, $(w_+,w_-)=(0.075,0.25)$,
and $(w_+,w_-)=(0.1,0.25)$ which, as shown, differ negligibly from each other.
The resulting bias function $a(\xi)$ 
is well described by the simple universal function
\begin{equation}
\label{weights}
    a(w_+,w_-;\xi,\tau)\approx a(\xi) \equiv
    -\tfrac{1}{4}\,\tanh \xi \,.
\end{equation}
To show the efficacy of the improved model (\ref{weighted_Gaussian})
and the improvement as compared with the simple model (\ref{eq:simple}),
we again compare in Fig.~\ref{fig_impr_model} the
proper energy density rapidity distributions
from asymmetric collisions along with the predictions of the above
improved model (\ref{weighted_Gaussian}) with bias function
(\ref{weights}),
for the same cases shown earlier in Fig.~\ref{plots}.
As one sees,
the curves are now essentially indistinguishable.

\section{Discussion}
\label{sec:discussion}

The goals of this work were twofold: On the one hand by studying
and quantitatively modeling asymmetric planar shock collisions
via holography,
we aim to help bridge the gap between descriptions of very early states
of a quark gluon plasma formed during heavy ion collisions and the
later hydrodynamic regime to which the system evolves.
On the other hand, we also hope that a relatively
didactic and detailed description of the computational techniques
and software construction will be useful to others.

By studying asymmetric collisions of planar shockwaves in AdS$_5$,
we found that the simple ``universal flow'' description of 
symmetric shock collisions, found in Ref.~\cite{Chesler:2015fpa},
generalizes very naturally to asymmetric shock collisions.
Within the hydrodynamic regime, the fluid flow is extremely close
to ideal boost invariant flow, while the proper energy density
has a Gaussian rapidity dependence.
Characterizing the dependence on the amplitude and widths of the
initial shocks enabled the construction of a simple model
for mapping initial state energy density distributions to
hydrodynamic initial data, valid to leading order in
transverse gradients and having potential applicability to
non-central collisions of highly relativistic nuclei.

The hydrodynamization time  was confirmed to be
very insensitive to the widths of the colliding shocks,
and dependent only on the CM frame energy density.
Viewing asymmetric collisions of planar shockwaves as models for
``pixels'' within non-central collisions of finite sized projectiles
with large aspect ratios, this result implies that the hydrodynamization
time, measured in the lab frame, increases towards
the fringes of the almond-shaped overlap region that forms the
post-collision quark-gluon plasma.
Suitably modeling the initial state transverse energy density
as a function of the distance to the center of the Lorentz-contracted
nuclei allows one to estimate the hydrodynamization time of
different layers of the quark-gluon plasma.

Possible topics for future work include the analysis of
non-local observables and entropy production during asymmetric
collisions of planar shocks,
explicit comparison of holographic results for localized shock collisions 
with our model for hydrodynamic initial data,
and systematic incorporation of higher terms in an expansion in
transverse gradients into this model.

\acknowledgments
    {%
    This work was supported, in part,
    by the U.~S.~Department of Energy grant DE-SC\-0011637.
    LY gratefully acknowledges the hospitality of the University of Regensburg 
    and generous support from the Alexander von Humboldt foundation.
    The work of SW was supported by the research scholarship program
    of the Elite Network of Bavaria.
    }

\appendix
\addtocontents{toc}{\protect\setcounter{tocdepth}{1}}

\section{Einstein equations for planar shocks}
\label{subsection:Appendix_EE}

In this section we write down explicit forms for the Einstein
equations~(\ref{eqn:Sigma})-(\ref{eqn:d+d+Sigma}) for planar shocks.
We parametrize the rescaled spatial metric $\hat{g}$ as
\begin{align}
	\hat{g} =
	\begin{pmatrix}
		e^B & 0 & 0 \\
		0 & e^B & 0  \\
		0 & 0 & e^{-2B}
	\end{pmatrix} ,
\end{align}
with a single anisotropy function $B(u,t,z)$.
(Recall that $u \equiv 1/r$.)
The time-space metric components $F_x$ and $F_y$ vanish
due to rotational invariance in the transverse plane and,
for brevity, we write just $F$ below in place of $F_z$ for 
the remaining time-space component.
The resulting Einstein equations in our infalling coordinates
have the schematic form:
\begin{subequations}
\begin{align}
\label{eq1}\left(\partial_r^2+Q_\Sigma[B] \right)\Sigma
	& = 0\,,
\\
\label{eq2}\left( \partial_r^2+P_F[B,\Sigma]\partial_r+Q_F[B,\Sigma] \right)F
	& = S_F[B,\Sigma] \,,
\\
\label{eq3}\left( \partial_r+Q_{d_+\Sigma}[\Sigma] \right)d_+\Sigma
	& = S_{d_+\Sigma}[B,\Sigma,F] \,,
\\
\label{eq4}\left(\partial_r+Q_{d_+B}[B,\Sigma]\right)d_+B	
	& = S_{d_+B}[B,\Sigma,F,d_+\Sigma] \,,
\\
\label{eq5}\partial_r^2A
	& = S_A[B,\Sigma,F,d_+\Sigma,d_+B] \,,
\\
\label{eq6}\left( \partial_r+Q_{d_+F}[B,\Sigma] \right)d_+F
	& = S_{d_+F}[B,\Sigma,F,d_+\Sigma,d_+B,A] \,,
\\
\label{eq7}
d_+\left(
d_+\Sigma\right)
	& = S_{d^2_+\Sigma}[B,\Sigma,F,d_+\Sigma,d_+B,A] \,,
\end{align}
\end{subequations}
which specialize the general infalling form (\ref{eq:einsteineqns})
to the case of planar shocks.
Denoting radial derivatives with primes,
$f' \equiv \partial f/\partial r$, and
$f_{,z} \equiv \partial f/\partial z$ for
longitudinal derivatives,
the explicit form of the various coefficient and source functions
are as follows:%
\allowdisplaybreaks%
\begin{subequations}%
\begin{align}
	Q_{\Sigma} &= \tfrac 12 {B'^2} \,,
\\
	P_F &= 2 B'+ \Sigma' \, \Sigma^{-1} \,,
\\
	Q_F &= 2 B''
	+ (6 B' \, \Sigma' + 4 \Sigma'' ) \, \Sigma^{-1} 
	+ 3 B'^2
	- 4 \Sigma'^2 \, \Sigma^{-2} \,,
\\
	S_F &= 2 B'_{,z}
	+ (4 \Sigma'_{,z} + 6 B' \, \Sigma_{,z} ) \,  \Sigma^{-1}
	+ 3 B_{,z} \, B'
	- 4 \Sigma' \, \Sigma_{,z} \, \Sigma^{-2} \,,
\\
	Q_{d_+\Sigma} &= 2 \Sigma' \, \Sigma^{-1} \,,
\\
	S_{d_+\Sigma} &= -2 \Sigma
	+\frac{e^{2 B}}{12 \Sigma ^3}
	    \bigg\{{}
		8 \Sigma  \big[
		    F \big(2 \Sigma '_{,z} + F' \, \Sigma '\big)
		    + F^2 \, \Sigma''
		    + F_{,z} \, \Sigma'
		    + \Sigma_{,{zz}}
		\big]
\nonumber \\& {}
		+2 \Sigma \, (F \, \Sigma'+\Sigma_{,z} )
		    \big( 8 (F \, B' + B_{,z} ) + F' \big)
		-4 \big( F \Sigma '+\Sigma_{,z} \big)^2
\nonumber \\ & {}
		+\Sigma^2 \Big[
		    2 F \big( 4 B_{,z}' + B' \, (7 B_{,z}+4 F') + F'' \big)
		    + 2 F'_{,z}+4 B' F_{,z}
\nonumber \\ & \qquad {}
		    + F^2 \big(4 B''+7 B'^2\big)
		    + 4 B_{,z} F'+7 B_{,z}^2
		    + 4 B_{,zz}
		    + F'^2
		\Big]
	    \bigg\} \,,
\\
	Q_{d_+B} &=  \tfrac{3}{2} \Sigma ' \, \Sigma^{-1} \,,
\\
	S_{d_+B} &= 
	    \tfrac {3}{2} B' \, d_+\Sigma \, \Sigma^{-1}
	    -\frac{e^{2 B}}{6\Sigma ^4}\bigg\{
		 \Sigma ^2 \big(
		     2 F'_{,z} + B' F_{,z} + B_{,z} F'
		     + B_{,z}^2 + B_{,zz} + F'^2 \big)
\nonumber \\ &{}
		+ F \Big[
		    \Sigma (4 \Sigma'_{,z}+B' \, \Sigma_{,z}
			+ B_{,z} \, \Sigma ' -2 F' \, \Sigma ')
		    + 2 \Sigma^2 \big( B'_{,z} + B' (B_{,z} + F') + F''\big)
\nonumber \\ &{}
		    - 8 \Sigma' \Sigma_{,z}
		\Big]
		+ F^2 \Big[
		    \Sigma  \big(B' \Sigma '+2 \Sigma ''\big)
		    + \Sigma ^2 \big(B''+B'^2\big)-4 \Sigma '^2
		\Big]
\nonumber \\ &{}
		+ \Sigma \big(
		    B_z \Sigma_{,z} - 4 F' \Sigma_{,z}
		    + 2 F_{,z} \Sigma' + 2 \Sigma_{,zz}
		\big)
		- 4 \Sigma_{,z}^2
		\bigg\} \,,
\\
	S_{A} &=  
		\tfrac{3}{2} d_+B \, B'
		- 6 d_+\Sigma \, \Sigma ' \, \Sigma ^{-2}
		+ 2
		+
		\frac{e^{2 B}}{4 \Sigma ^4}
		\bigg\{
		    -8 \Sigma \Big[
			F (\Sigma'_{,z} + F' \, \Sigma' + F \Sigma'')
\nonumber \\ &\qquad{}
			+ F \Sigma'_{,z}
			+ F_{,z} \, \Sigma'
			+ \Sigma_{,zz}
			+ 2 (F \, B'+B_{,z}) (F \, \Sigma '+\Sigma_{,z} )
		    \Big]
		    + 4 \big(F \Sigma '+\Sigma_z\big){}^2
\nonumber \\ &{}
		    +
		    \Sigma^2 \Big[
			- 7 (F \, B' + B_{,z})^2 + F'^2
			- 4 \big(
			    F \big(2 B'_{,z}+B' \, F'\big)
			    + F^2 \, B'' + B' \, F_{,z} + B_{,zz}
			\big)
		    \Big]
		\bigg\} \,,
\\
	Q_{d_+F} &=
		2 B' - 2 \Sigma ' \, \Sigma^{-1} \,,
\\
	S_{d_+F} &=
		- 2 ( A'_{,z} + F \, A'' + A' \, F' )
		- 2 (B'- \Sigma '\, \Sigma^{-1}) (F A'+A_{,z})
		+ A' F'
\nonumber \\ &{}
		- 3 d_+B \Big[
		    F \, B' + B_{,z}
		    + 2 (F \Sigma ' + \Sigma_{,z} )\, \Sigma^{-1}
		\Big]
		- 2 \big(F \, (d_+B)' + (d_+B)_{,z}\big)
\nonumber \\ &{}
		+ d_+\Sigma \, 
		    \big(3 \Sigma  F'
		    + 4 (F \Sigma '+\Sigma_{,z}) \big) \, \Sigma^{-2}
		- 4 \big(F \, (d_+\Sigma)' + (d_+\Sigma)_{,z} \big) \, \Sigma^{-1}
		\,,
\\
	S_{d_+^2\Sigma} &= 
	    -\frac{e^{2 B}}{3 \Sigma ^2} \bigg\{
		\Sigma 
		\Big[
		    F A'_{,z}+F \big(A'_{,z}+F A''+A' F'\big)
		    + 2 \big(F A'+A_{,z}\big) \big(F B'+B_{,z}\big)
\nonumber \\ &{}
		    + A' \, F_{,z}
		    + A_{,zz}
		    - 2 d_+F \, (F \, B'+B_{,z} )
		    - (d_+F)_{,z}
		    - F (d_+F)'
		\Big]
\nonumber \\ &{}
		+ \big(F \Sigma '+\Sigma _{,z}\big) \big(F A'+A_{,z}-d_+F\big)
	\bigg\}
	- A' \, d_+\Sigma
	+ \tfrac{1}{2} \Sigma  \, d_+B{}^2 \,.
\end{align}
\end{subequations}

The condition (\ref{eqn:horizonPosition}) that the apparent horizon lie
at a fixed radial position $r_h$ has the explicit form (\ref{findh}).
For planar shocks the horizon stationarity condition
(\ref{eqn:HorizonStationary}) becomes:
\begin{align}
     0
 &=  
    A_{,zz}
    + A_{,z} \left[
	    -F'
	    - 2F \left( B' - \frac{\Sigma'}{\Sigma} \right)
	    + 2 B_{,z}
	    + \frac{\Sigma_{,z}}{\Sigma }
	\right]
\nonumber\\ & {}
    + \tfrac 14 A \,
	\bigg\{
	    F'^2
	    - 2 F'_{,z}
	    - 2 F' \left( 2B_{,z} + \frac{\Sigma_{,z}}{\Sigma} \right)
	    - 4 F_{,z} \left( B' - 3 \frac{\Sigma'}{\Sigma} \right)
	    + F^2 \bigg[ \bigg( B' - \frac{4 \Sigma'}{\Sigma} \bigg)^2
			- \frac{6 \Sigma''}{\Sigma}
		    \bigg]
\nonumber\\ & \qquad {}
	    + 4F F' \left( B' - \frac{\Sigma'}{\Sigma} \right)
	    - 4F \left(
		    B'_{,z} 
		    + 2B' B_{,z} 
		    - 6 B_{,z} \, \frac{\Sigma'}{\Sigma}
		    + B' \, \frac{\Sigma_{,z}}{\Sigma}
		    - \frac{\Sigma'_{,z}}{\Sigma}
		    - 2 \frac{\Sigma_{,z} \Sigma'}{\Sigma^2}
		\right)
\nonumber\\ & \qquad {}
	    + 4 B_{,zz}
	    + 7 (B_{,z})^2
	    + 16 B_{,z} \frac{\Sigma_{,z}}{\Sigma}
	    + \frac{8 \Sigma_{,zz}}{\Sigma}
	    - \frac{4 (\Sigma_{,z})^2}{\Sigma^2}
	    + 24 \, e^{-2 B} \left( \Sigma' \, d_+\Sigma - \Sigma^2 \right)
	\bigg\}
\nonumber\\ & {}
    + F_{,z} \, \bigg( 2 d_+B - \frac{d_+\Sigma}{\Sigma} \bigg)
    - \tfrac 32 F^2 \left(
		d_+B \, B'
		- \frac{(d_+\Sigma)'}{\Sigma}
		+ 4
		- \frac{2d_+\Sigma}{\Sigma}
		    \left( B'+\frac{\Sigma'}{\Sigma}\right)
	    \right)
\nonumber\\ & {}
    - F \, \left(
		\frac{3(d_+\Sigma)_{,z}}{\Sigma}
		+ d_+B \bigg(B_{,z}-\frac{4 \Sigma_{,z}}{\Sigma }\bigg)
		- \frac{d_+\Sigma}{\Sigma}
		    \bigg(3 F'-2 B_{,z}+\frac{2 \Sigma_{,z}}{\Sigma}\bigg)
    \right)
\nonumber\\ & {}
    + \frac{e^{2B}}{4\Sigma^2}
	\left\{
	    - 6 \, (d_+B)^2 \, \Sigma^4
	    + F^4 \bigg(B'+\frac{2\Sigma'}{\Sigma}\bigg)^2
	    + 2F^3 \bigg(B'+\frac{2 \Sigma'}{\Sigma}\bigg)
		\bigg(2 F' + B_{,z} + \frac{2\Sigma_{,z}}{\Sigma} \bigg)
	\right.
\nonumber\\ & \qquad {} \left.
	    + F^2 \bigg(
		    F'^2 + 4 B_{,z} \, F' + (B_{,z})^2
		    + (2 F'+B_{,z}) \frac{4\Sigma_{,z}}{\Sigma}
		    + \frac{4(\Sigma_{,z})^2}{\Sigma^2}
		\bigg)
	\right\}
    .
    \label{stationary}
\end{align}

\section{Transformation to infalling coordinates}
\label{subsection:Appendix_CoordinateTransformation}

The metric
\begin{align}
	ds_{\rm FG}^2 = \tilde{\rho}^{-2}
	\left(
	-d\tilde{t}^2+d\tilde{\mathbf{x}}^2_{\bot}+d\tilde{z}^2+d\tilde{\rho}^2
	\right)
	+\tilde{\rho}^2 \, h(\tilde{x}_{-}) \, d\tilde{x}_{+}^2 \,,
\end{align}
with $\tilde{x}_\pm \equiv \tilde{t} \pm \tilde{z}$,
describes a single shock moving in the $+\tilde z$ direction
using Fefferman-Graham (FG) coordinates.
It gives a solution to the Einstein equations for any
longitudinal profile function $h(\tilde x_+)$,
To construct initial data decsribing two counter-propagating shocks,
we first transform a single shock solution to the infalling
Eddington-Finkelstein (EF) form,
\begin{align}\label{eqn:AppEFmetric}
	ds_{\rm EF}^2
	=
	-2 dt\left[ u^{-2} \, du +A \, dt + F \, dz \right]
	+ \Sigma^{2} \left[ e^B d\mathbf x_\perp^2 + e^{-2B} \, dz^2 \right]
\end{align}
with the metric functions $A$, $F$, $\Sigma$, and $B$ depending only
on $t{-}z$ and the inverted radial coordinate $u \equiv 1/r$.
In other words, the components
$g_{uA}$ all vanish except for $g_{ut} = -u^{-2}$.
We relate the FG and EF coordinates according to 
\begin{equation}\label{eqn:ApParametrizationCoordTrafo}
	\tilde{t} = t+u+\alpha(t{-}z,u) \,,\quad
	\tilde{z} = z-\gamma(t{-}z,u) \,,\quad
	\tilde{\rho} = u+\beta(t{-}z,u) \,,
\end{equation}
along with
$
	\tilde{\mathbf{x}}_{\bot} = \mathbf{x}_\bot 
$.
Demanding that this change of coordinates yields a metric
of the desired form, i.e.,
\begin{align}
	(g_{\rm EF})_{CD} = 
	\frac{\partial \tilde{x}^A}{\partial x^C} \,
	\frac{\partial \tilde{x}^B}{\partial x^D} \,
	(g_{\rm FG})_{AB} \,,
\end{align}
leads to the following equations for the transformation functions,
\begin{subequations}\label{eq:AppForm}%
\begin{align}
\label{eqn:AppFormgUU}
    0 &=
    -\alpha' \left(\alpha'{+}2\right)
    +\beta' \left(\beta'{+}2\right)
    +\gamma'^2
    +H (\beta {+}u)^4 \left(\alpha'{+}\gamma'{+}1\right)^2 ,
\\
\label{eqn:AppFormgUZ}
    0 &=
    -\left(\alpha'{+}1\right) \alpha_{,z}
    +\left(\beta'{+}1\right) \beta_{,z}
    -\gamma' \left(-\gamma_{,z}{+}1\right)
    +H (\beta {+}u)^4 \left(\alpha'{+}\gamma'{+}1\right)
	\left(\alpha_{,z}{+}\gamma_{,z}{-}1\right) ,
\\
\label{eqn:AppFormgUT}
    0 &=
    \gamma' \left(2\gamma_{,z}{+}1\right)
    + \beta^2/u^2 + 2\beta/u
    -\alpha'-\gamma' \,,
\end{align}
\end{subequations}
arising from the specified values of
$(g_{\rm EF})_{uu}$,
$(g_{\rm EF})_{uz}$, and
$(g_{\rm EF})_{ut}+(g_{\rm EF})_{uz}$, respectively.
Here primes denote radial derivatives
$\partial/\partial u$,
and $H \equiv h\left(t-z+u+\alpha+\gamma\right)$.
The dependence of the functions $H$, $\alpha$, $\beta$ and $\gamma$
on their two arguments of $t{-}z$ and $u$ is suppressed for brevity.
The desired solutions to Eqs.~(\ref{eq:AppForm}) have the near-boundary
behavior
\begin{equation}
    \alpha \sim -\lambda u^2  (1{+} \lambda u)^{-1} + O(u^5) \,,\quad
    \beta \sim -\lambda u^2 (1{+} \lambda u)^{-1} + O(u^5) \,,\quad
    \gamma \sim O(u^5) \,.
\label{eq:transformbc}
\end{equation}

Following Ref.~\cite{Chesler:2013lia}, it is helpful to redefine
$\alpha$ and $\beta$ in terms of two new functions $\delta$
and $\zeta$ via
\begin{align}\label{eqn:AppDefinitionsDeltaZeta}
	\alpha = -\gamma + \beta + \delta \,,\qquad
	\beta = -\frac{u^2\zeta}{1+u\zeta} \,.
\end{align}
Inserting these expressions into equations (\ref{eq:AppForm}) and taking
appropriate linear combinations of the results
leads to a pair of coupled equations for $\delta$ and $\zeta$,
\begin{align}\label{eqn:AppEqnDelta}
	\frac{\partial\delta}{\partial u}
	- \frac{u^2}{(1+u\zeta)^2}\frac{\partial \zeta}{\partial u} = 0
	\,,\quad
	\frac{1}{u^2}\left(u^2\frac{\partial\zeta}{\partial u }\right)
	+  \frac{2 u H}{(1+u\zeta)^5)} = 0 \,,
\end{align}
plus a single decoupled equation for $\gamma$,
\begin{align}
	\frac{\partial\gamma}{\partial u} - \frac{u^2}{(1+u\zeta)^2}\frac{\partial\zeta}{\partial u }
	+ \frac{u^4}{2(1+u\zeta)^2}\left(\frac{\partial\zeta}{\partial u }\right)^2 
	+ \frac{u^4 H}{2(1+u\zeta)^6} = 0
	\,.
\end{align}

Alternatively,
starting from the infalling form (\ref{eqn:AppEFmetric}),
it is easy to show that curves along which $r \equiv 1/u$ varies
with all other coordinates held fixed are null geodesics
(with $r$ as an affine parameter).
Since coordinate transformations are isometries,
the same curves must satisfy the
geodesic equation expressed in FG coordinates, i.e.
\begin{align}
	\frac{d^2 \widetilde{Y}^A}{dr^2}
	+ \widetilde{\Gamma}^A_{BC} \,
	\frac{d \widetilde{Y}^B}{dr} \, \frac{d \widetilde{Y}^C}{dr} = 0 \,,
\label{eq:FGgeodesic}
\end{align}
where $\widetilde{\Gamma}$ are the Christoffel symbols evaluated in FG
coordinates.
The solution $\widetilde Y^A(r)$ to this geodesic equation which begins
at boundary coordinates $x^\mu = (t, \mathbf x_\perp, z)$ 
with null tangent
$\frac d{dr} \widetilde Y^A(r) = (\delta^A_t +\delta^A_\rho)$ on the boundary
directly gives the FG coordinates corresponding to the event
with EF coordinates of $x^M = (t, \mathbf x_\perp, z, 1/r)$.
Parametrizing the resulting coordinate transformation using
Eq.~(\ref{eqn:ApParametrizationCoordTrafo}),
the non-trivial $t$, $z$ and $u$
components of the geodesic equation
(\ref{eq:FGgeodesic}) lead to second order equations
for the transformation functions,
\begin{subequations}
\begin{align}\label{eqn:AppGeoT}
	\alpha''
    &=
	-\frac{2 \left(\alpha'{+}1\right)}{u}
	+ \frac{2 (\alpha '{+}1) (\beta '{+}1)}{\beta {+}u}
\nonumber\\ &\qquad {}
	+
	\tfrac 12
	\left[
	    H' (\beta {+}u)^4 (\alpha '{+}\gamma '{+}1)^2
	    +8 H (\beta '{+}1) (\beta {+}u)^3 (\alpha '{+}\gamma'{+}1)
	\right] ,
\\[5pt]
	\gamma'' 
    &=
	-\gamma' \,  \frac{2(\beta -u \beta ')}{( u (\beta +u))}
	-\tfrac{1}{2}\left[H' (\beta {+}u)^4 \left(\alpha '{+}\gamma '{+}1\right)^2
	+8 H  (\beta '{+}1) (\beta {+}u)^3 (\alpha '{+}\gamma '{+}1)\right] ,
\\
	\beta''
    &=
	-\frac{2}{u} \, \beta'
	-\frac{2}{u}
	+\frac{1}{\beta}\left[
	H  (\beta +u)^4 (\alpha '+\gamma '+1)^2
	+ (\alpha ' (\alpha '+2) +\beta '^2-\gamma '^2 -u \beta '')\right] .
\end{align}
\end{subequations}

\subsection{Near-boundary expansions}
\label{subsec:AppendixNearBoundaryExpansion}

The transformation equations (\ref{eq:AppForm}),
with boundary conditions (\ref{eq:transformbc}),
may be solved order-by-order in $u$.
If one chooses the radial shift $\lambda$ to vanish, then
\begin{equation}
    \alpha = u^5 \sum_{i=0}^\infty a_i \, u^i \,,\quad
    \beta = u^5 \sum_{i=0}^\infty b_i \, u^i \,,\quad
    \gamma = u^5 \sum_{i=0}^\infty g_i \, u^i \,,\quad
\end{equation}
while with a non-vanishing radial shift $\lambda$ one instead has
\begin{align}
    &\alpha =
    \sum_{i=1}^\infty
	u^{i+1}\lambda^i
	- \sum_{i=0}^{\infty} \> a_i \, u^{i+5} \>
	    \sum_{j=-}^{\infty} \binom{4+j+i}{j} \, \lambda^j \, u^j \,,
\label{expansion2}
\\
    &\beta =
    \sum_{i=1}^\infty 
	u^{i+1} \lambda^i
	+ \sum_{i=0}^{\infty} \> b_i \, u^{i+5} \>
	    \sum_{j=0}^{\infty} \binom{4+j+i}{j} \, \lambda^j \, u^j \,,
\label{expansion1}
\\
    &\gamma =
    \sum_{i=0}^{\infty} g_{i} \, u^{i+5} \>
	\sum_{j=0}^{\infty} \>
		\binom{4+j+i}{j} \, \lambda^j \, u^j \,.
\label{expansion3}
\end{align}
For our choice of a Gaussian profile function (\ref{eq:h}),
the first six orders of expansions coefficients are:
\begin{subequations}
\begin{align}
    &a_{0}=\frac{4 \, e^{-\frac{z^2}{2 w^2}}}{15 \sqrt {2\pi} \, w}
    \,,
\\ &
    a_1=\frac{11 \, z \, e^{-\frac{z^2}{2 w^2}}}{60 \sqrt{2 \pi } \, w^3}
    \,,
\\&
    a_2=\frac {37 \, z \, (z^2 - 3 w^2 ) \, e^{-\frac {z^2} {2 w^2}} }
	{2016 \sqrt {2 \pi } \, w^7}
    \,,
 \\&
    a_3=\frac { 768 \, w^7 e^{-\frac {z^2} {w^2}}
	+ 23\sqrt {2\pi} \, (3 w^4 - 6 w^2 z^2 + z^4 ) \,
	    e^{-\frac {z^2} {2 w^2}} } {12096\pi \, w^9} \,,
\\&
    a_4=\frac {1896 \, z w^7 e^{-\frac {z^2} {w^2}}
	    + 7\sqrt {2\pi} \, z (15 w^4 - 10 w^2 z^2 + z^4) \,
		e^{-\frac {z^2} {2 w^2}}} {21600\pi \, w^{11}} \,,
\\&
    a_5=\frac {
	(-48456 w^9 + 89736 w^7 z^2) \, e^{-\frac {z^2} {w^2}}
	- 67\sqrt {2\pi}  \, (15 w^6 - 45 w^4 z^2 + 15 w^2 z^4 - z^6 ) \,
	    e^{-\frac {z^2} {2 w^2}} } {1425600\pi \, w^{13}} \,,
\\&
    b_0=\frac {e^{-\frac {z^2} {2 w^2}}} {6 \sqrt {2 \pi } \, w} \,,
\\&
    b_1=\frac {z e^{-\frac {z^2} {2 w^2}}} {10\sqrt {2\pi} \, w^3} \,,
\\&
    b_2=-\frac {(w^2 - z^2) \, e^{-\frac {z^2} {2 w^2}}}
	{30\sqrt {2\pi} \, w^5} \,,
\\&
    b_3=-\frac {z  (3 w^2 - z^2) \, e^{-\frac {z^2} {2 w^2}}}
	{126\sqrt {2\pi} \, w^7} \,,
\\&
    b_4=\frac {116 w^7 \, e^{-\frac {z^2} {w^2}}
	+ 3\sqrt {2\pi}  (3 w^4 - 6 w^2 z^2 + z^4 ) \,
	    e^{-\frac {z^2} {2 w^2}} } {4032\pi \, w^9} \,,
\\&
    b_5=\frac {312 z \, w^7 \, e^{-\frac {z^2} {w^2}}
	+ \sqrt {2\pi} z \, (15 w^4 - 10 w^2 z^2 + z^4 )
	    e^{\frac {z^2} {2 w^2}} } {8640\pi \, w^{11}} \,,
\\
    &g_0=-\frac {e^{-\frac {z^2} {2 w^2}}} {5 \sqrt {2 \pi } \, w} \,,
\\&
    g_1=-\frac {3 z \, e^{-\frac {z^2} {2 w^2}}} {20\sqrt {2\pi} \, w^3} \,,
\\&
    g_2=\frac {5  (w^2 - z^2 ) \, e^{-\frac {z^2} {2 w^2}}}
	{84\sqrt {2\pi} \, w^5} \,,
\\&
    g_3=\frac {-11 z  (z^2 - 3 w^2 ) \, e^{-\frac {z^2} {2 w^2}}}
	{672\sqrt {2\pi} \, w^7} \,,
\\&
    g_4=\frac { -32 w^7 \, e^{-\frac {z^2} {w^2}}
	- \sqrt {2\pi}  (3 w^4 - 6 w^2 z^2 + z^4 ) \,
	    e^{-\frac {z^2} {2 w^2}} } {576\pi \, w^9} \,,
\\&
    g_5=\frac {-3408 z \, w^7 \, e^{-\frac {z^2} {w^2}}
	- 13\sqrt {2\pi} \, z \,  (15 w^4 - 10 w^2 z^2 + z^4 ) \,
	    e^{-\frac {z^2} {2 w^2}} } {43200\pi \, w^{11}} \,.
\end{align}
\end{subequations}
Inserting these expansions into expression
(\ref{eq:B}) for the metric anisotropy function
yields its near boundary expansion,
$B \sim \sum_{i=4}^\infty B_i \, u^i$, with
\begin{subequations}
\begin{align}
    B_4&=-\frac{e^{-\frac{z^2}{2 w^2}}}{3 \sqrt{2 \pi }\,  w}	\,,
\\
    B_5&=\frac{e^{-\frac{z^2}{2 w^2}} \left(20 \lambda\,   w^2-3 z\right)}{15 \sqrt{2 \pi }\,  w^3} \,,
\\
    B_6&=\frac{e^{-\frac{z^2}{2 w^2}} \left(-50 \lambda ^2 \, w^4+w^2 \, (15 \lambda \,  z+1)-z^2\right)}{15 \sqrt{2 \pi }\,  w^5} \,,
\\
    B_7&=\frac{e^{-\frac{z^2}{2 w^2}} \left(2100 \lambda ^3\,  w^6-63 \lambda \,  w^4 \, (15 \lambda \,  z+2)+3 w^2 \, z \, (42 \lambda \,  z+5)-5 z^3\right)}{315 \sqrt{2 \pi }\,  w^7} \,,
\\
    B_8&=\frac{e^{-\frac{z^2}{2w^2}}}{10080 \pi \, w^9}
	\Big[\sqrt{2 \pi }\, 
	    \big(-58800 \lambda ^4\,  w^8+7056 \lambda ^2\,  w^6\,  (5 \lambda \,  z+1)-3 w^4\,  (2352 \lambda ^2\,  z^2
\nonumber\\&\qquad\qquad{}
    +560
   \lambda \,  z  +15)+10 w^2 \, z^2\,  (56 \lambda \,  z+9)-15 z^4 \big)
   -280 \, e^{-\frac{z^2}{2 w^2}} \, w^7\Big]
    \,,
\\
    B_9&=\frac{e^{-\frac{z^2}{2w^2}}}{30240 \pi \,  w^{11}}
	\Big[
	    \sqrt{2 \pi }\,
	    (282240 \lambda ^5\,  w^{10}
		-14112 \lambda ^3\,  w^8 (15 \lambda \, z +4)
\nonumber \\ &\qquad\qquad{}
   +72 \lambda \,  w^6 (784 \lambda ^2 \, z^2+280 \lambda \,  z+15)-15 w^4 \, z \, (448 \lambda ^2 \, z^2+144 \lambda \,  z+7)
\nonumber \\&\qquad\qquad{}
    +10 w^2 \, z^3\,  (36 \lambda \,  z+7)-7 z^5)
	    + 1120 \, e^{-\frac{z^2}{2 w^2}} \, (6 \lambda \, w^9-w^7 \, z)
   \Big]
    \,,
\\
    B_{10}&=\frac{e^{-\frac{z^2}{2w^2}}}{453600 \pi \,  w^{13}}
	\Big[\sqrt{2 \pi }\,
	\Big(-6350400 \lambda ^6 \, w^{12}
	    +1905120 \lambda ^4\,  w^{10} \, (3 \lambda \,  z+1)
\nonumber \\ &\qquad\qquad{}
	-1620 \lambda ^2 w^8 \,
	    (1176 \lambda ^2 \, z^2+560 \lambda \,  z+45)
\nonumber \\ &\qquad\qquad{}
	+15 w^6 \, (20160 \lambda ^3 \, z^3+9720 \lambda ^2 \, z^2+945 \lambda \,  z+14)
\nonumber \\ &\qquad\qquad{}
    -90 w^4 \, z^2\,  (270 \lambda ^2 \, z^2+105 \lambda \,  z+7)
    +105 w^2 \, z^4\,  (9 \lambda \,  z+2)-14 z^6\Big)
\nonumber \\ &\qquad\qquad{}
       -84 \, e^{-\frac{z^2}{2 w^2}}\,  w^7 \, (5400 \lambda ^2 \, w^4
	-4 w^2 \, (450 \lambda  \, z+19)+137 z^2)\Big]
   \,.
\end{align} 
\end{subequations}

\section{Pseudo-spectral methods}
\label{subsection:Appendix_PseudospectralMethods}

Pseudo-spectral methods are a class of mean weighted residual
approximation techniques.
These methods provide highly efficient techniques for constructing
accurate numerical approximations to linear differential equations
of the form
\begin{align}
	L f = g \,,
\label{eq:Lu=f}
\end{align}
with $L$ being a linear differential operator.
One approximates the solution
$f$ by a linear combination of a finite set of of basis functions,
$f^{(N)} = \sum_{m=0}^{N-1} \> c_m \, \phi_m $,
and defines the residual 
\begin{align}
	R^{(N)} \equiv L f^{(N)} - g  \,.
\end{align}
Given some scalar product $(\cdot,\cdot)$ for the function space
in which the basis functions $\phi_m$ reside, and a chosen
sequence $\{ \xi_m \}$ of test functions, one solves for the
coefficients $\{ c_m \}$ of the spectral approximation $f^{(N)}$ by
demanding that the residual vanish on these test functions,
\begin{align}\label{eqn:ResidualVanish}
	\left(\xi_m, R^{(N)}\right) = 0  \,,
\end{align}
for $m = 0, {\cdots}, N{-}1$.
Different weighted residual methods are distinguished by the
choice of the test functions.
So-called ``pseudo-spectral'' or ``collocation'' methods are a
subclass of mean weighted residual algorithms in which one chooses
the test functions to have point support.
In, for example, one dimensional problems one chooses
\begin{equation}
    \xi_m = \delta(x - x_m) \,,
\end{equation}
for some selected set of points $\{ x_m \}$.
In other words, in pseudo-spectral approximations,
one demands that the residual vanish identically on some discrete
set of grid points distributed across the computational domain.
For a given basis set $\{ \phi_m \}$, $m = 0,{\cdots},N{-}1$,
there is a corresponding optimal choice of grid 
$\{ x_m \}$, $m = 0,{\cdots},N{-}1$, namely the abcissas of
a Gaussian quadrature integration scheme associated with
this basis set \cite{Boyd:Spectral}.
Given a choice of $N$ basis functions $\{ \phi_m \}$
and associated spectral grid $\{ x_m \}$, 
it is convenient to define ``cardinal functions'' $\{ C_m \}$
which are uniquely defined as linear combinations of these basis functions
which take the value 1 on a given grid point while vanishing on all other
points,
\begin{equation}
    C_m(x_n) = \delta_{mn} \,,\qquad m,n = 0,{\cdots},N{-}1 \,.
\label{eqn:DeltaPropCardinal}
\end{equation}
The original spectral approximation $f^{(N)} = \sum_{m=0}^{N-1} c_m \, \phi_m$
is then exactly equivalent to a linear combination of cardinal functions,
\begin{equation}
    f^{(N)} = \sum_{m=0}^{N-1} \> f_m \, C_m \,,
\label{eq:cardinalsum}
\end{equation}
in which each coefficient is the value of the function approximation
on a given grid point,
$
    f_m \equiv f^{(N)}(x_m)
$.

For one dimensional problems on a finite interval,
the most commonly used basis functions are Chebyshev polynomials.
There are actually two corresponding sets of optimal spectral grids
differing in whether the endpoints of the interval are themselves
gridpoints.
It is easiest to deal with boundary conditions when endpoints are
included in the spectral grid in which case,
for the interval $[-1,1]$, the appropriate $N{+}1$ point grid
consists of the points
\begin{equation}
    x_m = \cos(m \pi/N) \,,\qquad m = 0,{\cdots},N \,.
\label{eq:Lobatto}
\end{equation}
This is sometimes referred to as a Chebyshev-Gauss-Lobatto grid.

For problems on a periodic interval, a truncated Fourier series
provides the most useful spectral approximation.
In this case, an appropriate $2N$ point spectral grid consists of 
$2N$ evenly spaced points around the periodic interval.
So, for the interval $[0,2\pi]$, one may use
\begin{equation}
    x_m = \pi \, m/N \,,\qquad m = 0,{\cdots},2N{-}1 \,.
\label{eq:Fouriergrid}
\end{equation}

If the differential equation of interest (\ref{eq:Lu=f})
involves an $M$-th order differential operator,
$
    L = \sum_{k=0}^{M} \> p_k(x)\frac{d^k}{dx^k}
$,
then computing the values of the residual $R$ on all grid points,
using the cardinal representation (\ref{eq:cardinalsum}),
requires the evaluation of up to $M$-th order derivatives of
each cardinal function at every point on the grid.
This computation need only be performed once, and defines
a set of ``spectral differentiation matrices''
with components
\begin{equation}
    (D^{(N)}_k)_{mn} \equiv \frac {d^k C_n(x_m)}{dx^k} \,,\qquad
    m,n = 0,{\cdots},N{-}1 \,.
\end{equation}
Given these matrices, the application of the differential operator
$L$ to the spectral approximation of some function reduces to the
application of the finite matrix $L^{(N)} \equiv \| L^{(N)}_{mn} \|$,
with
\begin{equation}
    L^{(N)}_{mn} = \sum_{k=0}^M \> p_k(x_m) \, (D^{(N)}_k)_{mn} \,,\qquad
    m,n = 0,{\cdots},N{-}1 \,,
\end{equation}
to the vector of function values on the spectral grid,
$
    (L f^{(N)})(x_m) = \sum_n L^{(N)}_{mn} \, f_n
$.
Solving for (the spectral approximation to) the solution of the
differential equation (\ref{eq:Lu=f}) then reduces to the standard
algebraic problem of solving a finite system of linear equations.

\subsection{Explicit expressions}

Analytic expressions for cardinal functions and differential matrix
components, for many different sets of basis functions,
may be found in appendix~F of Ref.~\cite{Boyd:Spectral}.
For a Chebyshev basis and the Chebyshev-Gauss-Lobatto grid (\ref{eq:Lobatto}),
cardinal functions satisfying (\ref{eqn:DeltaPropCardinal}) are given by
\begin{align}
	C_j(x) =
	(-1)^{j+1}\frac{\left(1-x^2\right)}{c_j \, N^2(x-x_j)} \,
	\frac{dT_N(x)}{dx} \,,
\end{align}
where $T_k(x)$ denote Chebyshev polynomials of the first kind and
$c_j \equiv 1$ for $0 < j < N$ while $c_0 = c_N \equiv 2$.
The interior grid points lie at extrema of $T_N(x)$.
Derivatives of these cardinal functions,
evaluated on the Gauss-Lobatto grid, can be evaluated explicitly.
For the first derivative one finds \cite{Boyd:Spectral}
\begin{align}
	(D^{(N+1)}_1)_{mn}
	=
	\frac{dC_n}{dx}\bigg\vert_{x = x_m}
	= 
	\begin{cases}
		\tfrac 16 (1+2N^2) \,,	& m = n = 0 ; \\
		-\tfrac 16 (1+2N^2) \,,	& m = n = N ; \\
		-\half \, x_n/(1-x_n^2) \,, & m=n \mbox{ with } 0 < n < N ;\\
		(-1)^{m+n} c_m/[c_n(x_m-x_n)] \,, & m\neq n \,.
	\end{cases}
\label{diffmat}
\end{align}
Higher derivatives are obtained by taking powers of this matrix,
$D^{(N+1)}_k = (D^{(N+1)}_1)^k$.

For the Fourier grid with endpoint (\ref{eq:Fouriergrid}),
cardinal functions can be expressed as
\begin{align}
	C_j(x)
	=
	\frac{1}{2N} \, 
	\sin [ N(x-x_j) ]
	\cot [ \tfrac{1}{2}(x-x_j) ] \,,
\label{eq:Fouriercards}
\end{align}
and the first two differentiation matrices are given by
\begin{align}
	(D^{(2N)}_1)_{mn}
	\equiv
	\frac{dC_n}{dx}\bigg\vert_{x=x_m}
	& =
	\begin{cases}
		0 \,		& m = n ;\\
		\frac{1}{2} (-1)^{i+j}\cot
		    \left[ \frac{1}{2}\left(x_i-x_j\right) \right] \,,
		& m\neq n ,
	\end{cases}
    \label{diff_Fourier}
\\
	(D^{(2N)}_2)_{mn}
	\equiv
	\frac{d^2C_j}{dx^2}\bigg\vert_{x=x_i}
	& =
	\begin{cases}
		-\tfrac 16 (1+2N^2) \,,  & m = n ;\\
		\tfrac{1}{2}(-1)^{i+j+1}
			\csc^2\left[ \frac{1}{2}(x_i-x_j) \right] \,,
		& m\neq n .
	\end{cases}
\end{align}
A linear transformation, $y = a x + b$, may be used to convert
the above expressions into forms appropriate for arbitrary finite intervals.

\subsection{Domain decomposition}

The error in an $N$-term spectral approximation to some function $u$ decreases
exponentially with increasing $N$, provided $u$ satisfies appropriate
analyticity conditions \cite{Boyd:Spectral}.
In practice, this desirable behavior only holds as long as
there is negligible round-off error from finite precision
numerical arithmetic.
Unfortunately, differentiation matrices become increasingly ill-conditioned
as $N$ increases and this leads to progressively worsening
numerical errors in the eventual solution of the linear system.
Moreover, with Chebyshev grids, the spacing between grid points
is non-uniform and near the endpoints of the interval the
grid spacing decreases as $1/N^2$.
This rapid decrease of grid spacing can lead to short wavelength
(so-called `CFL') instabilities in time evolution problems.

These difficulties can be alleviated by partitioning the computational
domain into multiple subdomains, inside each of which one constructs
an independent spectral approximation.
This is known as domain decomposition.
In effect, one solves the differential equation of interest
independently in each subdomain with boundary conditions
which enforce appropriate continuity conditions
connecting adjoining subdomains.
Differentiation matrices for the entire domain become
block-diagonal.

For a one dimensional problem,
if one partitions the full domain into  $M$ subdomains,
and uses an $N$-point Chebyshev grid containing endpoints
within each subdomain, then each interior subdomain boundary
will appear twice in the resulting complete list of
grid points (\ref{eq:ujk}).
For a second order differential equation,
before solving the resulting
linear system,
$
    L^{(MN)} \, f^{(MN)} = g^{(MN)}
$,
one simply replaces each pair of rows which represent the same 
interior subdomain boundary by a near pair of linear equations
which encode continuity of the function,
\begin{equation}
    f^{(MN)}_{i,N-1} - f^{(MN)}_{i+1,0} = 0 \,,
\end{equation}
and of its first derivative,
\begin{equation}
    \sum_{k=0}^{N-1} \>
    (D^{(N)}_1)_{N-1,k} \, f^{(MN)}_{i,k}
    -
    (D^{(N)}_1)_{0,k} \, f^{(MN)}_{i+1,k}
    =
    0 \,.
\end{equation}
For further detail refer to Ref.~\cite{Boyd:Spectral}.

\section{Filtering}\label{Filtering}

\subsection{Longitudinal filter}

Numerically filtering the propagating data,
namely the functions $\{B,a,f,\lambda\}$,
to remove small amplitude noise, specifically
cutoff-scale rapid variations in the
longitudinal direction, is essential to achieve
stable time evolution with low background energy density,
especially for narrow shock collisions where a very fine
longitudinal grid is required.
The reasons behind this, involving spectral blocking in non-linear
equations, are discussed in Refs.~\cite{Chesler:2013lia,Boyd:Spectral}.

Such filtering must be applied carefully.
To maintain consistency  of the solution
of the nested set of Einstein equations (\ref{eq:einsteineqns}),
we only filter at the end of each time step, not
within RK4 substeps and not in between solutions of the
nested radial differential equations.

There are many ways to implement a low-pass filter.
For our periodic functions of $z$,
we use a smooth multiplicative filter in $k$-space.
A cardinal function representation using the uniform grid
(\ref{eq:Fouriergrid}) and Fourier cardinal functions
(\ref{eq:Fouriercards}),
$
    \phi(z) = \sum_{j=0}^{2N-1} \> \phi_j \, C_j(z)
$,
is exactly equivalent to a truncated Fourier series,
\begin{equation}
    \phi(z) = \sum_{k=-N}^N \>  \tilde \phi_k \, e^{i k z} \,.
\end{equation} 
with
\begin{equation}
    \tilde \phi_k
    \equiv
    \frac{(1{-}\half\delta_{|k|}^N)}{2N} \>
    \sum_{j=0}^{2N-1} \phi(\pi j/N) \, e^{-{2 \pi i  j k}/N} \,.
\label{eq:fouriercoefs}
\end{equation}

\begin{figure}
\hspace{3.5cm}
\includegraphics[scale=0.3]{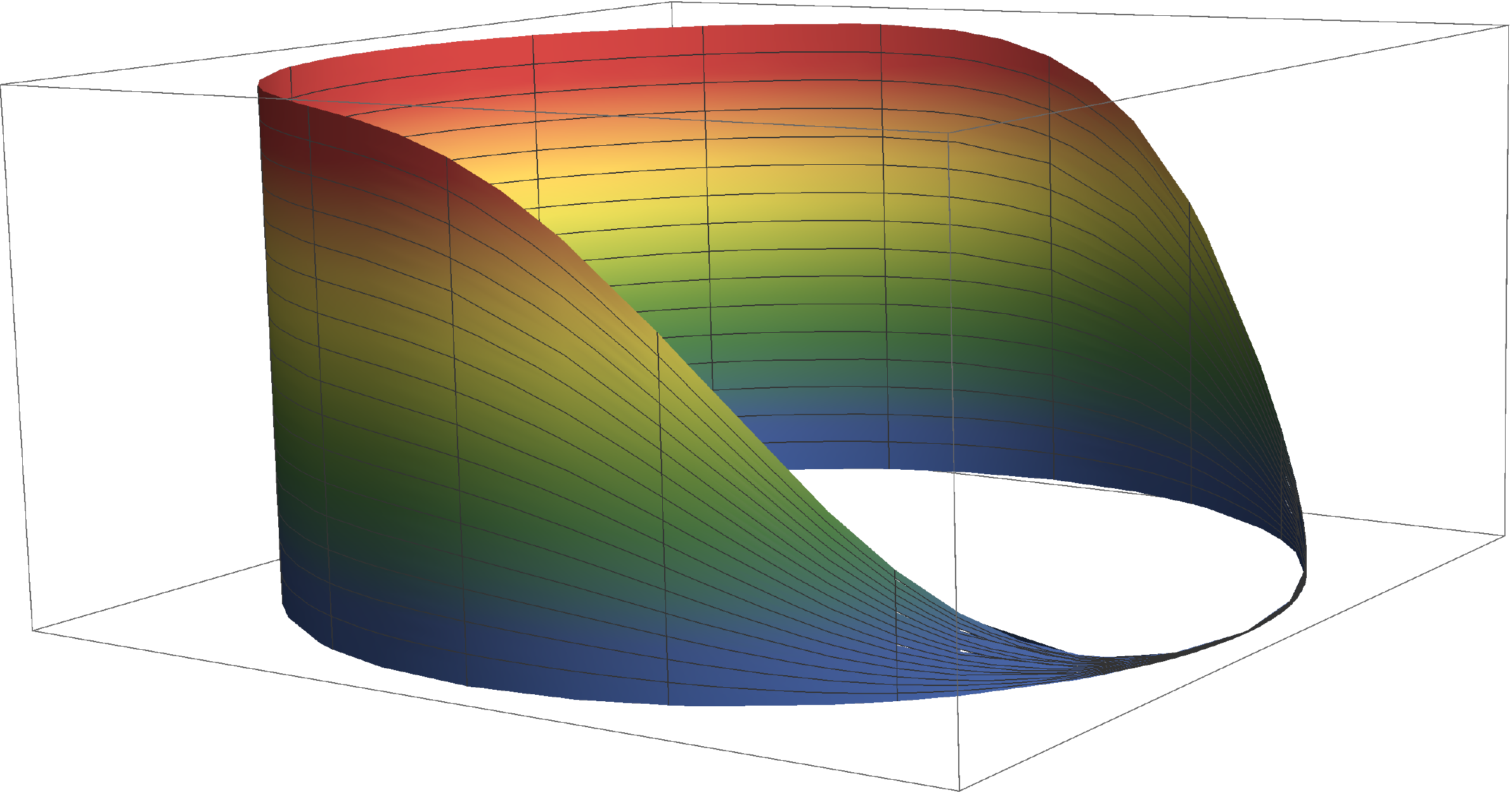}
\caption
    {%
    Visualization of our low-pass Fourier filter for periodic functions.
    Modes with wavevector $|k|/N \gtrsim \frac 23$ are suppressed.
    The filter width $\delta $ is chosen to be $1/2$.}
\end{figure}

We suppress the amplitude of modes with large $|k|$ by multiplying
the Fourier coefficients $\{ \tilde \phi_k \}$ by the filter function
\begin{equation}
    \widetilde F(k)
    \equiv
    \frac{1}{2}\left(
	1- \mathrm{erf}
	\left[\frac{2 \pi}{\delta} \left(\frac{|k|}{N}-\Lambda\right) \right]
	\right) .
\label{filter_function}
\end{equation}   
The parameter $\Lambda$ is the fractional bandwidth of the filter
while $N\delta/(2\pi)$ is the characteristic width in wavevectors
of the filter roll-off.
We chose to use $\Lambda = 2/3$ and $\delta =1/2$.
Transforming back to real space produces the smoothed function
\begin{equation}
    \overline\phi(z)
    \equiv \sum_{k=-N}^N \>  \widetilde F(k) \, \tilde \phi_k \, e^{i k z} \,.
\label{eq:filterphi}
\end{equation} 
In practice, it is convenient to compute the real-space form
of this filter by combining expressions (\ref{eq:fouriercoefs})
and (\ref{eq:filterphi}), yielding a convolution matrix which is
computed once, and then applied directly to function values on
the longitudinal grid to yield filtered functions,
\begin{equation}
	\overline \phi(z_m)
	=
	\sum_{n=0}^{2N-1} \> F_{n-m} \, \phi(z_n) \,.
\end{equation}

\subsection{Radial filter}
\label{Near boundary filter}

After transforming single shock solutions to infalling coordinates,
as described in Appendix 
\ref{subsection:Appendix_CoordinateTransformation},
we found that moderately high derivatives of the resulting anisotropy
function,
such as $\partial_u^3 \partial_z^3 b(u,z,t_0)$,
after the longitudinal filtering as just described, would still
show visible noise with rapid radial and longitudinal variation.
Such noise grows and becomes problematic upon time evolution.
To suppress such artifacts in the initial data, we perform
radial filtering on the initial anisotropy function in a matter
designed to suppress radial noise near the boundary while
simultaneously ensuring the correct near boundary asymptotic behavior.

We do this by first constructing, analytically, the near boundary
expansion of the transformation functions solving
Eqs.~(\ref{ansatz1})--(\ref{diffeqbc}), and thence the resulting
metric anisotropy function $B(u,z,t_0)$ via Eq.~(\ref{eq:B})
[or equivalently the rescaled function $b(u,z,t_0)$ defined in
Eq.~(\ref{redefinition_1})].
Explicit expressions for these near-boundary expansions appeared in
Appendix \ref{subsec:AppendixNearBoundaryExpansion}.
Let $b^{(K)}(u,z,t_0)$ denote the $K$-term partial sum of the near-boundary
expansion for the rescaled anisotropy function $b(u,z,t_0)$.
We define a correction function $\Delta^{(K)}(u,z)$ by the condition
that
\begin{equation}
    \left(\tfrac{\partial}{\partial u}\right)^m 
    \left[ b(u,z,t_0) + \Delta^{(K)}(u,z) \right]
    \Big|_{u=0}
    =
    \left(\tfrac{\partial}{\partial u}\right)^m 
	b^{(K)}(u,z,t_0) 
    \Big|_{u=0} \,,
\label{eq:Deltadef}
\end{equation}
for $m = 0,{\cdots},K{-}1$, while simultaneously requiring that
$\Delta^{(K)}(u,z)$, evaluated on the radial grid,
is only non-zero on the first $K$ radial grid points closest to the boundary.
On the left side of condition (\ref{eq:Deltadef}),
the radial derivatives are evaluated using the
spectral derivative matrix $D_u$ applied to the list of values
of $b$ and $\Delta^{(K)}$ on the radial grid.
These conditions uniquely determine the correction function $\Delta^{(K)}$
(represented on the spectral grid).
The corrected function
$
    b_{\rm improved} \equiv b + \Delta^{(K)}
$
coincides with the input function $b$ away from the boundary (by more
than $K$ grid points), while having corrected values of radial derivatives
up through order $K{-}1$ at the boundary.
Choosing $K = 7$, we find that this procedure is effective in
suppressing numerical noise in initial data up to quite high orders
of derivatives in both radial and longitudinal directions.
Unlike a conventional filter, the effect of this procedure is restricted
to a small region near the boundary.


\section{Runge-Kutta methods}
\label{RK4}

Given a first order differential equation for
some $\mathbb{R}^k$-valued function $\Phi(t)$,
\begin{equation}
    \frac d{dt} \, \Phi(t)= F(t, \Phi(t)) \,,
\label{RK_start}
\end{equation}
with initial condition $\Phi(t_0) = \Phi_0$,
the standard fourth order Runge-Kutta (RK4) algorithm iteratively constructs
an approximate solution $\tilde\Phi$
at times $t_n \equiv t_{n-1} + \delta t$,
via the recursion relation
\begin{equation}
    \tilde \Phi(t_{n+1}; \delta t) \equiv \tilde\Phi(t_{n})
    + \delta t \sum_{j=1}^4 \> b_j \, K_j(t_n) \,,
\end{equation}
where 
\begin{equation}
    K_j(t_n) \equiv
    F\big(t_n+\alpha_j \, \delta t,\,
	\tilde\Phi(t_n)+\alpha_j \, \delta t \, K_{j-1}(t_n)\big) \,,
\end{equation}
with $\tilde\Phi(t_0; \delta t) = \Phi_0$.
The coefficient vectors defining the RK4 ``substeps'' are given by
\begin{equation}
    \alpha = \left(0,\tfrac{1}{2},\tfrac{1}{2},1 \right) ,\qquad
    b = \left( \tfrac{1}{6},\tfrac{1}{3},\tfrac{1}{3},\tfrac{1}{6} \right) .
\label{RK_end}
\end{equation}

To convert this method to an adaptive stepsize integration method,
one needs a local error estimation,
i.e., some estimate of the difference between $\tilde\Phi(t_n)$ and
the desired solution $\Phi(t_n)$, assuming that $\tilde\Phi(t_{n-1})$
is correct,
together with an algorighm for decreasing or increasing the
time step $\delta t$ based on this error estimate.
The easiest way to achieve this is to compare the results of
performing a single RK4 step with timestep $\delta t$ versus
two RK4 steps with timestep $\delta t/2$.
The latter (more time consuming) calcuation will suffer from
less error due to timestep discretization and,
if $\delta t$ is sufficiently small, this difference will be a
decent approximation to the deviation from the true solution.
We define
\begin{equation}
    \text{err}(t{+}\delta t)
    =
    \big|
	\tilde\Phi(t{+}\delta t;\delta t)-\tilde\Phi(t{+}\delta t;\delta t/2)
    \big|,
\end{equation}
given a common starting value at time $t$.
For the choice of norm, we use an $L^\infty$ norm, or the maximum
over all components of $\tilde\Phi$.
If the goal of the numerical calculation is to achieve a relative
precision of $10^{-a}$, then
we adjust the time step according to
\begin{equation}
    \delta t_{n+1}= \delta t_n \bigg(\frac{10^{-a}}{\text{err}} \bigg)^{1/4}.
\label{adaptive_update}
\end{equation}
The ``learning rate" of this adaptive algorithm is governed by the
exponent ${1}/{4}$ in this rule.
This value reflects the fact that in the basic RK4 method, 
the error scales as $(\delta t)^4$ for sufficiently small timestep
$\delta t$.
In our code we did not impose minimum or maximum step sizes.
And in our specific application of transformation to infalling
coordinates, when starting with an initial step size of
$\delta u = 0.0001$ it turned out to be sufficient to
update the step size using Eq.~(\ref{adaptive_update}) and always
advance directly to the next slice without further adjustments.
More generally, it can be necessary to reject a trial step and repeat the
the calculation with a smaller step size
if the initial error exceeds the desired limit.


\begin{thebibliography}{99}

\bibitem{Chesler:2015fpa}
    P.~M.~Chesler, N.~Kilbertus and W.~van der Schee,
    {\it Universal hydrodynamic flow in holographic planar shock collisions,}
    \jhep{1511}{2015}{135},
    \arXivid{1507.02548} [hep-th].

\bibitem{Chesler:2013lia}
  P.~M.~Chesler and L.~G.~Yaffe,
  {\it Numerical solution of gravitational dynamics in asymptotically anti-de Sitter spacetimes,}
  \jhep {1407}{2014}{086}
  \arXivid{1309.1439}

\bibitem{Chesler:2010bi}
  P. M. Chesler, L. G. Yaffe,
  {\it Holography and colliding gravitational shock waves in asymptotically AdS$_5$ spacetime,}
  DOI: 10.1103/PhysRevLett.106.021601
  \arXivid{1011.3562} [hep-th]  

\bibitem{Heller:2012km}
  M.~P.~Heller, D.~Mateos, W.~van der Schee and D.~Trancanelli,
  {\it Strong coupling isotropization of non-Abelian plasmas simplified,}
  Phys.\ Rev.\ Lett.\  {\bf 108} (2012) 191601,
    \arXivid{1202.098} [hep-th].

\bibitem{Heller:2012je}
  M.~P.~Heller, R.~A.~Janik and P.~Witaszczyk,
  {\it A numerical relativity approach to the initial value problem in asymptotically Anti-de Sitter spacetime for plasma thermalization - an ADM formulation,}
  Phys.\ Rev.\ D {\bf 85} (2012) 126002,
   \arXivid{1203.0755} [hep-th].

\bibitem{Casalderrey-Solana:2013aba}
  J.~Casalderrey-Solana, M.~P.~Heller, D.~Mateos and W.~van der Schee,
  {\it From full stopping to transparency in a holographic model of heavy ion collisions,}
  Phys.\ Rev.\ Lett.\  {\bf 111} (2013) 181601,
  \arXivid{1305.4919} [hep-th].

\bibitem{Buchel:2015saa}
  A.~Buchel, M.~P.~Heller and R.~C.~Myers,
  {\it Equilibration rates in a strongly coupled nonconformal quark-gluon plasma,}
  Phys.\ Rev.\ Lett.\  {\bf 114} (2015) no.25,  251601,
   \arXivid{1503.07114} [hep-th].

\bibitem{Chesler:2015wra}
    P.~M.~Chesler and L.~G.~Yaffe,
    {\it Holography and off-center collisions of localized shock waves,}
    \jhep{1510}{2015}{070}
    \arXivid{1501.04644}

\bibitem{1307.2539}
W.~van der Schee, P.~Romatschke, S.~Pratt,
{\it A fully dynamical simulation of central nuclear collisions}
Phys.\ Rev.\ D {\bf 111} (2013) 222302,
\arXivid{1307.2539} [hep-th]

\bibitem{1507.08195}
W.~van der Schee, B.~Schenke,
{\it Rapidity dependence in holographic heavy ion collisions}
Phys.\ Rev.\ D {\bf C 92} (2015) 064907,
\arXivid{1507.08195} [hep-th]

\bibitem{1607.05273}
J.~Casalderrey-Solana, D.~Mateos, W.~van der Schee, M.~Trianae,
{\it Holographic heavy ion collisions with baryon charge}
   \jhep{1609}{2015}{108}
\arXivid{1607.05273 } [hep-th]

\bibitem{1609.03676}
C.~Ecker, D.~Grumiller, P.~Stanzer, S.~A.~Stricker, W.~van der Schee,
{\it Exploring nonlocal observables in shock wave collisions}
  \jhep{1611}{2016}{054},
\arXivid{1609.03676} [hep-th]

\bibitem{1506.02209}
P.~M.~Chesler,
{\it 
Colliding shock waves and hydrodynamics in small systems}
Phys.\ Rev.\ Lett {\bf 115} (2015) 241602,
\arXivid{1506.02209 } [hep-th]

\bibitem{1604.06439}
M.~Attems, J.~Casalderrey-Solana, D.~Mateos, D.~Santos-Oliván, C.~F. Sopuerta, M.~Triana, M.~Zilhão,
{\it Holographic collisions in non-conformal theories}
  \jhep{1701}{2016}{026},
\arXivid{1604.06439} [hep-th]

\bibitem{1601.01583}
P.~M.~Chesler,
{\it How big are the smallest drops of quark-gluon plasma?}
  \jhep{1603}{2016}{146},
\arXivid{1601.01583} [hep-th]

\bibitem{Fuini:2015hba}
  J.~F.~Fuini and L.~G.~Yaffe,
  {\it Far-from-equilibrium dynamics of a strongly coupled non-Abelian plasma with non-zero charge density or external magnetic field,}
  \jhep{1507}{2015}{116},
    \arXivid{1503.07148} [hep-th].

\bibitem{Buchel:2004di}
  A.~Buchel, J.~T.~Liu and A.~O.~Starinets,
  {\it Coupling constant dependence of the shear viscosity in N=4 supersymmetric Yang-Mills theory,}
  Nucl.\ Phys.\ B {\bf 707} (2005) 56,
  \arXivid{0406264} [hep-th].

\bibitem{Waeber:2018bea}
  S.~Waeber and A.~Sch\"afer,
  {\it Studying a charged quark gluon plasma via holography and higher derivative corrections,}
    \jhep{1807}{2018}{069},
    \arXivid{1804.01912} [hep-th].

\bibitem{Bondi:1960jsa}
  H.~Bondi,
  {\it Gravitational waves in general relativity,}
  Nature {\bf 186} (1960) no.4724, 535.

\bibitem{Sachs:1962wk}
  R.~K.~Sachs,
  {\it Gravitational waves in general relativity. 8. Waves in asymptotically flat space-times,}
  Proc.\ Roy.\ Soc.\ Lond.\ A {\bf 270} (1962) 103.

\bibitem{Boyd:Spectral}
  J.~P.~Boyd,
  {\it Chebyshev and Fourier Spectral Methods (Revised),} Dover Books on
Mathematics, Dover Publications, 2001

\bibitem{Arnold:2014jva}
  P.~Arnold, P.~Romatschke and W.~van der Schee,
  ``Absence of a local rest frame in far from equilibrium quantum matter,''
  JHEP {\bf 1410}, 110 (2014)
  \arXivid{1408.2518} [hep-th].
  
\bibitem{Janik}
  R.~A.~Janik, R.~Peschanski
  {\it Asymptotic perfect fluid dynamics as a consequence of AdS/CFT},
  Phys.\ Rev.\ D {\bf 73} (2006) 045013,
  \arXivid{hep-th/0512162}
\end{thebibliography}
\end{document}